\documentclass[a4paper,12pt]{article}

\usepackage{jcappub} 
\usepackage{aas_macros}
\usepackage[T1]{fontenc} 
\usepackage{ulem}
\usepackage{fancyvrb}
\usepackage{fancybox}

\newcommand{\bfq}{{\boldsymbol q}}
\newcommand{\bfy}{{\boldsymbol y}}
\newcommand{\bfx}{{\boldsymbol x}}
\newcommand{\bfr}{{\boldsymbol r}}
\newcommand{\bfs}{{\boldsymbol s}}
\newcommand{\bfk}{{\boldsymbol k}}
\newcommand{\bfv}{{\boldsymbol v}}
\newcommand{\bfpsi}{{\boldsymbol \psi}}
\newcommand{\deltalin}{\delta_{\rm lin}}
\newcommand{\deltaK}{\delta^{\rm K}}
\newcommand{\deltaE}{\delta^{\rm E}}
\newcommand{\deltas}{\delta_{\rm short}}

\newcommand{\deltag}{\delta_{\rm g}}
\newcommand{\deltab}{\delta_{\rm b}}
\newcommand{\gammaL}{\gamma^{\rm L}}

\newcommand{\tildePi}{\widetilde{\Pi}}
\newcommand{\hatPi}{\widehat{\Pi}}
\newcommand{\hatk}{\hat{k}}
\newcommand{\hatz}{\hat{z}}
\newcommand{\bL}{b^{\rm L}}
\newcommand{\cK}{c_{\rm K}^{\rm L}}
\newcommand{\cdeltaK}{c_{\delta{\rm K}}^{\rm L}}
\newcommand{\cKK}{c_{\rm K2}^{\rm L}}
\newcommand{\ct}{c_{\rm t}^{\rm L}}
\newcommand{\bK}{b_{\rm K}}
\newcommand{\bdeltaK}{b_{\delta {\rm  K}}}
\newcommand{\bKK}{b_{\rm K2}}
\newcommand{\bt}{b_{\rm t}}
\newcommand{\cov}{{\rm Cov}}
\newcommand{\ngal}{\overline{n}_{\rm g}}

\thisfancyput(14.5cm,0cm){{YITP-21-49}}

\title{\boldmath Lagrangian approach to super-sample effects on biased tracers at field level: galaxy density fields and intrinsic alignments}

\author[a,b]{Atsushi Taruya}
\author[c]{Kazuyuki Akitsu}
\affiliation[a]{ Center for Gravitational Physics, Yukawa Institute for Theoretical Physics, Kyoto University, Kitashirakawa Oiwakecho, Sakyo-ku, Kyoto 606-8502, Japan}
\affiliation[b]{Kavli Institute for the Physics and Mathematics of the Universe,  The University of Tokyo Institutes for Advanced Study, 
 The University of Tokyo, 5-1-5 Kashiwanoha, Kashiwa, Chiba 277-8583, Japan (Kavli IPMU, WPI)}
\affiliation[c]{ School of Natural Sciences, Institute for Advanced Study, 1 Einstein Drive, Princeton, NJ 08540, USA } 

\emailAdd{ataruya@yukawa.kyoto-u.ac.jp}
\emailAdd{kakitsu@ias.edu}

\abstract{It has been recognized that the observables of large-scale structure (LSS) is susceptible to long-wavelength density and tidal fluctuations whose wavelengths exceed the accessible scale of a finite-volume observation, referred to as the super-sample modes. The super-sample modes modulate the growth and expansion rate of local structures, thus affecting the cosmological information encoded in the statistics of galaxy clustering data. In this paper, based on the Lagrangian perturbation theory, we develop a new formalism to systematically compute the response of a biased tracer of LSS, which is expressed perturbatively in terms of the matter density field of sub-survey modes, to the super-sample modes at the field level. The formalism presented here reproduces the power spectrum responses that have been previously derived, and provides an alternative way to compute statistical quantities with super-sample modes.  
As an application, we consider the statistics of the intrinsic alignments of galaxies and halos, and derive the field response of the galaxy/halo shape bias to the super-sample modes. Possible impacts of the long-mode contributions on the covariance of the three-dimensional power spectra of the intrinsic alignment are also discussed, and the signal-to-noise ratios are estimated. 
}

\usepackage{amsmath}	
\begin{document}
\maketitle
\flushbottom

\section{Introduction}
\label{sec:intro}

The large-scale structure (LSS) observed via galaxy redshift surveys provides a wealth of cosmological information through its statistical properties. Among various LSS observables, the baryon acoustic oscillations (BAO) and redshift-space distortions (RSD) are the key to probe the cosmic expansion history and the growth of structure, with which one can not only clarify the nature of cosmic acceleration but also test the theory of gravity on cosmological scales (e.g., \cite{Weinberg_etal2013}). As increasing the statistical precision in ongoing and upcoming surveys, an accurate description of the large-scale structure is an important and critical issue toward unbiased estimation of cosmological parameters, taking also the systematics inherent in the observations into account.

Recently, it has been recognized that the galaxy distribution observed in a finite-volume survey is susceptible to long-wavelength density and tidal fluctuations whose wavelengths exceed the fundamental mode of the observed survey volume (e.g., \cite{Takada_Hu2013,Li_Hu_Takada2014a,Li_Hu_Takada2014b,Chiang_etal2014,Chiang_etal2015,Akitsu_Takada_Li2017}, see also Refs.~\cite{Hu:2002we,Hamilton_etal2006} for early works). These are called the super-sample modes, and are known to modulate the growth and expansion rate of local structures, leading to a non-trivial coupling between large- and small-scale fluctuations as a result of gravitational evolution. Then, the covariance matrix of the power spectrum is modified, and the off-diagonal components of the covariance appear non-vanishing, on top of the so-called non-Gaussian covariance, which is induced by the mode coupling due to the small-scale gravitational clustering. There has been numerous works investigating the impact of super-sample modes on the observed LSS and cosmological parameter estimation \cite{Takada_Jain2009,Takahashi2009,Akitsu_Takada_Li2017,Barreira:2017fjz,Akitsu_Takada2018,Li_Schmittfull_Seljak2018,Barreira_etal2018,KwanChan_etal2018,Akitsu_Sugiyama_Shiraishi2019,Castorina:2020blr,Digvijay_Scoccimarro2020}. However, most of these works has focused on the statistical quantities based on the Eulerian perturbation theory, and considered the power spectrum and its covariance.

Here, we are particularly interested in the responses of LSS observables to the super-sample modes at field level. This provides a basis not only to compute the responses of cross-power spectrum between different observables, but also to evaluate the higher-order statistics and their covariance matrices, furthermore giving a systematic way to calculate next-to-leading order corrections to the short-mode contributions. We shall present a perturbative framework for their systematic calculations based on the Lagrangian perturbation theory (LPT) \cite{Zeldovich1970,Shandarin_Zeldovich1989,Bouchet92,Buchert92,Catelan1995,Buchert93,Bouchet95,Bernardeau1994,Matsubara2015}. The higher-order LPT has been recently used for a consistent generation of the initial conditions including super-sample tidal fluctuations \cite{Stucker_etal2021,Akitsu_Li_Okumura2020}. We show in this paper that the systematic field-level LPT calculations reproduce the previous results in both real and redshift space using squeezed-limit $n$-point functions based on the Eulerian PT calculations.

Further, in this paper, the field-level calculation is applied to the super-sample effects on the intrinsic alignment (IA) of galaxies. The statistics of the shape and orientation of galaxies recently attract much attention as a powerful cosmological probe complementary to the conventional galaxy clustering statistics \cite{Chisari_Dvorkin2013,Schmidt_Chisari_Dvorkin2015,Chisari_etal2016,Taruya_Okumura2020,Biagetti_Orland2020,Akitsu_etal2021,Kurita_etal2021,Jingjing_etal2021,Schmidt_Jeong2012,Schmidt_Pajer_Zaldarriaga2014}. While the IA has been long thought to be a contaminant in measuring the weak gravitational lensing effect (Refs.~\cite{Troxel_Ishak2015,Joachimi_etal2015} for review), there is growing evidence that the spatial correlation of the IAs follows the gravitational tidal fields induced by the large-scale structures (e.g., \cite{Okumura_Jing_Li2009,Blazek_etal2011}), and hence it is expected to contain valuable information. Indeed, recent studies suggest that the statistics of the IAs not only provide a complementary probe, but also offer a clue to the early universe that is even difficult to probe with the galaxy clustering data \cite{Schmidt_Chisari_Dvorkin2015, Akitsu_etal2021}. Besides, based on $N$-body studies, a clear BAO feature has been found in various three-dimensional statistics related to the IAs \cite{Okumura_Taruya_Nishimichi2019,Faltenbacher_Li_Wang2012,Kurita_etal2021}. A measurement of BAO in the IA is thus beneficial, and combining it with conventional clustering statistics can significantly tighten the constraints on the cosmological parameters \cite{Taruya_Okumura2020}. Nevertheless, systematics associated with the measurement of the IA have not been fully explored. The influence of super-sample modes and its quantitative impact on the cosmological parameter estimation is one such issue to be clarified, especially in three dimension. Note that the super-sample effects on the IA have been partly considered in Ref.~\cite{Ansarifard_Movahed2020} in the context of lensing cosmology. They studied specifically the impact of the super-sample covariance for the lensing-IA angular cross correlation, including only the so-called growth effect arising from the super-sample overdensity, thus ignoring the dilation effect, the super-sample tidal field and the bias to relate the galaxy shape to the LSS. Here, taking consistently the higher-order bias for the IA into account, we derive the field-level expression of the IA including the super-sample modes, and compute the covariance of three-dimensional power spectra.

This paper is organized as follows. In Sec.~\ref{sec:formalism}, after briefly reviewing a perturbative description of the galaxy/halo density field as one of the biased tracers of large-scale structure, we present a prescription to compute the field-level response of the observable short modes to the long modes. As an explicit demonstration, we derive, in both real and redshift space, the leading-order expressions for the field-level responses of the galaxy/halo bias expansion to the super-sample modes, which are shown to consistently reproduce the previous results known in the form of power spectrum responses (see Sec.~\ref{subsubsec:real_space} and Appendix \ref{Appendix:pkred_SS-mode}). Based on the formalism, we consider the intrinsic alignments of galaxies in Sec.~\ref{sec:SSmodes_IA}, and compute perturbatively the galaxy/halo ellipticity field in three dimensional space, including the contributions from the super-sample modes (Sec.~\ref{subsec:IA_with_SS-modes}), with key expressions presented in Appendix \ref{Appendix:derivation_of_gamma_with_SS-modes}. Projecting these results onto the sky, the E-/B-mode decomposition is made (Sec.~\ref{subsec:E-/B-mode_ellipticities}), and the power spectrum responses relevant to the observations are derived (Sec.~\ref{subsec:pk_IA}). Finally, the resultant analytical expressions are used to estimate signal-to-noise ratios for the auto- and cross-power spectra of the galaxy ellipticity and density fields, including the covariance arising from the super-sample modes (Sec.~\ref{subsec:SSC_IA}). Conclusion of this paper is summarized in Sec.~\ref{sec:conclusion}, together with the discussion on possible future directions.

\section{Field-level response to super-sample modes}
\label{sec:formalism}

Throughout the paper, our primary interest is the response of the observable short-mode fluctuations to the long-wavelength perturbations which exceed the accessible scale of a finite-volume survey. In galaxy redshift surveys, the galaxy density field is a major observable, and through the perturbative description in terms of the matter density field of sub-survey modes, how such a tracer field responds to the super-survey modes at field level is the focus of this section. 
In Sec.~\ref{subsec:LPT}, we begin by reviewing the Lagrangian perturbation theory and galaxy bias expansion. Sec.~\ref{subsec:SSmodes_GG} considers the decomposition of long- and short-mode fluctuations in Lagrangian space, and presents a systematic way to compute the Eulerian-space quantities involving the long-mode contributions, keeping the long-mode contributions at the linear order. The procedure given here is then applied to the derivation of the field-level response of the galaxy bias expansion to the super-sample modes in both real (Sec.~\ref{subsubsec:real_space}) and redshift (Sec.~\ref{subsubsec:redshift_space}) space.

\subsection{Lagrangian perturbation theory}
\label{subsec:LPT}

The Lagrangian perturbation theory (LPT) is a framework to perturbatively deal with the gravitational evolution of density fields via the Lagrangian picture. The building block of the LPT is the displacement field of mass element, $\bfpsi$, which connects between the Eulerian position of each mass element, $\bfx$, and the Lagrangian position $\bfq$ (initial position of each mass element), and is given as a function of Lagrangian coordinate as follows: 
\begin{align}
\bfx=\bfq+\bfpsi(\bfq). 
\label{eq:Eulerian_Lagrangian}
\end{align}
Treating the displacement field as a small and perturbed quantity, one can expand it as  
\begin{align}
\bfpsi(\bfq)=\bfpsi^{(1)}(\bfq)+\bfpsi^{(2)}(\bfq)+\cdots. 
\label{eq:LPT_expansion}
\end{align}
Solving the equation of motion for a mass element, the $n$-th order displacement field, $\bfpsi^{(n)}$, is obtained by the recurrence relation, and in the late-time universe dominated by the growing mode, it is analytically expressed, up to the second order, as (e.g., \cite{Bouchet92,Catelan1995,Matsubara2015})
\begin{align}
 \psi_i^{(1)}(\bfq)&=-\frac{\partial_{q_i}}{\partial_q^2}\deltalin(\bfq),
\label{eq:psi_1st}
\\
 \psi_i^{(2)}(\bfq)&=-\frac{3}{14}\frac{\partial_{q_i}}{\partial_q^2}
\Bigl[\bigl\{\nabla_q\cdot\bfpsi^{(1)}(\bfq)\bigr\}^2-\partial_{q_k}\psi_\ell^{(1)}(\bfq)\partial_{q_\ell}\psi^{(1)}_k(\bfq)\Bigr],
\label{eq:psi_2nd}
\end{align}
where the operator $\partial_{q_i}$ represents the derivative with respect to the variable $q_i$, and $1/\partial_q^2$ indicates the inverse Laplacian for the variable $\bfq$ acting on the quantity at right hand side. Here and in what follows, we adopt the Einstein summation convention that the repetition of the same subscripts indicates the sum over the whole multiplet components. 

Given the mapping relation between Eulerian and Lagrangian space at Eq.~(\ref{eq:Eulerian_Lagrangian}), the density fields of the biased tracer defined in the Eulerian space, $\deltag$, is related to the one defined in the Lagrangian space, which we denote by $\deltag^{\rm L}$. The underlying assumption here is that the velocity of the tracer field follows that of the matter distribution and the number of the biased tracer is conserved in this mapping. Then, we have
\begin{align}
 \bigl\{ 1+\deltag(\bfx)\bigr\}d^3\bfx= \bigl\{ 1+\deltag^{\rm L}(\bfq)\bigr\}d^3\bfq, 
\label{eq:galaxy_Eulerian_Lagrangian}
\end{align}
which can be recast as follows:
\begin{align}
 1+\deltag(\bfx)=\int d^3\bfq \int\frac{d^3\bfk}{(2\pi)^3}\,e^{i\bfk\cdot\{\bfx-\bfpsi(\bfq)\}}\,\bigl\{1+\deltag^{\rm L}(\bfq)\bigr\}.
\label{eq:deltag_mapping}
\end{align}
The Lagrangian density field for the tracer, $\deltag^{\rm L}$, is not necessarily given by a simple linear relation to the (initial) linear density field, $\deltalin$. Rather, it is described by a general expansion form as
\begin{align}
 \deltag^{\rm L}(\bfq)=b_1^{\rm L}\,\deltalin(\bfq)+\frac{1}{2}b_2^{\rm L}\,\bigl\{\deltalin(\bfq)\bigr\}^2+\frac{1}{2}b_{s^2}^{\rm L}\,C_{ij}(\bfq)C_{ij}(\bfq)+\cdots
\label{eq:Lagrangian_bias_expansion}
\end{align}
with the scale-independent coefficients, $b_1^{\rm L}$, $b_2^{\rm L}$, and $b_{s^2}^{\rm L}$. The third term represents the tidally-induced contribution and we define
\begin{align}
 C_{ij}(\bfq)=\Bigl(\frac{\partial_{q_i}\partial_{q_j}}{\partial^2_q}-\frac{1}{3}\deltaK_{ij}\Bigr)\deltalin(\bfq).
  \label{eq:def_C_ij}
\end{align}
The Eulerian counterpart, $\deltag$, is also described perturbatively in a similar expansion form, but it is expressed in terms of the Eulerian mass density field $\delta(\bfx)$:
\begin{align}
 \deltag(\bfx)=b_1\,\delta(\bfx)+\frac{1}{2}b_2\,\bigl\{\delta(\bfx)\bigr\}^2+\frac{1}{2}b_{s^2}\,K_{ij}(\bfx)K_{ij}(\bfx)+\cdots
\label{eq:Eulerian_bias_expansion}
\end{align}
with the tidal tensor $K_{ij}$ defined by
\begin{align}
 K_{ij}(\bfx)=\Bigl(\frac{\partial_i\partial_j}{\partial^2_x}-\frac{1}{3}\deltaK_{ij}\Bigr)\delta(\bfx).
 \label{eq:def_K_ij}
\end{align}
In contrast to the Lagrangian bias expansion at Eq.~(\ref{eq:Lagrangian_bias_expansion}),  the expansion given at Eq.~(\ref{eq:Eulerian_bias_expansion}) is based on the {\it evolved} mass density field, $\delta$. This results in the non-trivial relation between Eulerian and Lagrangian bias coefficients (e.g., 
\cite{KwanChan_Scoccimarro_Sheth2012,Baldauf_Seljak_Desjacques_McDonald2012,Desjacques_Jeong_Schmidt2018}): 
\begin{align}
 (b_1,\,b_2,\,b_{s^2})=\Bigl(1+b_1^{\rm L},\,b_2^{\rm L}+\frac{8}{21}b_1^{\rm L},\,b_{s^2}^{\rm L}-\frac{4}{7}b_1^{\rm L}\Bigr).
\label{eq:bias_delta_Lagrangian_Eulerian}
\end{align}
Note that in describing the galaxy density field, there also appear the stochastic contributions that characterize the influence of small-scale perturbations on the galaxy formation \cite{Desjacques_Jeong_Schmidt2018}. Coupled to the deterministic bias terms, they produce new terms in the above expansion, but all the contributions are given in an additive manner. In Appendix \ref{Appendix:stochasticity}, taking the super-sample modes into account, we present extra terms arising from the stochasticity at the field level, and summarize their contributions to the power spectrum responses.

\subsection{Long- and short-mode decomposition}
\label{subsec:SSmodes_GG}

The expressions given in previous subsection generally involves contributions coming from both the long- and short-wavelength modes. In this paper, we are particularly interested in discriminating between these two contributions, and deriving relevant expressions for the sub-survey modes $(k\gtrsim 2\pi/L)$, involving explicitly the effect of super-survey modes $(k\lesssim 2\pi/L)$ whose wavelength exceeds a typical scale of the survey region. 


In the standard picture of structure formation, the super-survey modes are originated from the second derivative of  the large-scale gravitational potential (see  Ref.~\cite{Schmidt_Pajer_Zaldarriaga2014} for other possibilities). Their leading effects are a large-scale overdensity (or underdensity) and tidal field, both of which are coherent over the survey region. To be explicit, denoting the large-scale gravitational potential by $\Phi_{\rm long}$, the corresponding tidal tensor is decomposed into two pieces (e.g., \cite{Akitsu_Takada_Li2017}):
\begin{align}
 \partial_{q_i}\partial_{q_j}\Phi_{\rm long}(\bfq)=4\pi\,G\,\rho_{\rm m}\,a^2\Bigl(\frac{1}{3}\deltaK_{ij}\,\deltab+\tau_{ij}\Bigr)
\label{eq:tidal_tensor}
\end{align}
with the quantities $\deltab$ and $\tau_{ij}$ defined by
\begin{align}
 \deltab &= \frac{1}{4\pi\,G\rho_{\rm m}a^2}\nabla_q^2\Phi_{\rm long},
\label{eq:def_deltab}
\\
 \tau_{ij} &= \frac{1}{4\pi\,G\rho_{\rm m}a^2}\Bigl(\partial_{q_i}\partial_{q_j}-\frac{1}{3}\deltaK_{ij}\,\nabla_q^2\Bigr)\Phi_{\rm long},
\label{eq:def_tau_ij}
\end{align}
which are respectively the long-wavelength density and tidal perturbations. Throughout the paper, these modes are assumed to be constant over the survey region, and to follow the linear evolution, i.e., $\deltab,\,\tau_{ij}\propto\,D_+(t)$ with $D_+$ being the linear growth factor. Note that by definition, the tidal tensor $\tau_{ij}$ satisfies the traceless and symmetric conditions, i.e., $\tau_{ii}=0$ and $\tau_{ij}=\tau_{ji}$.

In the presence of long-mode contributions, $\deltab$ and $\tau_{ij}$, the quantities defined in Lagrangian space, i.e., $\deltalin$ and $C_{ij}$, are decomposed into  
\begin{align}
 \deltalin(\bfq)\longrightarrow \deltas(\bfq)+\deltab,
\label{eq:deltalin_short_long}
\end{align}
\begin{align}
 C_{ij}(\bfq)\longrightarrow C_{ij,{\rm short}}(\bfq)+\tau_{ij},
\label{eq:Cij_short_long}
\end{align}
where the subscript $_{\rm short}$ implies the sub-survey modes. The displacement field $\bfpsi$ also includes the contributions from super-survey modes, and is decomposed into 
\begin{align}
 \bfpsi(\bfq)\longrightarrow\bfpsi_{\rm short}(\bfq)+\bfpsi_{\rm long}(\bfq),
\label{eq:psi_short_long}
\end{align}
where, the field $\bfpsi_{\rm short}$ represents the displacement arising purely from the sub-survey modes. In general, short-mode contributions can become nonlinear through the gravitational evolution, but for our interest of the large-scale sub-survey modes, their nonlinear corrections to the displacement field are still mild, and can be perturbatively described by the LPT, with the expansion form given at Eq.~(\ref{eq:LPT_expansion}) and  solutions at Eqs.~(\ref{eq:psi_1st}) and (\ref{eq:psi_2nd}), where the linear density field $\deltalin$ in their expressions has to be interpreted  as the short-mode contribution, $\deltas$. On the other hand, while the super-survey modes $\deltab$ and $\tau_{ij}$ are dealt with linear theory, the long-mode contribution of the displacement field, $\bfpsi_{\rm long}$, receives corrections through the mode coupling with sub-survey modes. This is perturbatively described as follows:
\begin{align}
\bfpsi_{\rm long}&=\bfpsi_{\rm long}^{(1)}+\bfpsi_{\rm long}^{(2)}+\cdots;
\nonumber
\\
 \psi_{{\rm long},i}^{(1)}(\bfq)&=-\frac{1}{3}\deltab\,q_i-\tau_{ij}\,q_j,
\label{eq:psi_long1}
\\
 \psi_{{\rm long},i}^{(2)}(\bfq)&=-\frac{3}{7}\frac{\partial_{q_i}}{\partial_q^2}
\Bigl[\bigl(\nabla_q\cdot\bfpsi_{\rm long}^{(1)}\bigr)\bigl(\nabla_q\cdot\bfpsi_{\rm short}^{(1)}(\bfq)\bigr)-\partial_{q_k}\psi_{{\rm long},\ell}^{(1)}\partial_{q_\ell}\psi_{{\rm short},k}^{(1)}\Bigr]
\nonumber
\\
&=-\frac{3}{7}\frac{\partial_{q_i}}{\partial_q^2}\Bigl(\frac{2}{3}\deltab-\tau_{ij}\frac{\partial_{q_k}\partial_{q_\ell}}{\partial_q^2}\Bigr)\deltas(\bfq).
\label{eq:psi_long2}
\end{align}
In the above, Eq.~(\ref{eq:psi_long1}) leads to $\partial_j\psi_{{\rm long},i}^{(1)}=-(1/3)\deltab\,\deltaK_{ij}-\tau_{ij}$. Recall from Eq.~(\ref{eq:psi_1st}) that the first-order displacement, $\bfpsi^{(1)}$, is expressed in terms of the gravitational potential as $\psi_i^{(1)}(\bfq) =-1/(4\pi\,G\,\rho_{\rm m}\,a^2)\,\partial_{q_i}\Phi(\bfq)$, this is consistent with Eq.~(\ref{eq:tidal_tensor}). On the other hand, the expression of the second-order displacement field, $\psi_{{\rm long},i}^{(2)}$, is obtained from Eq.~(\ref{eq:psi_2nd}) by decomposing the first-order displacement into long- and short-mode contributions, and substituting Eq.~(\ref{eq:psi_long1}) into the long-wavelength contributions. In principle, given the expressions of $\bfpsi_{\rm long}^{(1)}$ and LPT expansion, this procedure can be applied to the computation of the higher-order displacement fields with long-modes. 
Note that in the above, we keep only the terms linearly proportional to $\deltab$ and $\tau_{ij}$, and the contributions of  $\mathcal{O}(\delta^2_{\rm b},\tau^2_{ij})$ are ignored, meaning in general that the $n$-th order displacement field, $\psi_{\rm long}^{(n)}$, consists of only the terms of  $\mathcal{O}(\delta_{\rm short}^{n-1}\times\delta_{\rm b})$ and $\mathcal{O}(\delta_{\rm short}^{n-1}\times\tau_{ij})$.

The expressions for the long-/short-mode decomposition given above are all the ingredients necessary to derive the perturbative expressions for the Eulerian observables with super-survey modes. Through the mapping relation given at Eq.~(\ref{eq:deltag_mapping}), we will below present an explicit calculation to derive the Eulerian real- and redshift-space tracer fields at leading order, ignoring the stochastic contributions, which are all given in an additive manner. The extra terms arising from the stochasticity, together with the corrections to the power spectrum responses, are presented in Appendix \ref{Appendix:stochasticity}.

\subsubsection{Real space}
\label{subsubsec:real_space}

Let us first consider the real-space case, and derive the expressions for a field-level response to the super-sample modes. 

To derive the leading-order expression, we perturbatively expand Eq.~(\ref{eq:deltag_mapping}). Taylor-expanding the displacement field in the exponent, the integrand of the right-hand side of Eq.~(\ref{eq:deltag_mapping}) becomes
\begin{align}
& e^{i\,\bfk\cdot\{\bfs-\bfq-\bfpsi(\bfq)\}} \, \{1+\delta_{\rm g}^{\rm L}(\bfq)\}
\nonumber
\\
&\quad \simeq e^{i\,\bfk\cdot(\bfx-\bfq)}\,\Biggl[
1+b_1^{\rm L}\deltalin(\bfq)-i\,\bfk\cdot\Bigl\{\bfpsi^{(1)}(\bfq)+\bfpsi^{(2)}(\bfq)\Bigr\}-\frac{1}{2}\bigl\{\bfk \cdot\bfpsi^{(1)}(\bfq)\bigr\}^2
\nonumber
\\
&\qquad -i\,b_1^{\rm L}\{ \bfk\cdot \bfpsi^{(1)}(\bfq)\}\deltalin(\bfq)
+\frac{1}{2}b_2^{\rm L}\,\{\deltalin(\bfq)\}^2+\frac{1}{2}b_{s^2}^{\rm L}\,C_{ij}(\bfq)C_{ij}(\bfq)+\cdots\Biggr], 
\label{eq:mapping_integrand}
\end{align}
which are relevant at the second order. Plugging this back into Eq.~(\ref{eq:deltag_mapping}), we have
\begin{align}
 1+\deltag(\bfx)&\simeq\int d^3\bfq \int\frac{d^3\bfk}{(2\pi)^3}\,e^{i\,\bfk\cdot(\bfx-\bfq)}
\nonumber
\\
&\times
\Biggl[1+b_1^{\rm L}\deltalin(\bfq)-\nabla_q\cdot\Bigl\{\bfpsi^{(1)}(\bfq)+\bfpsi^{(2)}(\bfq)\Bigr\}+\frac{1}{2}\partial_{q_i}\partial_{q_j}\Bigl\{\psi_i^{(1)}(\bfq)\psi_j^{(1)}(\bfq)\Bigr\}
\nonumber
\\
&\quad-b_1^{\rm L}\,\nabla_q\cdot\Bigl\{\bfpsi^{(1)}(\bfq)\,\deltalin(\bfq)\Bigr\}+\frac{1}{2}b_2^{\rm L}\,\{\deltalin(\bfq)\}^2+\frac{1}{2}b_{s^2}^{\rm L}\,C_{ij}(\bfq)C_{ij}(\bfq)+\cdots\Biggr].
\label{eq:deltag_mapping_expansion}
\end{align}
Here, the integration by parts has been partly performed for the terms having the explicit wavevector dependence, $\bfk$ [i.e., third, forth, and fifth terms in the bracket at Eq.~(\ref{eq:mapping_integrand})]. Then, we decompose each term in the integrand into long- and short-mode contributions, as described in previous subsection. Since we are interested in deriving the leading-order expressions including the super-survey modes, we retain only the contributions linearly proportional to $\deltas$. With a help of Eqs.~(\ref{eq:deltalin_short_long})--(\ref{eq:psi_long2}), the long-/short-mode decomposition of each term in the integrand of Eq.~(\ref{eq:deltag_mapping_expansion}) leads to 
\begin{align}
 b_1^{\rm L}\,\deltalin &\quad\longrightarrow \quad b_1^{\rm L}(\deltas+\deltab),
\\
-\nabla_q\cdot\Bigl\{\bfpsi^{(1)}+\bfpsi^{(2)}\Bigr\} &\quad \longrightarrow \quad
\deltas+\deltab+\frac{3}{7}\Bigl(\frac{2}{3}\deltab-\tau_{ij}\frac{\partial_{q_i}\partial_{q_j}}{\partial_q^2}\Bigr)\deltas,
\\
\frac{1}{2}\partial_{q_i}\partial_{q_j}\Bigl\{\psi_i^{(1)}(\bfq)\psi_j^{(1)}(\bfq)\Bigr\} &\quad\longrightarrow\quad \Bigl\{\deltab\Bigl(\frac{4}{3}+\frac{1}{3}\bfq\cdot\nabla_q\Bigr)
+\tau_{ij}\Bigl(q_i\partial_{q_j}+\frac{\partial_{q_i}\partial_{q_j}}{\partial_q} \Bigr)\Bigr\}\deltas,
\\
-b_1^{\rm L}\,\nabla_q\cdot\Bigl\{\bfpsi^{(1)}(\bfq)\,\deltalin(\bfq)\Bigr\}
&\quad\longrightarrow \quad 
b_1^{\rm L}\Bigl\{\deltab\Bigl(2+\frac{1}{3}\bfq\cdot\nabla_q\Bigr)+\tau_{ij}\,q_i\,\partial_{q_j}
\Bigr\}\deltas,
\\
\frac{1}{2}b_2^{\rm L}\{\deltalin\}^2&\quad\longrightarrow\quad b_2^{\rm L}\deltab\,\deltas,
\\
\frac{1}{2}\,b_{s^2}^{\rm L}C_{ij}C_{ij}&\quad\longrightarrow\quad
b_{s^2}^{\rm L}\,\tau_{ij}\,C_{ij}^{\rm short}.
\end{align}
Substituting these expressions into Eq.~(\ref{eq:deltag_mapping_expansion}), the integral over $\bfk$ is performed to give the Dirac delta function, $\delta_{\rm D}(\bfx-\bfq)$. Integrating further over $\bfq$,  we obtain the expression given as a function of real-space position, $\bfx$: 
\begin{align}
 \delta_{\rm g}(\bfx)&=(1+b_1^{\rm L})\Bigl\{1+\frac{34}{21}\deltab+\frac{4}{7}\tau_{ij}\frac{\partial_{x_i}\partial_{x_j}}{\partial_x^2}+\Bigl(\frac{\deltab}{3}x_i+\tau_{ij}x_j\Bigr)\partial_{x_i}\Bigr\}\deltas(\bfx)
\nonumber
\\
&+\Bigl\{\deltab\Bigl(b_2^{\rm L}+\frac{8}{21}b_1^{\rm L}\Bigr)\deltas(\bfx) + \tau_{ij}\bigl(b_{s^2}^{\rm L}-\frac{4}{7}b_1^{\rm L}\bigr)\frac{\partial_{x_i}\partial_{x_j}}{\partial_x^2}\Bigr\}\deltas(\bfx)+(1+b_1^{\rm L})\deltab.
\end{align}
Using the relation at Eq.~(\ref{eq:bias_delta_Lagrangian_Eulerian}),  the above expression is rewritten in terms of the Eulerian bias parameters [see Eq.~(\ref{eq:Eulerian_bias_expansion})]: 
\begin{align}
 \deltag(\bfx)&=b_1\Bigl\{1+\frac{34}{21}\deltab+\frac{4}{7}\tau_{ij}\,\frac{\partial_{x_i}\partial_{x_j}}{\partial_x^2}+\Bigl(\frac{\deltab}{3}\,x_i+\tau_{ij}x_j\Bigr)\frac{\partial}{\partial x_i}\Bigr\} \deltas(\bfx)
\nonumber
\\
&+\Bigl\{b_2\,\deltab+b_{s^2}\,\tau_{ij}\,\frac{\partial_{x_i}\partial_{x_j}}{\partial_x^2}\Bigr\}\deltas(\bfx)+b_1\deltab.
\label{eq:deltag_real_space}
\end{align}
Note here that the last term at the right-hand side is merely constant. While we will below omit it in the field-level expression, this DC mode contribution may not be negligible in general, and can affect the responses of the multi-point statistics to the super-sample modes. Keeping or dropping this DC mode contribution is related to how we choose the density estimator in practical measurements. We will come back this issue when we discuss the power spectrum responses in the local mean (see Sec.~\ref{subsec:pk_IA}).

Eq.~(\ref{eq:deltag_real_space}) describes how the short-mode fluctuations is modulated due to the  super-survey modes at the field level. This is a generalization of the result in Ref.~\cite{KwanChan_etal2018} to include the galaxy bias. To be precise, Ref.~\cite{KwanChan_etal2018} derived the Fourier-space expression for the matter fluctuations based on the Eulerian perturbation theory, in which apparently divergent shift terms arising from the super-sample modes need to be first identified, and to be removed in order to get a correct field-level expression. In our treatment, starting from the Lagrangian space and following the long- and short-mode decomposition rule in Sec. 2.2, there appear no divergent terms to be subtracted, and hence the calculation is straightforward to derive the field-level responses.

To see if the expression at Eq.~(\ref{eq:deltag_real_space}) is consistent with previous works, we consider the Fourier transform of Eq.~(\ref{eq:deltag_real_space}): 
\begin{align}
 \deltag(\bfk)=\Bigl[b_1+\deltab\,\widehat{a}_\delta(\bfk)+\tau_{ij}\,\widehat{b}_{\tau,ij}(\bfk)\Bigr]\deltas(\bfk)
\label{eq:deltag_real_SS-modes}
\end{align}
with the operators acting on $\deltas$, $\widehat{a}_\delta$ and $\widehat{b}_{\tau,{ij}}$, respectively given by 
\begin{align}
 \widehat{a}_\delta(\bfk)&= b_1 \Bigl\{\frac{34}{21}-\Bigl(1+\frac{1}{3}\bfk\cdot\nabla_k\Bigr)    \Bigr\} +b_2,
\label{eq:hat_a_delta}\\
 \widehat{b}_{\tau,ij}(\bfk)&=b_1\Bigl\{\frac{4}{7}\hatk_i\hatk_j-k_i \partial_{k_j}\Bigr\}+b_{s^2}\hatk_i\hatk_j. 
\label{eq:hat_b_tau}
\end{align}
Note that in deriving the above expression, we have used the following relation:
\begin{align}
 x_j\frac{\partial}{\partial x_i}\deltas(\bfx)=-\int\frac{d^3\bfk}{(2\pi)^3}\,e^{i\,\bfk\cdot\bfx}\,\Bigl\{\deltaK_{ij}+k_i\frac{\partial}{\partial k_j}\Bigr\}\deltas(\bfk)
\label{eq:eq:real-Fourier_x_j_x_i}.
\end{align}

We then compute the power spectrum defined by
\begin{align}
 \langle\deltag(\bfk)\deltag(\bfk')\rangle=(2\pi)^3\delta_{\rm D}(\bfk+\bfk')\,P_{\rm gg}(\bfk).
\label{eq:def_power_gg}
\end{align}
Keeping the terms in linear order of $\deltab$ and $\tau_{ij}$, substituting Eq.~(\ref{eq:deltag_real_SS-modes})  into the above leads to
\begin{align}
&  \langle\deltag(\bfk)\deltag(\bfk')\rangle
\nonumber
\\
&\quad=(2\pi)^3\delta_{\rm D}(\bfk+\bfk')\,\Biggl[b_1^2+2b_1\Bigl\{b_1\Bigl(\frac{34}{21}-1\Bigr)+b_2\Bigr\}\,\deltab+
2b_1\Bigl(\frac{4}{7}+b_{s^2}\Bigr)\hatk_i\hatk_j\tau_{ij}\Biggr]\,P_{\rm gg}(\bfk)
\nonumber
\\
&\quad\qquad
-b_1^2\Bigl\{\frac{1}{3}\,\deltab\,\bigl(\bfk\cdot\nabla_k+\bfk'\cdot\nabla_{k'}\bigr)+\tau_{ij}\,
\bigl(k_i\partial_{k_j}+k_i'\partial_{k_j'}\bigr)\Bigr\}\langle\deltas(\bfk)\deltas(\bfk')\rangle. 
\label{eq:real-space_pk_ensemble}
\end{align}
In Eq.~(\ref{eq:real-space_pk_ensemble}), the second line at the right-hand side include the operators acting on the ensemble-averaged quantities, for which we use the following relation (see Eq.~(\ref{eq:master_relation1}) with $A=1$ in Appendix \ref{Appendix:formulas}): 
\begin{align}
\bigl(k_i\partial_{k_j}+k_i'\partial_{k_j'}\bigr)\langle\deltas(\bfk)\deltas(\bfk')\rangle 
\,\,\longrightarrow \,\,
(2\pi)^3\delta_{\rm D}(\bfk+\bfk')\Bigl\{-\deltaK_{ij}+\hatk_i\hatk_j\frac{d\ln P_{\delta\delta}(k)}{d\ln k}\Bigr\}P_{\delta\delta}(k),
\label{eq:derivative_ensemble_formula1}
\end{align}
where the quantity $P_{\delta\delta}$ is the matter power spectrum of the sub-survey fluctuations $\deltas$, as is similarly defined at Eq.~(\ref{eq:def_power_gg}): 
\begin{align}
\langle\deltas(\bfk)\deltas(\bfk')\rangle=(2\pi)^3\delta_{\rm D}(\bfk+\bfk')\,P_{\delta\delta}(k). 
\label{eq:def_pk_matter}
\end{align}
Note that at the leading order, the spectrum $P_{\delta\delta}$ is nothing but the linear power spectrum. Making use of this relation, the leading-order (tree-level) expression of the power spectrum with super-sample modes is obtained, and we have 
\begin{align}
 P_{\rm gg}(\bfk) &= 
\Biggl[b_1^2+ b_1\,\Bigl\{\frac{47}{21}b_1+2\,b_2-\frac{b_1}{3}\frac{d\ln P_{\delta\delta}(k)}{d\ln k}\Bigr\}\,\deltab
\nonumber
\\
&
\qquad\qquad\qquad\qquad+b_1\Bigl\{\frac{8}{7}b_1+2\,b_{s^2}-b_1\frac{d\ln P_{\delta\delta}(k)}{d\ln k}\Bigr\}\,\tau_{ij}\hatk_i\hatk_j\Biggr]\,P_{\delta\delta}(k),
\label{eq:real-space_pk_SS-modes}
\end{align}
which consistently reproduces the result given in previous works (e.g., \cite{Akitsu_Takada2018, Akitsu_Sugiyama_Shiraishi2019, Akitsu_Takada_Li2017, Li_Schmittfull_Seljak2018, Chiang:2018mau, Schmidt_Pajer_Zaldarriaga2014}). This is the power spectrum of the galaxy number density field normalized by the global mean density. As we will discuss later, there appears additional correction when we define the galaxy density field with the local mean (see Sec.~\ref{subsec:pk_IA}).

\subsubsection{Redshift space}
\label{subsubsec:redshift_space}

The procedure given in previous subsection is also extended to the galaxy density field defined in redshift space. The redshift-space position $\bfs$ is related to the real-space position $\bfx$ through
\begin{align}
 \bfs = \bfx + \frac{1}{a\,H}\,(\bfv\cdot\hat{z})\,\hat{z}.
\end{align}
Throughout the paper, we will work with the distant-observer limit, in which the line-of-sight direction is fixed to a specific direction given by the unit vector $\hat{z}$.  Recall that the velocity field $\bfv$ is given by $\bfv=a\,(d\bfpsi/dt)$, the above relation is expressed in terms of the displacement field as 
\begin{align}
 \bfs = \bfq + \bfpsi^{\rm S}(\bfq)\,;\quad
\bfpsi^{\rm S}(\bfq)\equiv \bfpsi(\bfq)+\frac{1}{H}\frac{d\psi_z(\bfq)}{dt}\,\hatz.
\label{eq:displacement_z-space}
\end{align}
Thus, the perturbative expansion of the redshift-space displacement field $\bfpsi^{\rm S}$ leads to 
\begin{align}
 \psi_i^{{\rm S}\,(n)}(\bfq)&=\bigl(\deltaK_{ij}+n\,f\,\hatz_i\hatz_j\bigr)\psi_j^{(n)}(\bfq)
\nonumber
\\
&\equiv R_{ij}^{(n)}\psi_j^{(n)},
\label{eq:n-th_order_displacement_z-space}
\end{align}
where the quantity $f$ is the linear growth rate defined by $f=d\ln D_+/d\ln a$. With this displacement field, the redshift-space galaxy density field, which we denote by $\deltag^{\rm(S)}$, is expressed as follows, as similarly given in real space at Eq.~(\ref{eq:deltag_mapping}):
\begin{align}
1+ \deltag^{\rm(S)}(\bfs) =\int d^3\bfq \int \frac{d^3\bfk}{(2\pi)^3}\,e^{i\,\bfk\cdot\{\bfs-\bfpsi^{\rm S}(\bfq)\}}\,\bigl\{1+\deltag^{\rm L}(\bfq)\bigr\}.
\label{eq:deltag_redshift}
\end{align}

Then, taking Eq.~(\ref{eq:deltag_redshift}) as a new starting point,  we perturbative evaluate the right-hand side, and decompose all the fields in the expression into long- and short-mode contributions, as we demonstrated in the real-space case. The calculation is slightly intricate but rather straightforward. We present the derivation in Appendix \ref{sec:deltag_redshift_SS-mode}.  The resultant redshift-space galaxy density field at the leading order, involving the super-sample modes, is expressed in Fourier space as \footnote{Here, the DC mode contributions, which appear in the derivation (see Eq.~(\ref{eq:redshift_space_density_with_DC_modes})), are dropped in the final expression.}
\begin{align}
\deltag^{\rm (S)}(\bfk)=\Bigl[b_1+f\,\mu_k^2+\deltab\bigl\{ \widehat{a}_\delta(\bfk)+f\,\widehat{a}^{\rm(S)}_\delta(\bfk)\bigr\}+\tau_{ij}\bigl\{
\widehat{b}_{\tau,ij}(\bfk)+f\,\widehat{b}_{\tau,ij}^{\rm(S)}(\bfk) \bigr\}\Bigr]\deltas(\bfk)
\label{eq:deltag_redshift_SS-modes}
\end{align} 
with the quantity $\mu_k$ being the directional cosine between the wavevector and the line-of-sight direction, i.e., $\mu_k=\hat{k}\cdot\hat{z}$. 
Here, the operators $\widehat{a}_\delta$ and $\widehat{b}_{\tau,ij}$ are those given in the real-space density field, at Eqs.~(\ref{eq:hat_a_delta}) and (\ref{eq:hat_b_tau}). The operators $\widehat{a}_{\delta}^{\rm(S)}$ and $\widehat{b}_{\tau,ij}^{\rm(S)}$ are the new contributions in redshift space, defined by 
\begin{align}
&\widehat{a}_{\delta}^{\rm(S)}(\bfk)=\mu_k^2\Bigl\{\frac{5}{21}+\frac{f}{3}(2\mu_k^2-1)\Bigr\}
+b_1\Bigl(\mu_k^2-\frac{1}{3}k_z\partial_{k_z}\Bigr)
-\frac{\mu_k^2}{3}\Bigl\{(\bfk\cdot\nabla_k)+f\,k_z\partial_{k_z}\Bigr\},
\label{eq:Y_hat}
\\
&\widehat{b}_{\tau,ij}(\bfk) = -(b_1+f\,\mu_k^2)\,\deltaK_{iz}k_z\partial_{k_j}+2\,f\,\mu_k^3\hatk_i\deltaK_{jz}+\mu_k^2\Bigl\{-f\,\deltaK_{iz}\deltaK_{jz}+\frac{8}{7}\hatk_i\hatk_j-k_i\partial_{k_j}\Bigr\}.
\label{eq:Q_ij}
\end{align}
The expression given above is one of the new results in this paper. The result at the field level consistently reproduces the power spectrum responses to the super-sample modes known in previous works (e.g., \cite{Akitsu_Takada2018, Li_Schmittfull_Seljak2018, Chiang:2018mau,Akitsu_Sugiyama_Shiraishi2019}). For the sake of the completeness, in Appendix \ref{Appendix:pkred_SS-mode}, we present the derivation of the power spectrum expression based on Eq.~(\ref{eq:deltag_redshift_SS-modes}).

Note that the calculation and procedure presented in this section essentially gives the same expressions for the power spectrum response as those previously shown based on the squeezed limit of the bispectrum or the collapsed trispectrum. In this respect, the results presented so far are not quite new. However, one advantage of our treatment is to derive directly the field-level expression, from which we can compute any statistical quantity involving the super-sample modes. Making use of this advantage, in next section, we will consider the intrinsic alignment of galaxies, and compute the cross-power spectra between the intrinsic alignments and density fields.

\section{Impacts of super-sample modes on galaxy ellipticity fields}
\label{sec:SSmodes_IA}

In this section, we apply the procedure in previous section to the intrinsic alignments of galaxies, and compute their response to the super-sample modes. The intrinsic alignment (IA) of our interest here is quantified by the ellipticity field, $\gamma_{ij}$, defined by the traceless part of the second moment of the specific intensity, $I_{ij}$. To be precise, we write down the symmetric second moment observed at a position $\bfx$. For a type of galaxies, it is generally expressed as \cite{Vlah_Chisari_Schmidt2020,Vlah_Chisari_Schmidt2020b}
\begin{align}
I_{ij}(\bfx) =\mathcal{I}\,\Bigl[\,\frac{1}{3}\deltaK_{ij}\Bigl\{1+\delta_{\rm s}(\bfx)\Bigr\}+\gamma_{ij}(\bfx)\,\Bigr]
\label{eq:3D_second_moment}
\end{align}
with $\mathcal{I}$ being obtained by taking the ensemble average, $\mathcal{I}=\langle\mbox{Tr}\,I_{\ell m}\rangle$\footnote{We implicitly assume that $\gamma_{ij}$ is traceless, and has (global) zero mean together with $\delta_{\rm s}$, i.e., $\langle\gamma_{ij}\rangle=0=\langle\delta_s\rangle$. }. Here, the quantity $\delta_{\rm s}$ describes the fluctuation in the trace part, called the size fluctuation. The symmetric and traceless tensor $\gamma_{ij}$ is the quantity which we shall now focus on.

Note that Eq.~(\ref{eq:3D_second_moment}) characterizes the three-dimensional galaxy shape, and thus subscripts $i,j$ run over $x$, $y$, and $z$. Strictly, to obtain a relevant observable related to $\gamma_{ij}$, the second moment $I_{ij}$ has to be projected onto the celestial sphere \cite{Kogai_etal2018,Vlah_Chisari_Schmidt2020,Vlah_Chisari_Schmidt2020b}, taking the traceless part. Further, it should be appropriately normalized, e.g., by the trace part of the second moment for each galaxy or its mean. In this respect, how the observable responds to the fluctuation $\gamma_{ij}$ depends on the estimator of IA, and in practical measurement of the galaxy shape, one may not purely isolate $\gamma_{ij}$ from the size fluctuation $\delta_{\rm s}$\footnote{A simple example is the estimator using the three-dimensional shape, $g_{ij}\equiv [{\rm Tr}\,I_{\ell m}]^{-1}\bigl\{I_{ij}-(\deltaK_{ij}/3){\rm Tr}\,I_{\ell m} \bigr\}$, which gives $g_{ij}=\gamma_{ij}/(1+\delta_{\rm s})$.}. Nevertheless, the size fluctuation appears as a higher-order correction, and its effect on the super-sample modes can be separately treated at leading order. Furthermore, as long as we consider the distant-observer or plane-parallel limit in which we take the $z$-axis to be the line-of-sight direction, the observed IA is shown to be proportional to the $x,\,y$-components of $\gamma_{ij}$. Hence, we hereafter work only with the three-dimensional quantity $\gamma_{ij}$, and will later evaluate its projection in computing the power spectra relevant to the observations. The impact of the super-sample modes on the size fluctuation will be left for future work\footnote{As noted in Ref.~\cite{Vlah_Chisari_Schmidt2020}, the size fluctuation is the scalar quantity and can be described similarly to the galaxy density field $\deltag$. In this respect, the same treatment as we did in the density field can be applied, and the derivation of the long-mode contributions to the size field would be rather straightforward. } .

Similar to the galaxy density field, the galaxy ellipticity field $\gamma_{ij}$ is considered as a biased tracer of the tidal fields induced by the large-scale matter inhomogeneities. A general expansion scheme to perturbatively describe the ellipticity field has been recently discussed, including also the stochasticity \cite{Vlah_Chisari_Schmidt2020}. In what follows, we focus on the deterministic part. Since the stochasticity only yields the additive contributions at leading order, we separately discuss it in Appendix \ref{Appendix:stochasticity}.  Then, up to the second order, the (Eulerian) ellipticity field in real space is expressed in the following expansion form \cite{Schmitz_etal2018,Blazek_etal2019}: 
\begin{align}
&\gamma_{ij} (\bfx) = \Bigl\{1+\delta_{\rm g}(\bfx)\Bigr\}
\nonumber
\\
&\quad\times \Biggl[\bK K_{ij}(\bfx)+\bdeltaK\,\delta(\bfx)K_{ij}(\bfx) + 
 \bKK \Bigl\{K_{i\ell}(\bfx)K_{\ell j}(\bfx)-\frac{1}{3}\deltaK_{ij}\,
[{\rm Tr}\,K_{\ell m}(\bfx)]^2\Bigr\}
\nonumber
\\
&\qquad +\bt\,t_{ij}(\bfx) + \cdots\Biggr],
\label{eq:gamma_expansion}
\end{align}
where the coefficients $\bK$, $\bdeltaK$, $\bKK$ and $\bt$ are the parameters characterizing the linear and nonlinear response of the shape to the tidal fields of large-scale structure. 
In the above,  the factor $1+\deltag$ implies that the observed ellipticity field is given as a density-weighted quantity. Here, the tensor $K_{ij}$ is given by Eq.~(\ref{eq:def_K_ij}), and $t_{ij}$ is defined by
\begin{align}
 t_{ij}(\bfx)&\equiv \Bigl(\frac{\partial_{x_i}\partial_{x_j}}{\partial^2_x}-\frac{1}{3}\deltaK_{ij}\Bigr)\Bigl\{\theta(\bfx)-\delta(\bfx)\Bigr\},
\label{eq:def_t_ij}
\end{align}
with the field $\theta$ being the dimensionless velocity-divergence field defined by $\theta=-(\nabla\cdot\bfv)/(f\,a\,H)$. Note that all of the objects at right-hand side, i.e., $K_{ij}$, $t_{ij}$, and $\delta$ as well as $\deltag$,  are evolved fields. In particular, a perturbative calculation of $t_{ij}$ shows that it becomes zero at linear order, but becomes  non-vanishing at second order.

\subsection{Field-level expression with super-sample modes}
\label{subsec:IA_with_SS-modes}

We now derive the explicit expression for the IA including the contributions from super-sample modes. Here, we are interested in the quantities in redshift space, relevant to the spectroscopic galaxy samples. In analogy to the density field in Sec.~\ref{subsubsec:redshift_space},  the IA in redshift space, which we denote by $\gamma_{ij}^{\rm(S)}$, is expressed in terms of the quantities defined in Lagrangian space: 
\begin{align}
  \gamma_{ij}^{\rm (S)}(\bfs) &= \int d^3\bfq \int\frac{d^3\bfk}{(2\pi)^3}\,e^{i\,\bfk\cdot\{\bfs-\bfq-\bfpsi^{\rm S}(\bfq)\}}\,\gammaL_{ij} (\bfq),
\label{eq:gamma_redshift_integral_form}
\end{align}
where the Lagrangian IA, $\gammaL_{ij}$, is related to the one in Eulerian space given at Eq.~(\ref{eq:gamma_expansion}). A relevant expansion form of the Lagrangian IA, expressed in terms of the Lagrangian linear density $\deltalin(\bfq)$, is given by
\begin{align}
& \gammaL_{ij} (\bfq) = \Bigl\{1+\deltag^{\rm L}(\bfq)\Bigr\}\Biggl[-\cK C_{ij}(\bfq) + \cdeltaK\,\deltalin(\bfq) C_{ij}(\bfq) +
 \cKK \Bigl\{C_{i\ell}(\bfq)C_{\ell j}(\bfq)
\nonumber
\\
&\quad -\frac{1}{3}\deltaK_{ij}\,
[{\rm Tr}\,C_{\ell m}(\bfq)]^2\Bigr\} 
+\ct\,\Bigl(\frac{\partial_{q_i}\partial_{q_j}}{\partial_q^2}-\frac{1}{3}\deltaK_{ij}\Bigr)
\Bigl\{-\frac{4}{21}\deltalin(\bfq)^2+\frac{2}{7}[{\rm Tr}\,C_{\ell m}(\bfq)]^2\Bigr\}\Biggr],
\label{eq:gammaL_2nd_order}
\end{align}
which is valid at the second order. Here, the field $C_{ij}$ is the linear tidal tensor given at Eq.~(\ref{eq:def_C_ij}). The Lagrangian shape bias  parameters $\cK$, $\cdeltaK$, $\cKK$, and $\ct$ are related to the Eulerian shape bias introduced at Eq.~(\ref{eq:gamma_expansion}) through \cite{Schmitz_etal2018} 
\begin{align}
 \Bigl(\bK,\,\bdeltaK,\, \bKK,\, \bt\Bigr)=\Bigl(-\cK,\,\cdeltaK+\frac{2}{3}\cK,\,\cKK+\cK,\,\ct-\frac{5}{2}\cK\Bigr).
\label{eq:bias_delta_gamma}
\end{align}
For clarity, in Appendix \ref{Appendix:shape_bias}, the link between the shape bias parameters in Eulerian and Lagrangian space is explicitly shown, and the derivation of Eq.~(\ref{eq:bias_delta_gamma}) is given.

Provided the setup and basic ingredients to perturbatively describe the IA, we proceed to the explicit calculations based on the procedure in Sec.~\ref{sec:formalism}. Below we summarize each step to derive the field-level response to the super-sample modes:

\begin{enumerate}
 \item First, we plug the expansion form of $\gammaL_{ij}$ given at Eq.~(\ref{eq:gammaL_2nd_order}) into Eq.~(\ref{eq:gamma_redshift_integral_form}). Then, substituting the expansion form at Eqs.~(\ref{eq:n-th_order_displacement_z-space}), we Taylor-expand the redshift-space displacement in the exponent of Eq.~(\ref{eq:gamma_redshift_integral_form}) to obtain the expression valid at the second order in linear density and displacement fields. A part of the integrand having an explicit dependence of the wavevector $\bfk$ is rewritten by performing the integration by part, as demonstrated in Eqs.~(\ref{eq:mapping_integrand}) and (\ref{eq:deltag_mapping_expansion}). 

 \item Long-/short-mode decomposition is then applied to the expanded form of Eq.~(\ref{eq:gamma_redshift_integral_form}). We use Eqs.~(\ref{eq:deltalin_short_long}) and (\ref{eq:Cij_short_long}) to decompose the linear density ($\deltalin$) and tidal fields ($C_{ij}$). For the displacement field, the first-order displacement is the only relevant contribution, and the long-/short-mode decomposition given at Eq. ~(\ref{eq:psi_short_long}) with (\ref{eq:psi_long1}) is applied. 

\item Keeping the linear-order terms in short-mode density and tidal fields\footnote{There are also DC modes that only depend on the super-sample modes, $\deltab$ and $\tau_{ij}$. While we will neglect these contributions in the field-level expressions, the treatment of them may have to be carefully considered. We will discuss these contributions for the power spectrum responses in the local mean in Sec.\ref{subsec:pk_IA}.},
 the integration over $\bfk$ and $\bfq$ is performed. Then, the expression of $\gamma_{ij}^{\rm(S)}$ is now explicitly given as a function of redshift-space position $\bfs$. Rewriting further the Lagrangian bias parameters with the Eulerian counterparts through the relation (\ref{eq:bias_delta_gamma}), the expression involving the super-sample modes is finally obtained at the field level. 
\end{enumerate}

Calculations at each step given above are rather straightforward and have no ambiguity, but for ease of derivation, we summarize in Appendix \ref{Appendix:derivation_of_gamma_with_SS-modes} the key equations. The final expression for the IA field involving the super-sample modes, given in Fourier space, becomes (we again drop the DC-mode contributions from the expression at Eq.~(\ref{eq:gamma_red_with_SS-modes}))
\begin{align}
  \gamma_{ij}^{\rm(S)}(\bfk)&= \Bigl[\bK\,\tildePi_{ij}(\bfk)\,
+ \deltab\,\Bigl\{\widehat{A}_{ij}(\bfk)\, + f\,\widehat{A}_{ij}^{\rm (S)}(\bfk)\Bigr\}
+\tau_{\ell m} \Bigl\{
\widehat{B}_{ij\ell m}(\bfk)+f\,
\widehat{B}_{ij\ell m}^{\rm (S)}(\bfk) \Bigr\}
\Bigr]\deltas(\bfk)
\label{eq:gamma_with_SS-mode_z-space}
\end{align}
with the quantity $\tildePi_{ij}$ defined by
\begin{align}
 \tildePi_{ij}(\bfk)\equiv \hatk_i\hatk_j-\frac{1}{3}\,\deltaK_{ij}.
\end{align}
Here, the operators $\widehat{A}_{ij}$ and $\widehat{B}_{ij\ell m}$ are 
defined by
\begin{align}
 \widehat{A}_{ij}(\bfk)&=\tildePi_{ij}(\bfk)\Bigl\{\bK\Bigl(b_1+\frac{13}{21}\Bigr)+\bdeltaK-\frac{8}{21}\bt
-\frac{1}{3}\bK\,(\bfk\cdot\nabla_k)\Bigr\},
\label{eq:def_Aij}
\\
 \widehat{B}_{ij\ell m}(\bfk)&=\frac{1}{2}\,\Bigl\{\bK\Bigl(b_1+\frac{2}{3}\Bigr)+\bdeltaK\Bigr\}\bigl(\deltaK_{i\ell}\deltaK_{jm}+ \deltaK_{im}\deltaK_{j\ell}\bigr)
\nonumber
\\
&-\frac{1}{2}\,\bK\,\Bigl\{\hatk_j(\hatk_\ell\deltaK_{mi}+\hatk_m\deltaK_{\ell i})+\hatk_i(\hatk_\ell\deltaK_{mj}+\hatk_m\deltaK_{\ell j})-4\hatk_i\hatk_j\hatk_\ell\hatk_m\Bigr\}
\nonumber
\\
&+(\bKK+\bK)\Bigl\{\tildePi_{\ell j}(\bfk)\deltaK_{im}+\tildePi_{i\ell}(\bfk)\deltaK_{jm}-\frac{2}{3}\deltaK_{ij}\,\tildePi_{\ell m}(\bfk)\Bigr\}
\nonumber
\\
&
+\tildePi_{ij}(\bfk)\Bigl\{\frac{4}{7}\Bigl(\bt-\frac{5}{2}\bK\Bigr)\hatk_\ell\hatk_m-\bK\,k_\ell\,\partial_{k_m}\Bigr\}.
\label{eq:def_Bijlm}
\end{align}
Similarly to the case of the density field, even at the leading-order, the dependence of the higher-order shape bias ($\bKK$, $\bdeltaK$, and $\bt$) as well as the density bias ($b_1$) becomes manifest through the super-sample contributions. Also, the effect of the redshift-space distortions (i.e., non-zero $f$) appears through the super-sample modes, with the operators $\widehat{A}_{ij}^{\rm (S)}$ and $\widehat{B}_{ij\ell m}^{\rm (S)}$ given by
\begin{align}
\widehat{A}_{ij}^{\rm (S)}&=
-\frac{1}{3}\,\bK\,\Bigl\{\hatk_z(\hat{z}_i\hatk_j+\hat{z}_j\hatk_i-2\mu\,\hatk_i\hatk_j)
+\tildePi_{ij}(\bfk)\,k_z\partial_{k_z}\Bigr\},
\label{eq:def_Aij_z-space}
\\
\widehat{B}_{ij\ell m}^{\rm (S)}&=\frac{1}{2}\,\bK\,\Bigl[\hatk_z^2
(\deltaK_{i\ell}\deltaK_{jm}+\deltaK_{im}\deltaK_{j\ell})
\nonumber
\\
&\quad -\,\hatk_z
\Bigl\{ (\hat{z}_\ell\deltaK_{mi}+\hat{z}_m\deltaK_{\ell i})\hatk_j 
+(\hat{z}_\ell\deltaK_{mj}+\hat{z}_m\deltaK_{\ell j})\hatk_i 
-2(\hat{z}_\ell\hatk_m+\hat{z}_m\hatk_\ell)\hatk_i\hatk_j\Bigr\}
\nonumber
\\
&\quad -\tildePi_{ij}(\bfk)k_z\bigl(\hat{z}_\ell\partial_{k_m}+\hat{z}_m\partial_{k_\ell}\bigr)\Bigr].
\label{eq:def_Bijlm_z-space}
\end{align}
Note that in deriving the Fourier-space expression at Eq.~(\ref{eq:gamma_with_SS-mode_z-space}), we use the following relation:
\begin{align}
& s_m\,\partial_{s_k}K_{ij}^{\rm short}(\bfs)=-\int\frac{d^3\bfk}{(2\pi)^3}
\,e^{i\bfk\cdot\bfs}
\nonumber
\\
&\qquad \times\Bigl[
\deltaK_{km}\tildePi_{ij}(\bfk)
+\hatk_k\bigl(\deltaK_{im}\hatk_j+\deltaK_{jm}\hatk_i-2\hatk_i\hatk_j\hatk_m\bigr)
+\tildePi_{ij}(\bfk)\,k_k\partial_{k_m}
\Bigr]\deltas(\bfk).
\label{eq:x_derivative_Kij}
\end{align}

\subsection{Projection and E-/B-mode decomposition}
\label{subsec:E-/B-mode_ellipticities}

The expression given at Eq.~(\ref{eq:gamma_with_SS-mode_z-space}) describes how the three-dimensional shape in redshift space responds to the super-sample modes. Here, to make a direct link with observables, we consider the two-dimensional shape projected onto the sky, and define the 
two-component ellipticity field, $(\gamma_+,\,\gamma_\times)$: 
\begin{align}
\Bigl(
\begin{array}{c}
\gamma_+ \\
\gamma_\times 
\end{array} 
\Bigr) \equiv \Bigl(
\begin{array}{c}
 \gamma_{xx}-\gamma_{yy}\\
 2\gamma_{xy}
\end{array}
\Bigr),
\end{align}
where we work with the flat-sky limit, and take the line-of-sight direction to be the $z$-axis\footnote{A general projection can be expressed as $\gamma^{\rm 2D}_{ab} = {\cal P}^{i}_{\ a}{\cal P}^{j}_{\ b}\gamma_{ij}$ where ${\cal P}_{ij}\equiv \delta^{\rm K}_{ij} - \hat{n}_i\hat{n}_j$ is the projection tensor with $\hat{n}_i$ being the line-of-sight. Setting $\hat{n}_i = \hat{z}_i$ results in $\gamma_{ab}^{\rm 2D} = 
\begin{pmatrix}
\gamma_{xx} & \gamma_{xy} & 0 \\
\gamma_{yx} & \gamma_{yy} & 0 \\
0 & 0 & 0 \\
\end{pmatrix}
$. }. 
Note that while we shall below consider the two-dimensional projected ellipticity for each galaxy, we do not project the galaxy distribution. That is, we suppose that the ellipticity field defined above is still given in the three dimensional space, as similarly considered in Refs.~\cite{Okumura_Taruya_Nishimichi2019, Okumura_Taruya2019,Okumura_Taruya_Nishimichi2020, Kurita_etal2021, Jingjing_etal2021} (see also e.g., Ref.~\cite{Okumura_Jing_Li2009,Singh_Mandelbaum2016} for actual measurements of the three-dimensional correlation).  
In the weak lensing measurement, a more convenient way to characterize the projected ellipticity fields is known as the E-/B-mode decomposition, which gives a rotationally invariant decomposition \cite{Kamionkowski_etal1998,Crittenden_etal2002}. Denoting the E-/B-mode ellipticity field by $\gamma_{\rm E/B}$, this is defined by 
\begin{align}
\Biggl(
\begin{array}{c}
\gamma_{\rm E} \\
\gamma_{\rm B} 
\end{array} 
\Biggr)(\bfk) \equiv \mathbf{R}(\phi_k)\,\Biggl(
\begin{array}{c}
\gamma_+ \\
\gamma_\times
\end{array} 
\Biggr)(\bfk), 
\label{eq:E-/B-mode_decomposition}
\end{align}
with the quantity $\mathbf{R}$ being the rotation matrix given by 
\begin{align}
\mathbf{R}(\phi_k)\equiv \Biggl(
\begin{array}{cc}
 \cos(2\,\phi_k) & \sin(2\,\phi_k) \\
-\sin(2\,\phi_k) & \cos(2\,\phi_k)
\end{array}\Biggr). 
\label{eq:rotation_matrix}
\end{align}
Here, the angle $\phi_k$ is the azimuthal angle of the wavevector projected on the sky, measured from the $x$-axis. To be explicit, we write the wavevector $\bfk$ as  
\begin{align}
 \bfk=k\,(\sqrt{1-\mu_k^2}\cos\phi_k,\,\sqrt{1-\mu_k^2}\sin\phi_k,\,\mu_k)
\label{eq:wavevector_phi_k}
\end{align}
with $\mu_k$ being the directional cosine between the line-of-sight direction and wavevector, $\mu_k=\hatk\cdot\hatz$. Substituting Eq.~(\ref{eq:def_Bijlm_z-space}) into Eq.~(\ref{eq:E-/B-mode_decomposition}), we obtain the leading-order expression for the E-/B-mode ellipticity fields in redshift space: 
\begin{align}
&\Biggl(
\begin{array}{c}
 \gamma_{\rm E}^{\rm(S)} \\
 \gamma_{\rm B}^{\rm(S)}
\end{array}
\Biggr)(\bfk) 
= \Biggl[\bigl(1-\mu_k^2\bigr)\,\Bigl[
\bK +\deltab\,\Bigl\{\widehat{\alpha}_\delta(\bfk)+f\,\widehat{\alpha}_\delta^{\rm(S)}(\bfk) \Bigr\}
\nonumber
\\
&\qquad\qquad + \tau_{\ell m}\,\Bigl\{\widehat{\beta}_{\tau,{\ell m}}(\bfk)+f\,\widehat{\beta}_{\tau,{\ell m}}^{\rm(S)}(\bfk)\Bigr\} \Bigr]
\Biggl(
\begin{array}{c}
 1  \\
 0
\end{array}
\Biggr)+\tau_{\ell m}\,\mathbf{M}_{\ell m}(\bfk)\,\Biggr]\,\deltas(\bfk).
\label{eq:E_B-mode_ellipticity_SS-modes_z-space}
\end{align}
At right-hand side of Eq.~(\ref{eq:E_B-mode_ellipticity_SS-modes_z-space}), the first term in the bracket represents a pure E-mode contribution, which include the terms coming from the super-sample modes, $\deltab$ and $\tau_{ij}$. 
In the absence of the super-sample modes, this is reduced to the one obtained from the linear alignment model \cite{Catelan_etal2001,Hirata_Seljak2004,Okumura_Taruya2019,Okumura_Taruya_Nishimichi2020,Kurita_etal2021}. 
The operators acting on $\deltas$, i.e,. $\widehat{\alpha}_{\delta}$, $\widehat{\alpha}_\delta^{\rm(S)}$, $\widehat{\beta}_{\tau,{\ell m}}$ and $\widehat{\beta}_{\tau,\ell m}^{\rm(S)}$, are respectively defined as follows:
\begin{align}
\widehat{\alpha}_\delta(\bfk)& =
\bK\Bigl(b_1+\frac{13}{21}\Bigr)+\bdeltaK -\frac{8}{21}\bt\,-\frac{1}{3}\bK\,(\bfk\cdot\nabla_k),
\label{eq:alpha_delta}
\\
\widehat{\beta}_{\tau,\ell m}(\bfk)& =\frac{4}{7}\bigl(\bK+\bt\bigr)\hatk_\ell\hatk_m-\bK \,k_\ell\,\partial_{k_m},
\label{eq:alpha_tau}
\\
 \widehat{\alpha}_\delta^{\rm(S)}(\bfk)&=\frac{1}{3}\,\bK\,\bigl(2\mu_k^2-k_z\,\partial_{k_z}\bigr),
\label{eq:alpha_s_z-space}
\\
 \widehat{\beta}_{\tau,\ell m}^{\rm(S)}(\bfk)&=\bK\Bigl\{\hat{z}_\ell\hatk_m+\hat{z}_m\hatk_\ell-\frac{k_z}{2}\,\bigl(\hat{z}_\ell\, \partial_{k_m}+\hat{z}_m\,\partial_{k_\ell}\bigr)\Bigr\}.
\label{eq:alpha_tau_z-space}
\end{align}
Similar to the density field, we see from Eqs.~(\ref{eq:alpha_tau})  and (\ref{eq:alpha_tau_z-space}) that the super-sample tidal fields induce the additional quadrupolar anisotropies to the ellipticity fields. On the other hand, in Eq.~(\ref{eq:E_B-mode_ellipticity_SS-modes_z-space}), the second term in the bracket, $\tau_{\ell m}\mathbf{M}_{\ell m}$, describes another contribution from the super-sample tidal field $\tau_{ij}$, which produces both non-vanishing E- and B-modes. The explicit form of it is given by   
\begin{align}
 \tau_{\ell m}\mathbf{M}_{\ell m}(\bfk) &=\mathbf{R}(\phi_k)\,\Biggl\{2\bKK\,
\Biggl(
\begin{array}{c}
 \hatk_x\hatk_\ell\tau_{\ell x}-\hatk_y\hatk_\ell\tau_{\ell y}  \\
 \hatk_\ell\hatk_y\tau_{\ell x}+\hatk_x\hatk_\ell\tau_{\ell y} 
\end{array}
\Biggr)
- 2\,f\,\bK\,\mu_k\,\Biggl(
\begin{array}{c}
\hatk_x\tau_{xz}-\hatk_{y}\tau_{yz} \\
\hatk_y\tau_{xz}+\hatk_x\tau_{yz}
\end{array}
\Biggr)
\nonumber
\\
&
\quad + \Bigl\{\bK b_1+\bdeltaK-\frac{2}{3}\bKK+f\,\bK\,\mu_k^2\Bigr\}
 \Biggl(
\begin{array}{c}
\tau_{xx}-\tau_{yy} \\
2\tau_{xy}
\end{array}
\Biggr)\,\Biggr\}.
\label{eq:tau_M_ell_m} 
\end{align}
The non-vanishing B-mode is an interesting consequence of the super-sample tidal field that modulates the sub-survey modes through the mode coupling. Observationally, however, this is a spurious contribution, and is shown to have no impact on the statistics of IA. As we will see below, the B-mode contribution of Eq.~(\ref{eq:tau_M_ell_m}) becomes vanishes, and only the E-mode contribution proportional to the parameter $\bKK$ survives when considering the auto- and cross-power spectra of ellipticity and density fields.

\subsection{Power spectra}
\label{subsec:pk_IA}

We are in position to compute the statistical quantities of IAs including the effect of super-sample modes. First consider the two-point statistics of E-/B-mode ellipticity and density fields, defined by
 \begin{align}
&\frac{1}{2}\,\langle\gamma_{\rm X}^{\rm(S)}(\bfk)\gamma_{\rm Y}^{\rm(S)}(\bfk')+\gamma_{\rm X}^{\rm(S)}(\bfk')\gamma_{\rm Y}^{\rm(S)}(\bfk)\rangle =(2\pi)^3\,\delta_{\rm D}(\bfk+\bfk')\,P_{\rm XY}^{\rm(S)}(\bfk),
\label{eq:def_pk_EB}
\\
&\frac{1}{2}\,\langle\deltag^{\rm(S)}(\bfk)\gamma_{\rm X}^{\rm(S)}(\bfk')+\deltag^{\rm(S)}(\bfk')\gamma_{\rm X}^{\rm(S)}(\bfk)\rangle =(2\pi)^3\,\delta_{\rm D}(\bfk+\bfk')\,P_{\rm gX}^{\rm(S)}(\bfk),
\label{eq:def_pk_gB}
 \end{align}
where subscripts $X$ and $Y$ stand for the E-/B-mode ellipticity fields. There are thus five power spectra, i.e., $P_{\rm EE}^{\rm(S)}$, $P_{\rm EB}^{\rm(S)}$, $P_{\rm BB}^{\rm(S)}$, $P_{\rm gE}^{\rm(S)}$, and $P_{\rm gB}^{\rm(S)}$, among which the EB-mode and gB cross spectra usually become zero if the parity symmetry is preserved on the sky.

It is interesting to note that in the presence of the large-scale tidal field $\tau_{ij}$, the parity symmetry is apparently broken, leading to the non-vanishing EB-mode and gB cross spectra. On the other hand, the BB-mode auto spectrum is still zero in the linear-order calculation for the super-sample modes. For concreteness, we show the expressions of their power spectra. Substituting the E-/B-mode ellipticities at Eq.~(\ref{eq:E_B-mode_ellipticity_SS-modes_z-space}) and galaxy density field at Eq.~(\ref{eq:deltag_redshift_SS-modes}) into Eq.~(\ref{eq:def_pk_EB}), the ensemble average over the short-mode density field yields
\begin{align}
  P_{\rm EB}^{\rm(S)}(\bfk) &=(1-\mu_k^2)\,\bK\,\tau_{\ell m}\mathcal{F}_{\ell m}(\bfk)\,P_{\delta\delta}(k),
\label{eq:pkred_EB_wo_averaging}
\\
  P_{\rm gB}^{\rm(S)}(\bfk) &=(b_1+f\,\mu_k^2)\,\tau_{\ell m}\mathcal{F}_{\ell m}(\bfk)\,P_{\delta\delta}(k),
\label{eq:pkred_gB_wo_averaging}
\end{align}
where the common factor involving the super-sample tidal field, $\tau_{\ell m}\mathcal{F}_{\ell m}$, is expressed as
\begin{align}
\tau_{\ell m}\mathcal{F}_{\ell m}&= 2\bKK \Bigl\{ 
-\bigl(\hatk_x\hatk_\ell\tau_{\ell x}-\hatk_y\hatk_\ell\tau_{\ell y}\bigr)\sin(2\phi_k) + \bigl(\hatk_\ell\hatk_y\tau_{\ell x}+\hatk_x\hatk_\ell\tau_{\ell y}\bigr)\cos(2\phi_k)
\Bigr\}
\nonumber
\\
&-2\,f\,\bK\,\mu_k
\Bigl\{ 
-\bigl(\hatk_x\tau_{xz}-\hatk_y\tau_{yz}\bigr)\sin(2\phi_k) + \bigl(\hatk_y\tau_{xz}+\hatk_x\tau_{yz}\bigr)\cos(2\phi_k)
\Bigr\}
\nonumber
\\
&
+\bigl(\bK\,b_1+\bdeltaK-\frac{2}{3}\bKK+f\,\bK\,\mu_k^2\bigr)
\Bigl\{-(\tau_{xx}-\tau_{yy})\sin(2\phi_k)+2\tau_{xy}\cos(2\phi_k)\Bigr\}.
\end{align}
Here, we keep only the linear-order terms in $\tau_{ij}$.

The non-zero EB-mode and gB cross power spectra given above are a direct manifestation that the modulation due to the super-sample modes affects the statistical nature of sub-survey modes. While such an effect could be in principle imprinted on the observed ellipticity and density fields, the expressions given at Eqs.~(\ref{eq:pkred_EB_wo_averaging}) and (\ref{eq:pkred_gB_wo_averaging}) are the spectra characterized by the three-dimensional wave vector, and a measurement of such spectra would produce a large error from the finite-volume surveys, due largely to a limited number of available Fourier modes. Rather, what can be practically measured would be the quantities averaged over certain Fourier modes, and taking care of the anisotropies inherent in the ellipticity and density fields along the line-of-sight, a relevant observable would be the quantities taking the angle average on the sky (see also Ref.~\cite{Akitsu_Takada2018}): 
\begin{align}
 \overline{P}_{\rm XY}^{\rm (S)}(\bfk)= \int_0^{2\pi}\frac{d \phi_k}{2\pi}\,P_{\rm XY}^{\rm (S)}(\bfk), \qquad
 \overline{P}_{\rm gX}^{\rm (S)}(\bfk)= \int_0^{2\pi}\frac{d \phi_k}{2\pi}\,P_{\rm gX}^{\rm (S)}(\bfk),
\label{eq:def_averaged_pkred}
\end{align}
where the angle $\phi_k$ is defined on the plane perpendicular to the line-of-sight, given at Eq.~(\ref{eq:wavevector_phi_k}). Substituting Eqs.~(\ref{eq:pkred_EB_wo_averaging}) and (\ref{eq:pkred_gB_wo_averaging}) into Eq.~\eqref{eq:def_averaged_pkred}, using the explicit expression of $\bfk$ at Eq.~(\ref{eq:wavevector_phi_k}) immediately leads to  
\begin{align}
 &\overline{P}_{\rm EB }^{\rm(S)}(\bfk)=0,
 \\
 &\overline{P}_{\rm gB }^{\rm(S)}(\bfk)=0,
\end{align}
Also, we have
\begin{align}
 \overline{P}_{\rm BB }^{\rm(S)}(\bfk)=0. 
\end{align}
That is, the angle-averaged power spectra involving the B-mode ellipticity become all vanishing. This is true as long as we consider the leading order. On the other hand, the spectra involving the E-mode ellipticity, i.e., $P^{\rm (S)}_{\rm EE}$ and $P^{\rm (S)}_{\rm gE}$, become non-vanishing even after the angle average. The resultant EE-mode auto power spectrum is expressed as
\begin{align}
 \overline{P}^{\rm (S)}_{\rm EE}(\bfk)=\bK^2(1-\mu^2)^2\,P_{\delta\delta}(k)
+\frac{\partial \overline{P}^{\rm (S)}_{\rm EE}(\bfk)}{\partial \deltab}\,\deltab+ \frac{\partial \overline{P}^{\rm (S)}_{\rm EE}(\bfk)}{\partial \tau_{zz}}\,\tau_{zz},
 \label{eq:pk_EE_w_SS-modes}
\end{align}
where the quantities $\partial \overline{P}^{\rm (S)}_{\rm EE}/\partial \deltab$ and $\partial \overline{P}^{\rm (S)}_{\rm EE}/\partial \tau_{zz}$ represent 
the linear responses to the super-sample modes. Their explicit expressions are respectively given by
\begin{align}
& \frac{\partial \overline{P}^{\rm (S)}_{\rm EE}(\bfk)}{\partial \deltab}
=\bK\,(1-\mu_k^2)^2\,P_{\delta\delta}(k)\,
\nonumber
\\
&\times\Bigl[\Bigl(\frac{47}{21}+2b_1\Bigr)\bK-\frac{16}{21}\bt+2\bdeltaK-\frac{1}{3}\,\bK\frac{d\ln P_{\delta\delta}(k)}{d\ln k}
+\frac{f}{3}\,\bK
\Bigl\{4\mu_k^2+1-\mu_k^2\frac{\partial\ln P_{\delta\delta}(k)}{\partial \ln k}\Bigr\}\Bigr],
\label{eq:dpkEE_ddeltab}
\\
& \frac{\partial \overline{P}^{\rm (S)}_{\rm EE}(\bfk)}{\partial \tau_{zz}}
=\bK\,(1-\mu_k^2)^2\,P_{\delta\delta}(k)\,
\nonumber
\\
&\times\Bigl[
\frac{3\mu_k^2-1}{2}\Bigl\{\frac{8}{7}(\bK+\bt)-\bK\frac{\partial\ln P_{\delta\delta}(k)}{\partial \ln k}\Bigr\}-2\bKK
+f\,\bK\,\Bigl\{4\mu_k^2+1-\mu_k^2\frac{\partial\ln P_{\delta\delta}(k)}{\partial \ln k}\Bigr\}
\Bigr].
\label{eq:dpkEE_dtauzz}
\end{align}
Also, the expression of the averaged gE cross power spectrum becomes
\begin{align}
& \overline{P}^{\rm (S)}_{\rm gE}(\bfk)=\bK\,(b_1+f\,\mu_k^2)(1-\mu_k^2)\,P_{\delta\delta}(k)
+\frac{\partial \overline{P}^{\rm (S)}_{\rm gE}(\bfk)}{\partial \deltab}\,\deltab+ \frac{\partial \overline{P}^{\rm (S)}_{\rm gE}(\bfk)}{\partial \tau_{zz}}\,\tau_{zz} 
 \label{eq:pk_gE_w_SS-modes}
\end{align}
with the response to the super-sample modes given by
\begin{align}
&\frac{\partial \overline{P}^{\rm (S)}_{\rm gE}(\bfk)}{\partial \deltab}
\nonumber
\\
&=(1-\mu_k^2)\Bigl[b_1\bK\Bigl\{\frac{47}{21}-\frac{1}{3}\frac{\partial \ln P_{\delta\delta}(k)}{\partial\ln k}\Bigr\}+b_1^2\,\bK\,+b_1\,\bdeltaK-\frac{8}{21} b_1\bt+ b_2\,\bK\Bigr]\,P_{\delta\delta}(k)
\nonumber
\\
&\quad+(1-\mu_k^2)\Bigl[\,\frac{1}{3}b_1\bK(1+8\,\mu_k^2)+\Bigl(\bdeltaK+\frac{13}{7}\bK-\frac{8}{21}\bt\Bigr)\mu^2
\nonumber
\\
&\quad\qquad\qquad\qquad\qquad\qquad\qquad\qquad\qquad
-\frac{\bK}{3}\,(1+b_1)\mu_k^2
\frac{\partial\ln P_{\delta\delta}(k)}{\partial\ln k}\Bigr]\,f\,P_{\delta\delta}(k)
\nonumber
\\
&\quad -\frac{\bK}{3}\,\mu_k^4\,(1-\mu_k^2)\Bigl\{-4+\frac{\partial\ln P_{\delta\delta}(k)}{\partial\ln k}\Bigr\}\,f^2 \,P_{\delta\delta}(k),
\label{eq:dpkgE_ddeltab}
\\
&\frac{\partial \overline{P}^{\rm (S)}_{\rm gE}(\bfk)}{\partial \tau_{zz}}
\nonumber
\\
&=(1-\mu_k^2)\Bigl[\frac{3\mu_k^2-1}{2}\Bigl\{b_1\bK\Bigl(\frac{8}{7}-\frac{\partial\ln P_{\delta\delta}(k)}{\partial\ln k}\Bigr)+\frac{4}{7}b_1\bt+b_{s^2}\bK\Bigr\}-b_1\bKK\Bigr]\,P_{\delta\delta}(k)
\nonumber
\\
&\quad+(1-\mu_k^2)\Bigl[
b_1\bK(1+2\,\mu_k^2)-\bKK\,\mu_k^2+\frac{2}{7}(\bt+3\bK)(3\mu_k^2-1)\mu_k^2
\nonumber
\\
&\qquad\qquad\qquad\qquad\qquad\qquad
+\frac{\bK}{2}\,\mu_k^2(1-2b_1-3\mu_k^2)\,\frac{\partial\ln P_{\delta\delta}(k)}{\partial\ln k}
\Bigr]\,f\,P_{\delta\delta}(k)
\nonumber
\\
&\quad-\bK\,\mu_k^4(1-\mu_k^2)\Bigl\{-4+\frac{\partial\ln P_{\delta\delta}(k)}{\partial\ln k}\Bigr\}\,f^2\,P_{\delta\delta}(k).
\label{eq:dpkgE_dtauzz}
\end{align}

Expressions given at Eqs.~(\ref{eq:pk_EE_w_SS-modes})-(\ref{eq:dpkgE_dtauzz}) are one of the important results in this paper. Ignoring the super-sample modes, these are reduced to the power spectra obtained from the linear alignment model (see e.g., Ref.~\cite{Kurita_etal2021}). Note that the above results are valid strictly for the galaxy density field defined with the global mean. That is, Eqs.~(\ref{eq:dpkgE_ddeltab}) and (\ref{eq:dpkgE_dtauzz}) are relevant to the number density of galaxies normalized by its global mean, which is practically un-observable \cite{dePutter_etal2012}. Rather, the observed density fluctuations are defined with the local mean measured in the survey region. In this case, taking the DC mode dropped in the field-level expression into account, the contribution of the super-sample modes is changed to \cite{Akitsu_Sugiyama_Shiraishi2019}
\begin{align}
 \deltag^{\rm (S)}(\bfk) \longrightarrow \Bigl\{1- \Bigl(b_1+\frac{f}{3}\Bigr)\deltab - f\,\tau_{zz}\Bigr\}\,\deltag^{\rm (S)}(\bfk).
 \label{eq:corrections_local_average}
\end{align}
At the leading order, this leads to a slight change in the power spectrum response as follows:
\begin{align}
\frac{\partial \overline{P}^{\rm (S)}_{\rm gE}(\bfk)}{\partial \deltab}
&\longrightarrow \frac{\partial \overline{P}^{\rm (S)}_{\rm gE}(\bfk)}{\partial \deltab} - \Bigl(b_1+\frac{f}{3}\Bigr)\bK(b_1+f\,\mu_k^2)(1-\mu_k^2),
\label{eq:dpkgE_ddeltab_local}
\\
\frac{\partial \overline{P}^{\rm (S)}_{\rm gE}(\bfk)}{\partial \tau_{zz}} 
&\longrightarrow \frac{\partial \overline{P}^{\rm (S)}_{\rm gE}(\bfk)}{\partial \tau_{zz}} - f\,\bK(b_1+f\,\mu_k^2)(1-\mu_k^2).
\label{eq:dpkgE_dtauzz_local}
\end{align}

Note that Eq.~(\ref{eq:corrections_local_average}) corresponds to the cases adopting the simple density estimator in Refs.~\cite{dePutter_etal2012,Akitsu_Sugiyama_Shiraishi2019,Taruya_Nishimichi_Jeong2021}. Strictly, the corrections due to the local mean depends on the definition of the estimator (e.g., see Ref.~\cite{Digvijay_Scoccimarro2020} for the FKP estimator). This is also the case for the galaxy ellipticity field. That is, depending on the choice of the estimator, the corrections arising from the local mean potentially appear in the case of galaxy ellipticity field. In this paper, we suppose that the ellipticity field is measured with a hypothetical estimator that does not produce such corrections. 

\begin{figure}[tb]
\begin{center}
\includegraphics[width=0.51\linewidth]{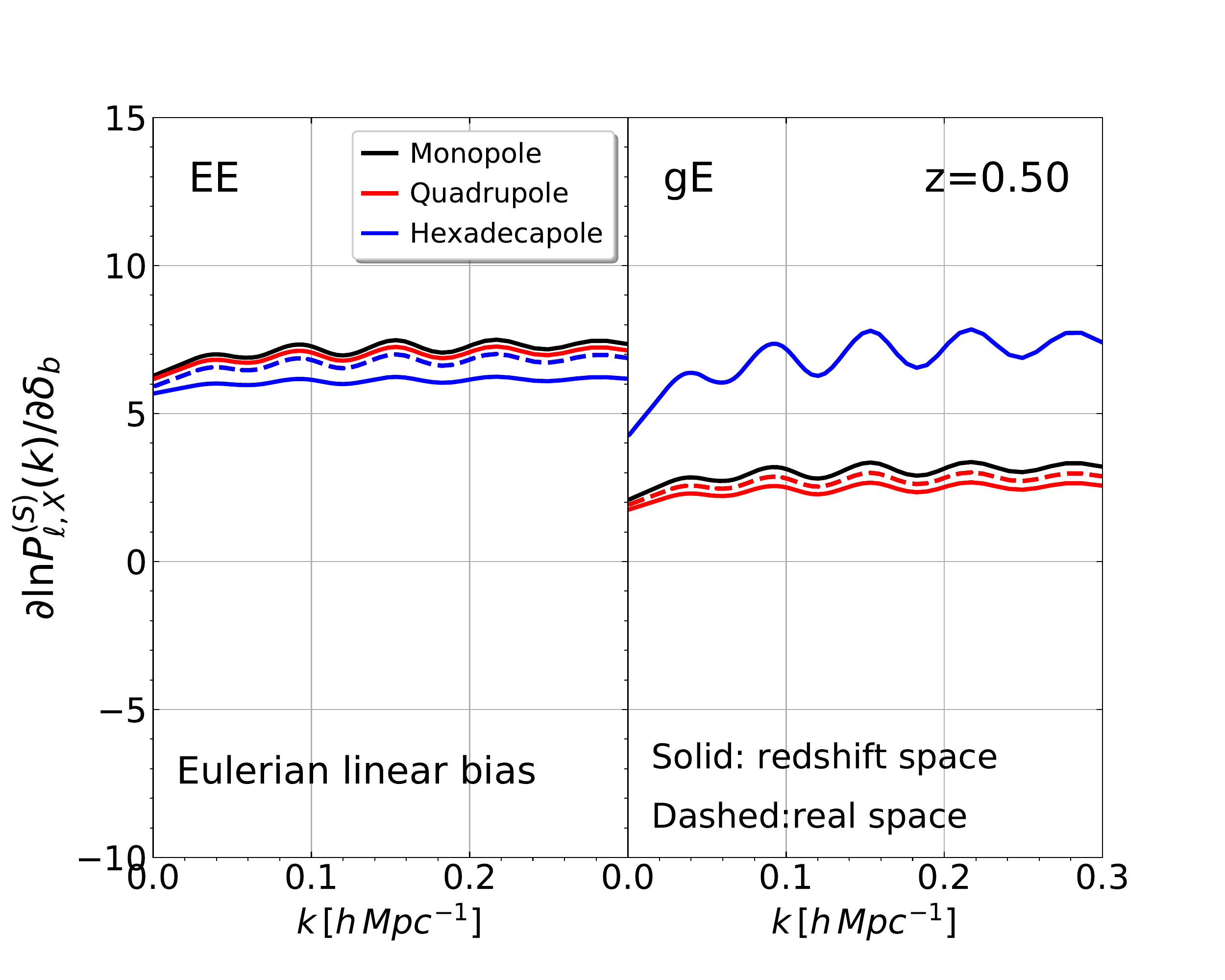}
\hspace*{-1.0cm}
\includegraphics[width=0.51\linewidth]{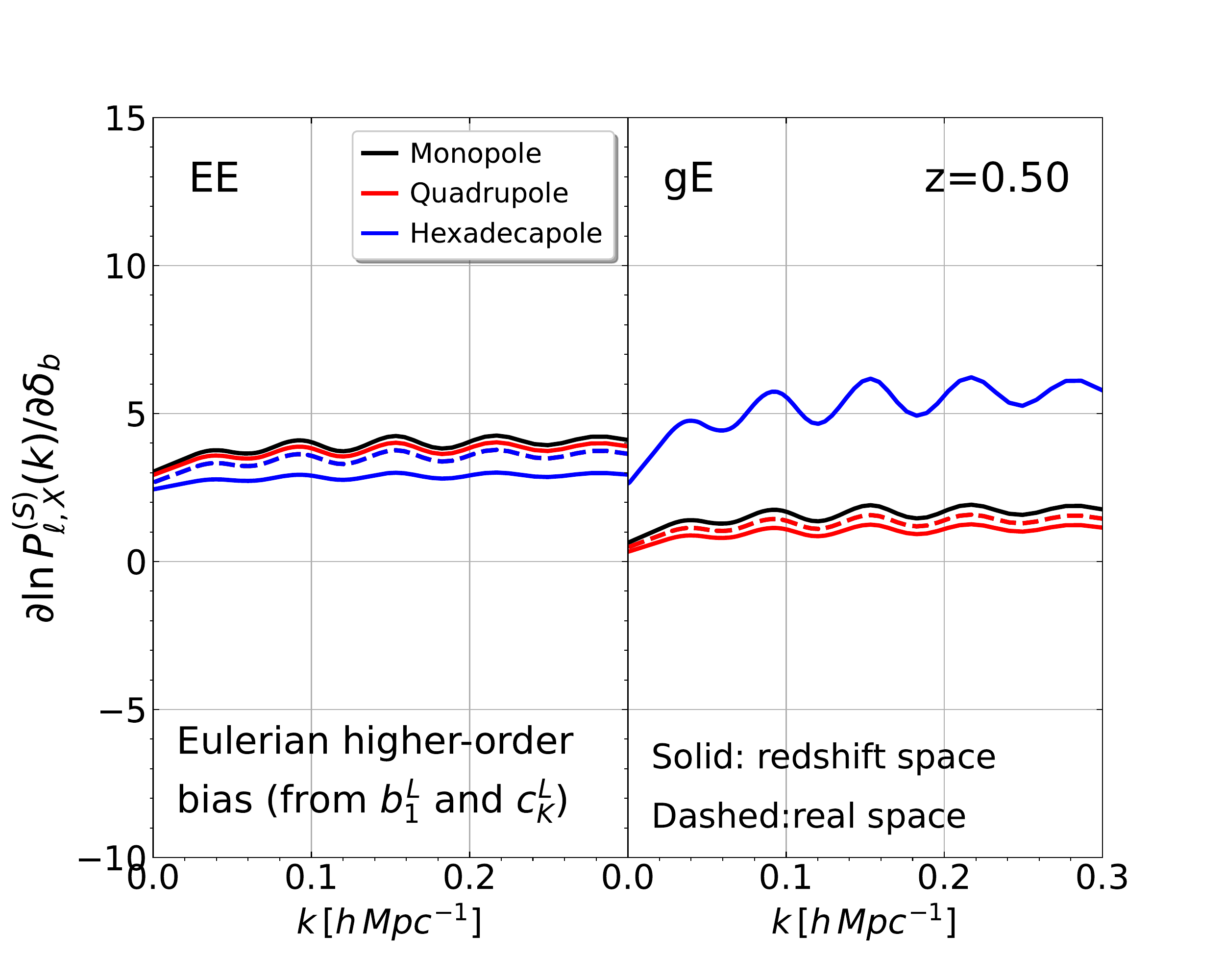}
\end{center}

\vspace*{-0.5cm}

    \caption{Logarithmic response of the power spectrum multipoles to the super-sample mode $\deltab$, given by $\partial\ln \overline{P}_{\ell,{\rm X}}^{\rm(S)}/\partial\deltab$. The results for the monopole ($\ell=0$, black), quadrupole ($\ell=2$, red), and hexadecapole ($\ell=4$, blue) moments taking the local average corrections at Eq.~(\ref{eq:dpkgE_ddeltab_local}) into account are particularly shown at $z=0.5$, assuming the Eulerian linear (left) and higher-order (right) bias. In both cases, the stochastic contributions are ignored (but see Appendix \ref{Appendix:stochasticity}). Here, we take the linear bias parameters to be $b_1=2$ and $\bK=-0.1$ in both cases, and set all the higher-order parameters to zero in the case of Eulerian linear bias, while we set $b_2=(8/21)(b_1-1)$, $b_{s^2}=-(4/7)(b_1-1)$, $\bdeltaK=-(2/3)\bK$, $\bKK=-\bK$ and $\bt=(5/2)\bK$ in the case of Eulerian higher-order bias, meaning that non-vanishing higher-order bias parameters are generated from the Lagrangian linear bias parameters $\bL_1$ and $\cK$. (see the main text in detail). In each panel, left and right plots summarize the results for EE auto- and gE cross-power spectra, respectively. Solid lines represent the response of the redshift-space power spectrum. Dashed lines are the response of the real-space power spectrum, obtained by setting the linear growth rate $f$ to zero. Note that in real space, the logarithmic responses of the EE auto-power spectrum for $\ell=0$, $2$, and $4$ become identical, and only the single line is plotted in left panel. While this is also the case for the gE cross-power spectrum, the hexadecapole moment $(\ell=4)$ is shown to be zero, and is not plotted here. 
    \label{fig:response_deltab}}
\end{figure}
\begin{figure}[tb]
\begin{center}
\includegraphics[width=0.51\linewidth]{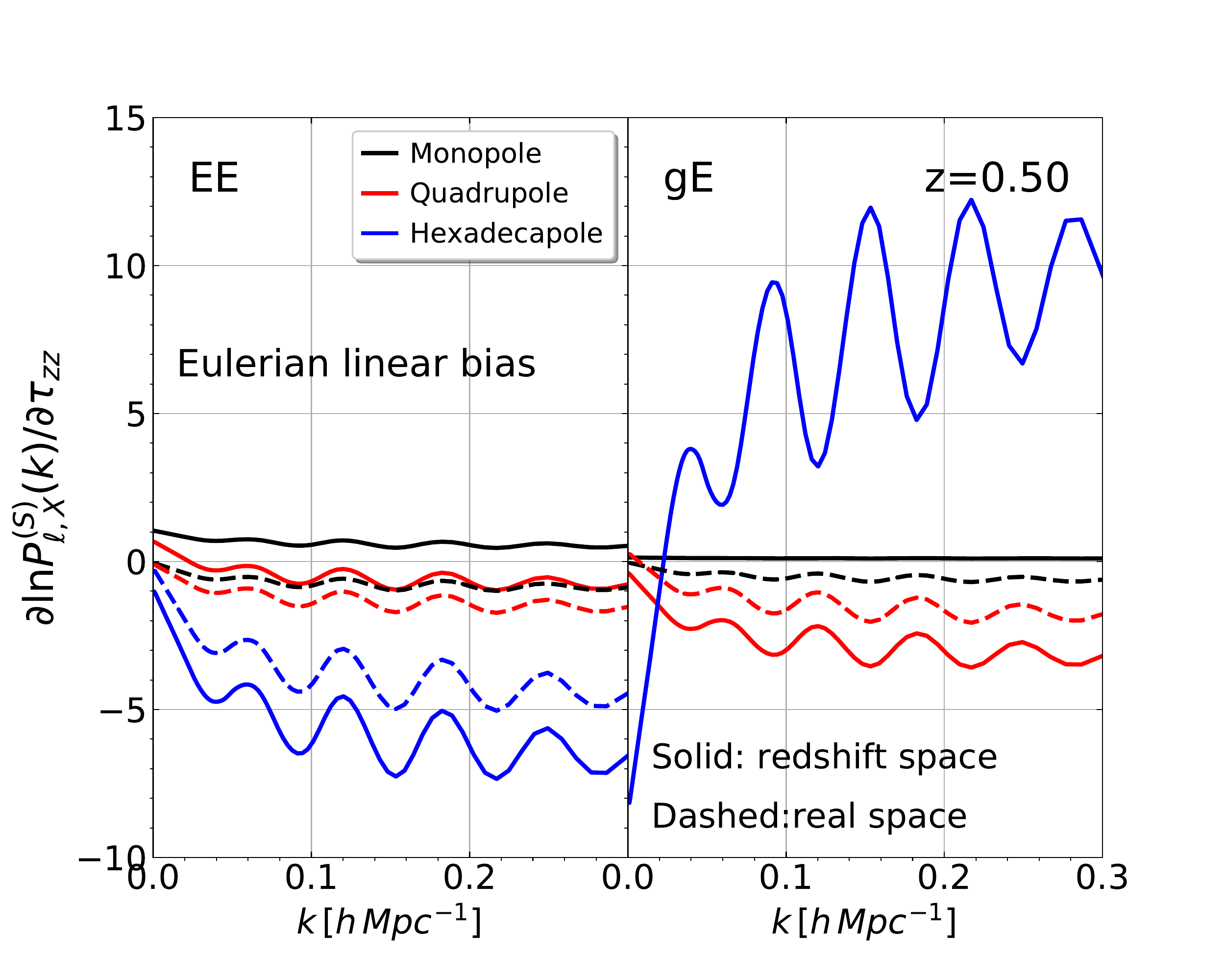}
\hspace*{-1.0cm}
\includegraphics[width=0.51\linewidth]{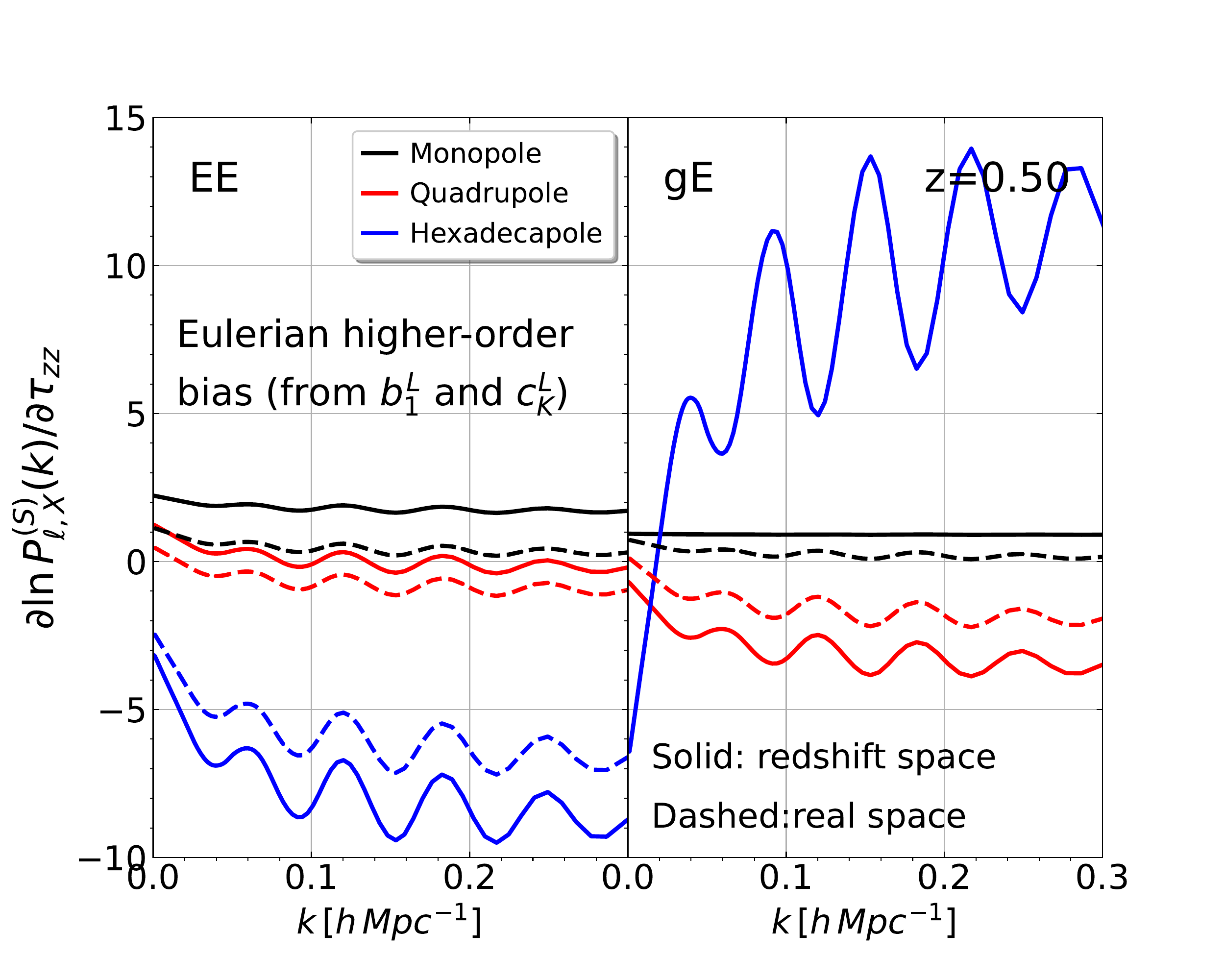}
\end{center}

\vspace*{-0.5cm}

    \caption{Same as Fig.~\ref{fig:response_deltab}, but the results for the logarithmic response to the super-sample tidal field, $\partial\ln \overline{P}_{\ell,{\rm X}}^{\rm(S)}/\partial \tau_{zz}$, are shown, taking the local average correction at Eq.~(\ref{eq:dpkgE_dtauzz_local}) into account. Note that the real-space result of the hexadecapole ($\ell=4$) moment for the gE cross-power spectrum, whose response actually gives non-zero values, is not shown because the spectrum itself vanishes (see Eq.~(\ref{eq:EE_gE_spectrum_multpoles})). 
    \label{fig:response_tauzz}}
\end{figure}

The expressions for the angle-average EE-mode auto and gE cross power spectra given above are the main results in this paper. As a result of the angle average, only the specific components of the super-sample modes, i.e., $\deltab$ and $\tau_{zz}$, affect the observed power spectra. Still, the effects of super-sample modes are evident. On top of the anisotropies inherent in the projected ellipticity and redshift-space density fields, the super-sample modes introduce another type of anisotropies, which result in the non-vanishing multipoles higher than $\ell=4$ (see Appendix \ref{Appendix:multipoles}). This is true even at the leading order. While the impact of this effect is supposed to be small in the measurement of power spectra, the presence of their anisotropies would affect the statistical error estimation through the covariance, which we will discuss in next subsection.

Finally, to elucidate how the super-sample modes modulate the power spectrum measured in the local universe, the power spectrum responses to $\deltab$ and $\tau_{zz}$ are respectively shown in Figs.~\ref{fig:response_deltab} and \ref{fig:response_tauzz}. To characterize their anisotropic nature, we plot the logarithmic responses of the power spectrum multipoles, $\partial \ln \overline{P}_{\ell,{\rm X}}^{\rm(S)}/\partial \deltab$ and $\partial \ln \overline{P}_{\ell,{\rm X}}^{\rm(S)}/\partial \tau_{zz}$, with the power spectrum multipoles $\overline{P}_{\ell,{\rm X}}^{\rm(S)}$ defined by
\begin{align}
 \overline{P}_{\ell,{\rm X}}^{\rm(S)}(k)=\frac{2\ell+1}{2}\int_{-1}^1 d\mu_k\,
\overline{P}_{\rm X}(\bfk)\,\mathcal{P}_\ell(\mu_k), \quad ({\rm X}={\rm EE}\,\,\, \mbox{or}\,\,\,{\rm gE}).
\label{eq:power_spectrum_multipoles}
\end{align}
The function $\mathcal{P}_\ell$ is the Legendre polynomials. Substituting the expressions at Eqs.~(\ref{eq:pk_EE_w_SS-modes})-(\ref{eq:dpkgE_dtauzz}) into the above, the non-vanishing power spectrum multipoles and their responses are analytically derived for even multipoles at $\ell\leq6$, whose explicit expressions are presented in Appendix \ref{Appendix:multipoles}. Note that the non-zero tetrahexacontapole ($\ell=6$) is originated purely from the redshift-space distortions, and it becomes vanishing in real space.

In Figs.~\ref{fig:response_deltab} and \ref{fig:response_tauzz}, the logarithmic responses for the monopole ($\ell=0$, black), quadrupole ($\ell=2$, red) and hexadecapole ($\ell=4$, blue) are computed in real (dashed) and redshift (solid) space, adopting the $\Lambda$ cold dark matter model with its cosmological parameters determined by Planck \cite{Planck2015_XIII}. For the gE cross power spectrum, we take into account the local average corrections given at Eqs.~(\ref{eq:dpkgE_ddeltab_local}) and (\ref{eq:dpkgE_dtauzz_local}) (see Eqs.~(\ref{eq:dpkgE0_ddeltab_local})-(\ref{eq:dpkgE4_dtauzz_local}) for the expressions of the multipole moments). Then, the results at $z=0.5$ are particularly shown in the following two cases, with the linear bias parameters commonly setting to $b_1=2$ and $\bK=-0.1$, based on Refs.~\cite{Okumura_Jing_Li2009,Kurita_etal2021,Jingjing_etal2021}: 

\begin{itemize}
 \item {\bf Eulerian linear bias}: higher-order bias parameters defined in Eulerian space, i.e.,  $b_2$, $b_{s^2}$, $\bdeltaK$, $\bKK$ and $\bt$, are all set to zero. 
 \item {\bf Eulerian higher-order bias (from $\bL_1$ and $\cK$)}: higher-order bias parameters defined in Lagrangian space, i.e., $b_2^{\rm L}$, $b_{s^2}^{\rm L}$, $\cdeltaK$, $\cKK$, and $\ct$, are all set to zero, but the linear-order parameters $\bL_1$ and $\cK$ are kept non-zero. This implies that through the mapping relations at Eqs.~(\ref{eq:bias_delta_Lagrangian_Eulerian}) and (\ref{eq:bias_delta_gamma}), the non-vanishing higher-order Eulerian bias parameters are generated, and we have $b_2=(8/21)(b_1-1)$, $b_{s^2}=-(4/7)(b_1-1)$, $\bdeltaK=-(2/3)\bK$, $\bKK=-\bK$, and $\bt=(5/2)\bK$.  
\end{itemize}

Figs.~\ref{fig:response_deltab} and \ref{fig:response_tauzz} show that the logarithmic responses exhibit a prominent oscillatory feature. This comes from the so-called dilation effect (e.g., \cite{Li_Hu_Takada2014a,Li_Schmittfull_Seljak2018}) through the logarithmic derivative of the power spectrum, in which the baryon acoustic oscillations are clearly visible. On top of these characteristic features, logarithmic responses to the super-sample overdensity $\deltab$ have a positive offset with comparable values in both EE auto- and gE cross-power spectra, indicating the enhancement of power spectrum amplitude in the presence of positive $\deltab$. On the other hand, the responses to $\tau_{zz}$ exhibit both positive and negative offsets with a rather clear acoustic feature. In particular, the response of the hexadecapole gE cross spectrum shows a notable behavior having a rather large oscillation with opposite phase\footnote{Note that the terms responsible for the dilation effect in the power spectrum response change their overall sign depending on the multipoles. Hence, the phase of the oscillation can become opposite, and with a dominant contribution of the dilation effect, the logarithmic response $\partial\ln \overline{P}_{4,{\rm gE}}^{\rm(S)}/\partial\tau_{zz}$ exhibit a rather prominent oscillatory behavior.}, indicating that the responses to the tidal fields tend to be dominated by the dilation effect.  
These results show that the super-sample modes introduce an anisotropic deformation, and modulates the primary anisotropies inherent in the EE auto- and gE-cross power spectra, similarly to the case of galaxy power spectrum, $P_{\rm gg}^{\rm (S)}$ (e.g., \cite{Akitsu_Takada_Li2017,Akitsu_Takada2018,Li_Schmittfull_Seljak2018}). It is to be noted that these behaviors also depend on the choice of bias parameters, and are, in particular, sensitive to the higher-order parameters. This point will be also discussed in next subsection.

\subsection{Super-sample covariance}
\label{subsec:SSC_IA}

As one of the important effects discussed in the literature, the uncertainty of the amplitude of super-sample modes induces the new term in the covariance matrix of the power spectrum measured from a finite-volume survey, referred to as the super-sample covariance \cite{Takada_Hu2013}. 
Since this effect comes from a non-trivial modulation of the mode coupling, the super-sample covariance has the non-vanishing off-diagonal components.
Given the power spectrum response as derived in previous section, we are now able to evaluate quantitatively the super-sample covariance for the IA statistics, and its impact on the signal-to-noise ratio.

As we have seen, the EE auto- and gE cross-power spectra exhibit anisotropic nature, and are given as a function of $k$ and $\mu_k$. This is also the case in real space. The multipole expansion given at Eq.~(\ref{eq:power_spectrum_multipoles}) thus provides a convenient basis to characterize their anisotropies. Consider a finite-volume survey with survey window function, $W(\bfx)$. For simplicity, we assume that the function $W(\bfx)$ takes either $1$ or $0$, depending on whether the position $\bfx$ is inside the survey region or not. We then adopt the following estimator for the multipole moments of the power spectra, $\overline{P}^{\rm (S)}_{\rm EE}$ and $\overline{P}^{\rm (S)}_{\rm gE}$, which we respectively denote by $\hat{P}_{\ell,{\rm EE}}^{\rm(S)}$ and $\hat{P}_{\ell,{\rm gE}}^{\rm(S)}$ (e.g., Refs.~\cite{Scoccimarro1999,Takada_Hu2013,Akitsu_Takada_Li2017,Akitsu_Takada2018,Li_Schmittfull_Seljak2018}): 
\begin{align}
 \hat{P}_{\ell,{\rm EE}}(k_i)&=\frac{2\ell+1}{V_{\rm W}}\int_{\bfk \in k_i}\frac{d^3\bfk}{V_{k_i}}\,\gamma_{\rm E,W}^{\rm(S)}(\bfk)\gamma_{\rm E,W}^{\rm(S)}(-\bfk)\mathcal{P}_\ell(\mu_{k}),
\label{eq:estimator_pk_EE}
\\
 \hat{P}_{\ell,{\rm gE}}(k_i)&=\frac{2\ell+1}{V_{\rm W}}\int_{\bfk \in k_i}\frac{d^3\bfk}{V_{k_i}}\,\mbox{Re}\,\bigl[\delta_{\rm g,W}^{\rm(S)}(\bfk)\gamma_{\rm E,W}^{\rm(S)}(-\bfk)\bigr]\mathcal{P}_\ell(\mu_{k}),
\label{eq:estimator_pk_gE}
\end{align}
where the function $\mathcal{P}_\ell(\mu)$ is the Legendre polynomials, with its argument $\mu_k$ given by $\mu_k=\hatk_z$. The integral is taken over a shell in Fourier space of the width $\Delta k$ and volume $V_{k_i}\simeq 4\pi k_i^2\Delta k$ for $\Delta k/k_i\ll1$. Here, the fields with subscript W, i.e., $\delta^{\rm(S)}_{\rm g,W}$ and $\gamma^{\rm(S)}_{\rm E,W}$, imply those convolved with the Fourier transform of the survey window function, $W(\bfk)$. The quantity $V_{\rm W}$ represents the effective survey volume defined by   
\begin{align}
 V_{\rm W}=\int d^3\bfx~ W(\bfx).
\end{align}
Note that the estimators given at Eqs.~(\ref{eq:estimator_pk_EE}) and (\ref{eq:estimator_pk_gE}) are regarded as the un-biased estimator, and for the mode $k$ with $k\gg 2\pi/V_{\rm W}^{1/3}$ of our interest,  we have $\langle\hat{P}_{\ell,{\rm X}}^{\rm(S)}(k)\rangle\simeq \overline{P}_{\ell,{\rm X}}^{\rm(S)}(k)$, with the function $\overline{P}_{\rm X}$ being the angle-averaged power spectrum multipole defined at Eq.~(\ref{eq:power_spectrum_multipoles}).

Given the power spectrum estimators, the covariance matrix of the multipole moment of the power spectrum is defined by
\begin{align}
 \cov_{\ell,\ell'}^{\rm X,X'}(k,k')\equiv\langle\hat{P}_{\ell,{\rm X}}(k)\hat{P}_{\ell',{\rm X'}}(k')\rangle-\langle\hat{P}_{\ell,{\rm X}}(k)\rangle\langle\hat{P}_{\ell,{\rm X'}}(k')\rangle, ({\rm X}={\rm EE}\,\,\, \mbox{or}\,\,\,{\rm gE})
\end{align}
which is generally decomposed into three contributions:
\begin{align}
 \cov_{\ell,\ell'}^{\rm X,X'}(k,k')=\, ^{\rm G}\cov_{\ell,\ell'}^{\rm X,X'}(k,k') +\, ^{\rm non\mbox{-}G}\cov_{\ell,\ell'}^{\rm X,X'}(k,k')+\, ^{\rm SSC}\cov_{\ell,\ell'}^{\rm X,X'}(k,k').
\label{eq:cov_G-NG-SSC}
\end{align}
The first term at the right-hand side represents the Gaussian contribution. For the sub-survey modes of $k\gg 2\pi/V_{\rm W}^{1/3}$, the expression is simplified, including the shot noise and shape noise contributions characterized by the number density of galaxies $\overline{n}_{\rm gal}$ and scatter in the intrinsic shape per component $\sigma_\gamma$ (see Eqs.~(\ref{eq:Gauss_Cov_EE_EE}) and (\ref{eq:Gauss_Cov_gE_gE})). We can derive analytical formulas for the Gaussian covariances, $^{\rm G}\cov_{\ell,\ell'}^{\rm EE,EE}$ and $^{\rm G}\cov_{\ell,\ell'}^{\rm gE,gE}$, and in Appendix \ref{Appendix:multipoles}, we summarize their expressions especially in the cases of $\ell=\ell'$ up to $\ell=4$. 

In Eq.~(\ref{eq:cov_G-NG-SSC}), the second and third terms at right-hand side describe the contributions arising from the mode coupling, which results in the non-vanishing off-diagonal components of the covariance matrix. To be precise, the second term characterizes the non-Gaussian contribution originated from the coupling between sub-survey modes. The third term is the super-sample covariance that includes the super-sample modes, and represents the coupling between super-sample and sub-survey modes. In practice, a proper way to estimate the signal-to-noise ratio needs all the three contributions, among which the non-Gaussian covariance is known to give a non-negligible impact at small scales and low redshifts in the case of matter power spectra (e.g., Refs.~\cite{Takahashi:2009bq,Blot2015,Taruya_Nishimichi_Jeong2021}). Nevertheless, its impact on the IA statistics is not fully explored, and one needs a further study by using numerical simulations (but see Ref.~\cite{Kurita_etal2021}). In what follows, simply ignoring the non-Gaussian contribution, we shall below estimate the signal-to-noise ratio, focusing especially on large-scale sub-survey modes, for which the non-Gaussian contribution is expected to be mild.

\begin{figure}[tb]
\begin{center}
\includegraphics[width=0.53\linewidth]{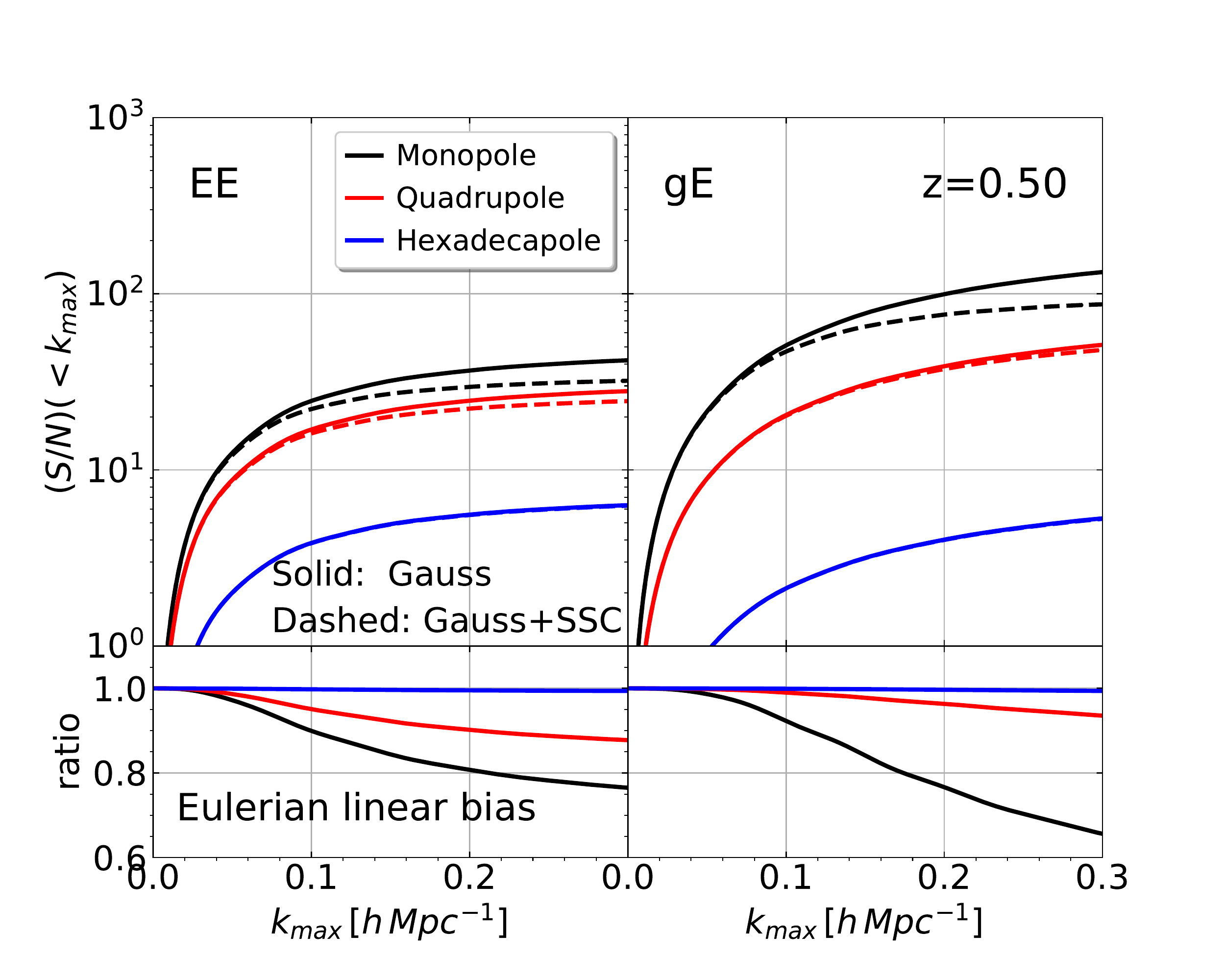}
\hspace*{-1.2cm}
\includegraphics[width=0.53\linewidth]{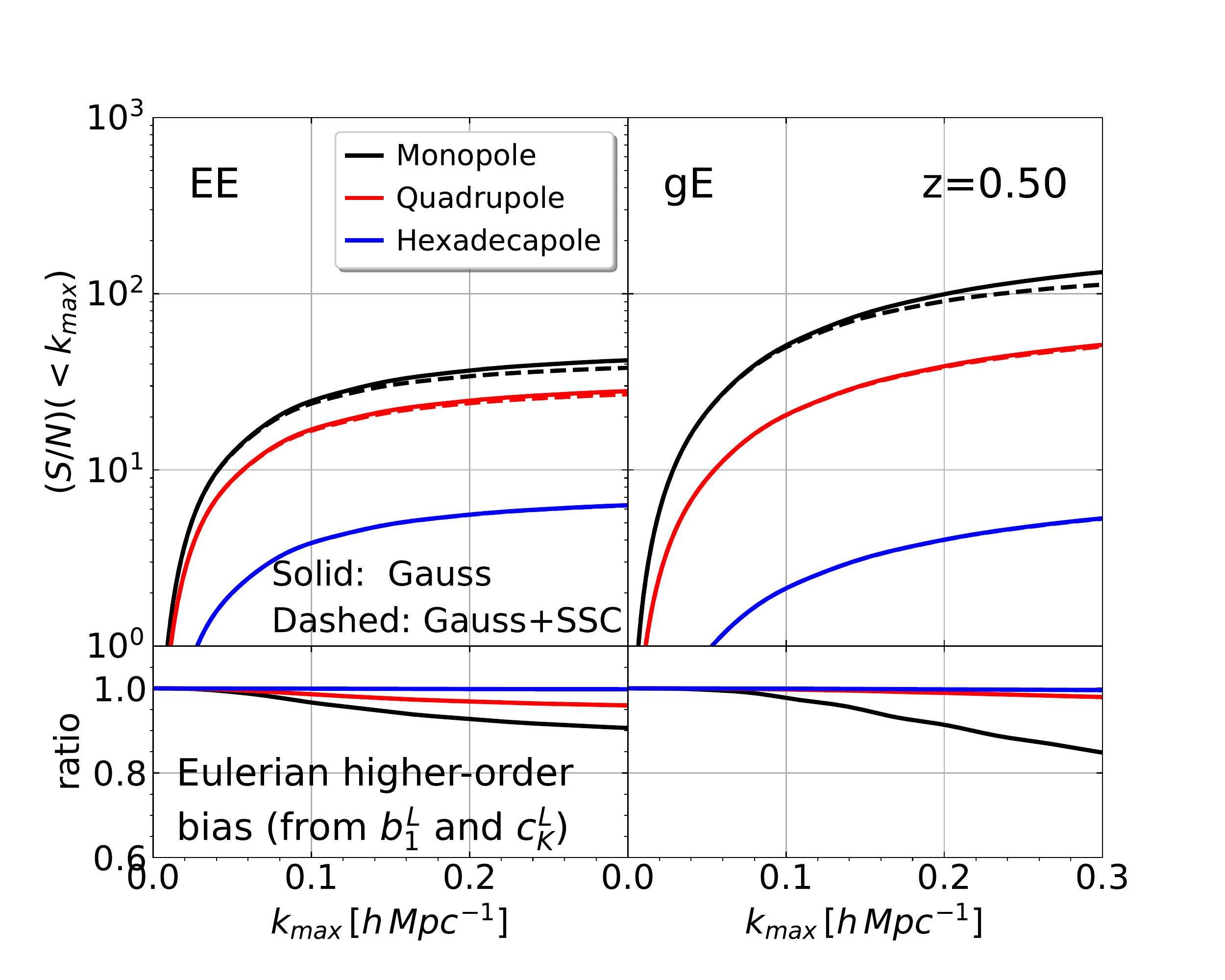}
\end{center}

\vspace*{-0.5cm}

    \caption{Cumulative signal-to-noise ratio, $({\rm S/N})(\leq k_{\rm max})$, for monopole (black), quadrupole (red) and hexadecapole (blue) moments of the redshift-space power spectra, $P_{\rm EE}^{\rm(S)}$ (left) and $P_{\rm gE}^{\rm(S)}$ (right).  Upper panels summarize the results with (dashed) and without (solid) the super-sample covariance (SSC), assuming the Gaussian covariance. Lower panels show the ratio of $(S/N)(\leq k_{\rm max})$ including SSC to that without SSC. Here, we consider a hypothetical survey of the volume $V_{\rm W}=1\,h^{-3}\,$Gpc$^3$ having a spherical survey geometry with the galaxy number density, $n_{\rm gal}=5\times10^{-4}\,(h^{-1}\mbox{Mpc})^{-3}$, and the scatter in the intrinsic shape per component, $\sigma_\gamma=0.2$. Left and right panels respectively plot the results assuming the Eulerian linear and higher-order bias as we examined in Figs.~\ref{fig:response_deltab} and \ref{fig:response_tauzz}, adopting also the same bias parameters (see also the main text in Sec.~\ref{subsec:pk_IA}). In both cases, we neglect the stochastic bias contributions (but see Appendix \ref{Appendix:stochasticity}).
    \label{fig:SNR_red}}
\end{figure}
\begin{figure}[tb]
\begin{center}
\includegraphics[width=0.53\linewidth]{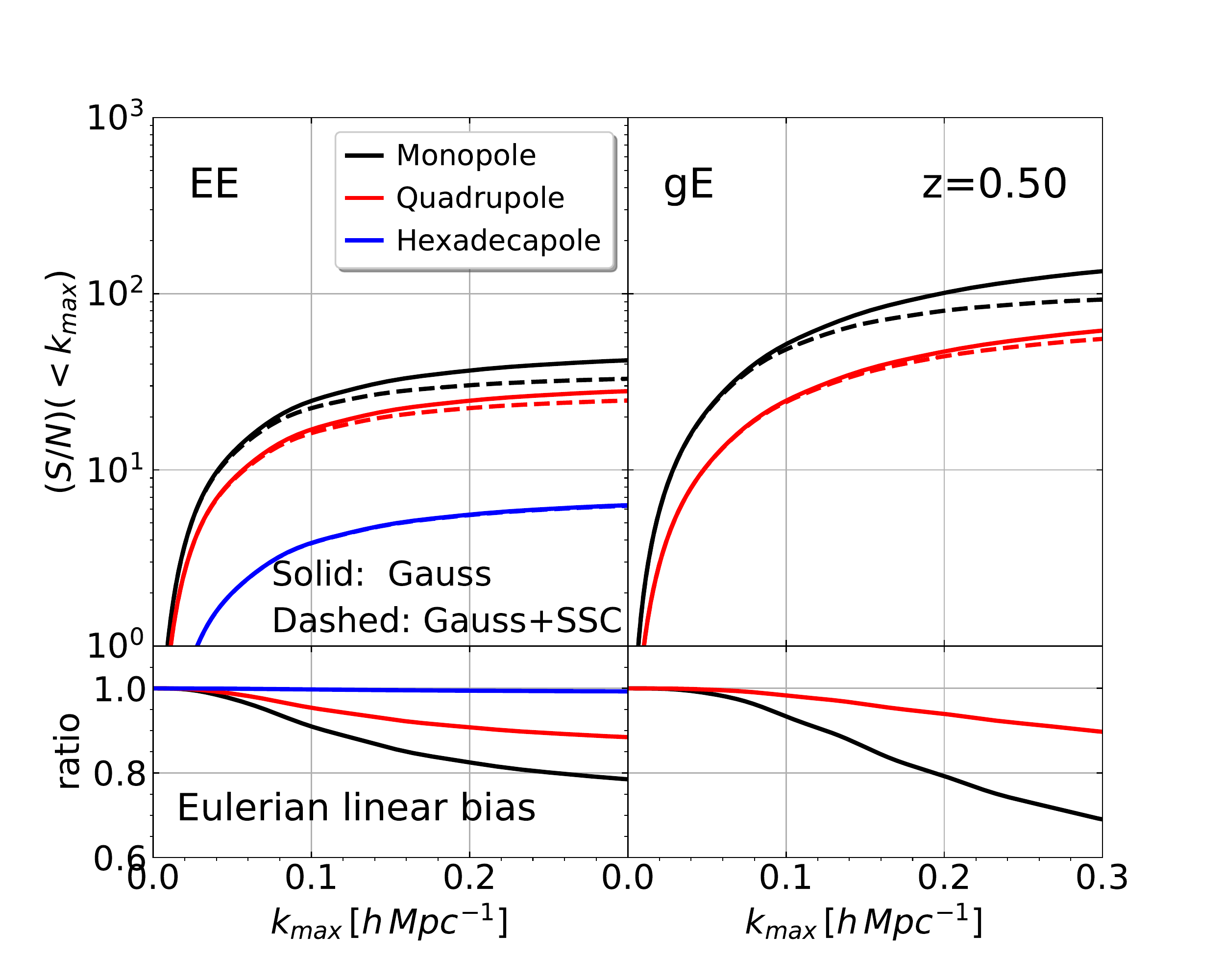}
\hspace*{-1.2cm}
\includegraphics[width=0.53\linewidth]{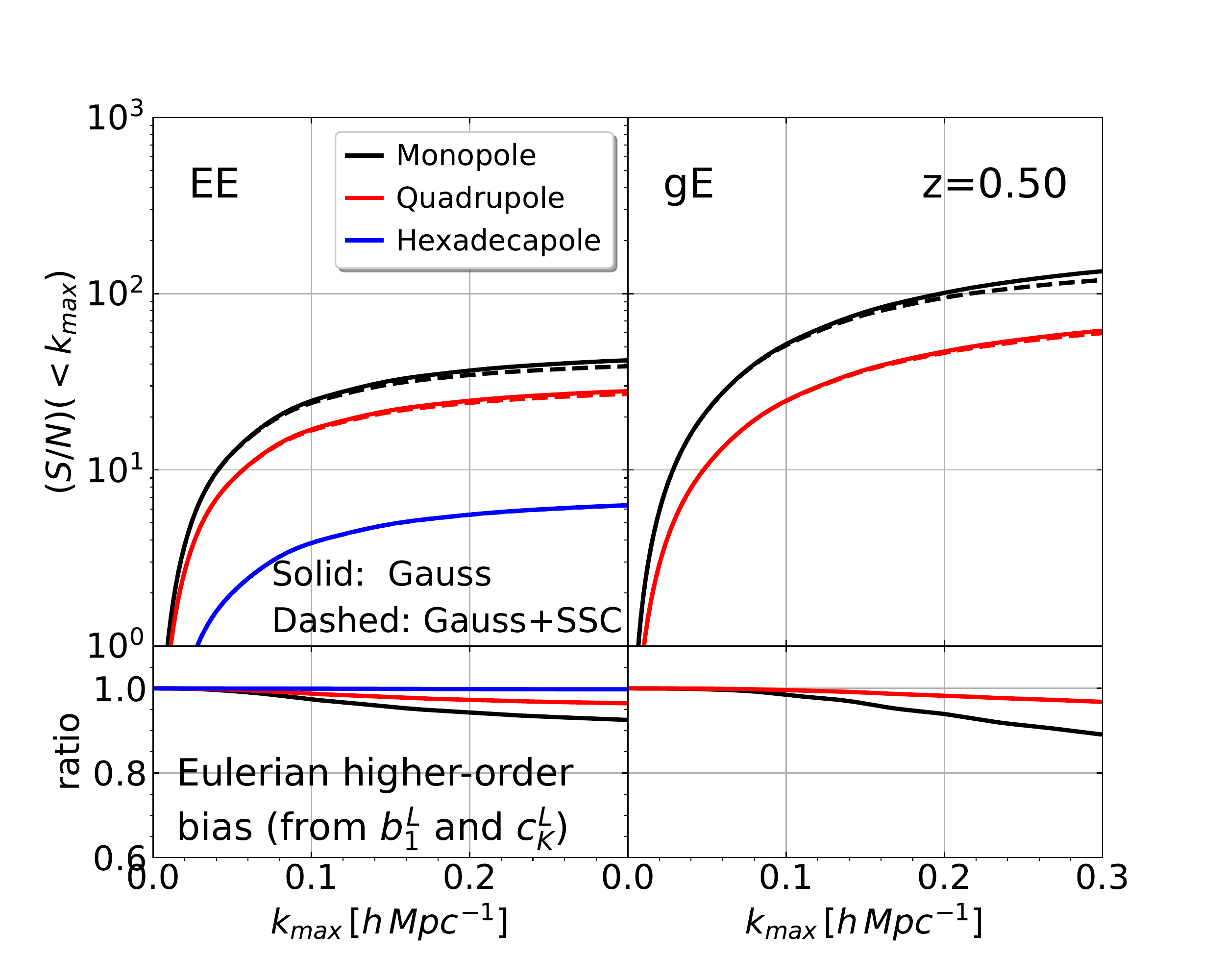}
\end{center}

\vspace*{-0.5cm}

    \caption{Same as Fig.~\ref{fig:SNR_red}, but the results for real-space spectra are shown, setting simply the linear growth rate $f$ to zero. Note that the hexadecapole moment of the cross spectrum $P_{\rm gE}$ becomes vanishing at leading order in real space, and we do not show its signal-to-noise ratio. 
    \label{fig:SNR_real}}
\end{figure}

As in the case of the galaxy power spectrum in real space discussed in the literature \cite{Takada_Hu2013}, the super-sample covariance of the power spectrum multipoles is shown to be expressed in terms of the power spectrum response \cite{Li_Schmittfull_Seljak2018,Digvijay_Scoccimarro2020}. 
Ignoring a possible correlation between $\deltab$ and $\tau_{zz}$ (i.e., $\langle\deltab\tau_{zz}\rangle=0$)\footnote{The correlation $\langle\deltab\tau_{zz}\rangle$ becomes exactly zero for a spherically symmetric survey window function. In general, it could have a non-zero value, but the size of correlation is negligibly small, in most of the cases, compared to $\langle\deltab^2\rangle$ and $\langle \tau_{zz}^2\rangle$.}, the super-sample covariances for EE auto- and gE cross-power spectra are respectively given as follows: 
\begin{align}
^{\rm SSC}\cov_{\ell,\ell'}^{\rm X,X'}(k,k') &=\sigma_b^2\,\frac{\partial \overline{P}_{\ell,{\rm X}}^{\rm(S)}(k)}{\partial \deltab}\frac{\partial \overline{P}_{\ell',{\rm X'}}^{\rm(S)}(k')}{\partial \deltab}+\sigma_{\tau}^2\,\frac{\partial \overline{P}_{\ell,{\rm X}}^{\rm(S)}(k)}{\partial \tau_{zz}}\frac{\partial \overline{P}_{\ell',{\rm X'}}^{\rm(S)}(k')}{\partial \tau_{zz}},\quad ({\rm X}, {\rm X'}={\rm EE\,\, or\,\, \rm gE}),
\label{eq:cov_SSC_XX'}
\end{align}
with the non-vanishing multipole of the responses, $\partial\overline{P}^{\rm(S)}_{\ell,X}/\partial\deltab$ and $\partial\overline{P}^{\rm(S)}_{\ell,X}/\partial\tau_{zz}$, summarized in Appendix \ref{Appendix:multipoles}. The quantities $\sigma_b^2$ and $\sigma_\tau^2$ are the dispersion of the super-sample modes, $\sigma_b^2\equiv\langle\delta_b^2\rangle$ and $\sigma_\tau^2\equiv\langle\tau_{zz}^2\rangle$:
\begin{align}
\sigma_b^2&=\int\frac{d^3\bfk}{(2\pi)^3}\,|W(\bfk)|^2P_{\delta\delta}(k),
\label{eq:sigma2_b}\\
\sigma_\tau^2&=\int\frac{d^3\bfk}{(2\pi)^3}\,\Bigl(\hatk_z^2-\frac{1}{3}\Bigr)^2|W(\bfk)|^2P_{\delta\delta}(k).
\label{eq:sigma2_tau}
\end{align}

To illustrate the impact of the super-sample covariance, we estimate the signal-to-noise ratio of the EE auto- and gE cross-power spectra, defined by
\begin{align}
 \Bigl(\frac{\rm S}{\rm N}\Bigr)_{\ell,{\rm X}}(<k_{\rm max})=
\Biggl[\sum_{i,j}P_{\ell,{\rm X}}^{\rm (S)}(k_i)
\bigl\{\cov_{\ell,\ell'}^{\rm X,X}(k_i,k_j)\bigr\}^{-1}P_{\ell,{\rm X}}^{\rm(S)}(k_j)\Biggr]^{1/2}, \quad ({\rm X}={\rm EE}\,\,\mbox{or}\,\,{\rm gE}).
\end{align}
Figs.~\ref{fig:SNR_red} and \ref{fig:SNR_real} respectively show the results of the signal-to-noise ratios in redshift and real space. Ignoring the non-Gaussian contributions $^{\rm non\mbox{-}G}\cov_{\ell,\ell'}^{\rm X,X'}$, we consider specifically a hypothetical survey of the volume $V_{\rm W}=1\,h^{-3}$\,Gpc$^3$ with a spherical survey geometry, assuming the number density of galaxies, $n_{\rm gal}=5\times10^{-4}\,h^3$\,Mpc$^{-3}$, and the scatter of IA, $\sigma_\gamma=0.2$. The plotted results are the signal-to-noise ratio at $z=0.5$ in the cases of the Eulerian linear (left) and higher-order (right) bias relations as we examined in Figs.~\ref{fig:response_deltab} and \ref{fig:response_tauzz},  adopting also the same cosmological model and bias parameters. Note that the estimated signal-to-noise ratios here are the results for the specific survey volume, but these are found to approximately scale as the square root of the survey volume\footnote{To be precise, the Gaussian covariance is inversely proportional to $V_{\rm W}$ (see Eqs.~(\ref{eq:Gauss_Cov_EE_EE}) and (\ref{eq:Gauss_Cov_gE_gE})), but the super-sample covariance depends on the dispersions $\sigma_b^2$ and $\sigma_\tau^2$, whose scaling is not trivial in general. Nevertheless, with the power spectrum given by $\Lambda$CDM model, Refs.~\cite{Akitsu_Takada_Li2017,Takada_Hu2013} found that these are mostly proportional to the inverse of survey volume. That is, as long as we consider the universe close to $\Lambda$CDM model, all the contributions to the covariance matrix approximately scale as ${\rm Cov}_{\ell,\ell'}^{\rm X,X'}\,\propto\, 1/V_{\rm W}$, and thus the signal-to-noise ratio is approximately proportional to the square root of the survey volume. }, i.e., $({\rm S}/{\rm N})\propto\,V_{\rm W}^{1/2}$. Thus,  the ratio of these signal-to-noise ratios, shown in lower panel, would remain unchanged irrespective of the survey volume.

Figs.~\ref{fig:SNR_red} and \ref{fig:SNR_real} show that the super-sample covariance suppresses the signal-to-noise ratios for both EE auto- and gE cross-power spectra, and the amount of the suppression depends on the bias prescription and the choice of parameters\footnote{ The value of the shape bias varies depending on the choice of the estimator. Because some estimators contain not only the ellipticity fields but also size fluctuations, the choice of the estimator may also affect the response functions themselves (See the discussion in Sec.~\ref{sec:SSmodes_IA}). However, taking into account the size perturbations leads to just adding some new terms to the results presented in this paper. Hence, even for such estimators our results can be applied with a little modification.}. These are essentially the same as those seen in the galaxy power spectrum. In particular, unlike a naive anticipation from Figs.~\ref{fig:response_deltab} and \ref{fig:response_tauzz}, the impact of the super-sample covariance is larger for lower multipoles. This is because the Gaussian covariance tends to dominate 
the estimation of signal-to-noise ratios for higher multipoles. As a result, the impact of the  super-sample covariance is especially prominent for the monopole gE cross-power spectrum assuming the Eulerian linear bias.  Strictly, the results shown here are based on the leading-order calculations, and they would not be quantitatively trusted if one goes to smaller scales $k\gtrsim 0.2\,h$\,Mpc$^{-1}$. Nevertheless, Figs.~\ref{fig:SNR_red} and \ref{fig:SNR_real} indicate that a sizable amount of the contribution from the super-sample covariance is expected, and a proper treatment of the super-sample modes would be crucial in cosmological parameter estimation from the observables related to the IA statistics.

\section{Conclusion and discussion}
\label{sec:conclusion}

In this paper, we have presented a formalism to systematically compute the responses of the tracers of matter fluctuations (i.e., galaxy density and ellipticity fields) to the long modes whose wavelengths exceed the observable scales of a galaxy survey, referred to as the super-sample modes. Unlike previous works, the formalism developed here provides a way to directly compute the contributions from the super-sample modes at the field level. This is based on the Lagrangian treatment, and all the effects of super-sample modes are encapsulated in the quantities defined in Lagrangian space, among which the displacement field $\bfpsi$ plays a key role, as well as to obtain the Eulerian-space observables. Including both the super-sample overdensity and tidal field, we explicitly show that the present formalism reproduces the results of the power spectrum responses known in the literature. Then, as an application, we consider the intrinsic alignment of galaxies, and derive the expressions of the galaxy ellipticity field including the contributions of super-sample modes.

After the E-/B-mode decomposition for the projected ellipticity, the resultant expression for the ellipticity field given in the three-dimensional space yields a non-zero B-mode ellipticity, and hence the EB and gB cross-power spectra are apparently generated. This is a direct manifestation that the super-sample modes affects the statistical nature of sub-survey modes, and violate the parity symmetry. However, these spectra, characterized with the three-dimensional wave vector, might not be directly observables due to a limited number of Fourier modes in a finite-volume survey. Rather, what are practically measurable would be the spectra taking the angle average over the sky. It is then shown that only the EE-mode auto- and gE cross-power spectra become non-vanishing. Their explicit expressions are summarized at Eqs.~(\ref{eq:pk_EE_w_SS-modes})-(\ref{eq:dpkgE_dtauzz}) (see also Appendix \ref{Appendix:multipoles} for their multipole moments). The leading-order expressions of the power spectrum responses involve higher-order bias parameters of the galaxy density and shape biases, and their amplitude sensitively changes depending on the choice of parameters. As a result, the signal-to-noise ratio is suppressed in the presence of super-sample covariance, typically by $5-15\%$ for the EE auto-power spectrum and $20-30\%$ for the gE cross-power spectrum at $k\sim0.2\,h$\,Mpc$^{-1}$, compared with those ignoring the super-sample covariance.

Note, however, that the signal-to-noise ratio estimation given here is based on the leading-order calculations, ignoring also the non-Gaussian contribution to the covariance. A more realistic estimation 
needs non-linear short-mode contributions to the super-sample covariance, taking further the non-Gaussian covariance into account. In this respect, the present estimation just serves as a guideline, and may be used for a consistency check for a more quantitative estimation with $N$-body simulations. Nevertheless, the leading-order results in this paper are still useful in estimating the higher-order bias parameters. Comparing the power spectrum responses measured in the separate universe simulations \cite{Li_Hu_Takada2014a,Wagner_etal2015,Baldauf_etal2016,Li_Hu_Takada2016,Lazeyras_etal2016,Masaki_Nishimichi_Takada2020,Akitsu_Li_Okumura2020,Stucker_etal2021}, our analytical formulas valid at large scales provide a way to disentangle the dependence of bias parameters, leading, in particular, to an accurate determination of the higher-order shape bias parameters.

Finally, a great emphasis on the present Lagrangian-based formalism is that the calculations of the field-level responses to the super-sample modes are rather straightforward at each step, and have no ambiguous points, compared to the Eulerian field-level treatment, in which the apparently divergent shift terms arising from the long-mode contributions need to be removed in order to get a correct result  (e.g., \cite{KwanChan_etal2018}). While our focus here is to derive the field-level response at leading order, it would be interesting to derive the corrections to the sub-survey modes at next-to-leading order, in particular, at the field level. This would open a way to compute the bispectrum response as well as the power spectrum response of the sub-survey modes at one-loop order. These are left to our future work.

\acknowledgments
This work was supported in part by MEXT/JSPS KAKENHI Grant Nos. JP17H06359, JP20H05861, and 21H01081 (AT). AT also acknowledges the support from JST AIP Acceleration Research Grant No. JP20317829, Japan. KA is supported by JSPS Overseas Research Fellowships. KA also acknowledges the Center for Computational Astrophysics, National Astronomical Observatory of Japan, for providing the computing resources of Cray XC50.

\appendix
\section{Redshift-space density fields with super-sample modes: derivation of field-level expression}
\label{sec:deltag_redshift_SS-mode}

In this Appendix, we present the derivation for the redshift-space galaxy density field with super-sample modes, given at Eq.~(\ref{eq:deltag_redshift_SS-modes}). 

Starting with the expression given at Eq.~(\ref{eq:deltag_redshift}), we first expand the displacement and Lagrangian density fields
$\bfpsi^{{\rm S}}$ and $\deltag^{\rm L}$ in the integrand. Substituting Eqs.~(\ref{eq:Lagrangian_bias_expansion}) and (\ref{eq:n-th_order_displacement_z-space}) into Eq.~(\ref{eq:deltag_redshift}), expanding up to the second order becomes
\begin{align}
1+ \deltag^{\rm(S)}(\bfs)&\simeq \int d^3\bfq\int\frac{d^3\bfk}{(2\pi)^3}\,e^{i\,\bfk\cdot(\bfs-\bfq)}
\,\Biggl[
1+b_1^{\rm L}\deltalin(\bfq)-\partial_{q_j}\bigl\{R_{jk}^{(1)}\psi_k^{(1)}(\bfq)\bigr\}
\nonumber
\\
&\quad
+\frac{1}{2}R_{i\ell}^{(1)}R_{jm}^{(1)} \partial_{q_i}\partial_{q_j}\bigl\{\psi_\ell^{(1)}(\bfq)\psi_m^{(1)}\bigr\}^2
-b_1^{\rm L}\partial_{q_j}\bigl\{ R_{jk}^{(1)} \psi_k^{(1)}(\bfq)\deltalin(\bfq)\bigr\} 
\nonumber
\\
&\quad
-\nabla_{q_j}\bigl\{R_{jk}^{(2)}\psi_k^{(2)}(\bfq)\bigr\}
+\frac{1}{2}b_2^{\rm L}\,\{\deltalin(\bfq)\}^2+\frac{1}{2}b_{s^2}^{\rm L}\,C_{ij}(\bfq)C_{ij}(\bfq)+\cdots\Biggr].
\label{eq:deltaz_expanded_2nd-order}
\end{align}
Here, similarly to the real-space case, the integration by parts has been partly performed for the terms involving the spatial derivatives. 
Then, each term in the bracket is decomposed into short- and long-mode contributions according to the prescription in Sec.~\ref{subsec:SSmodes_GG}. We have
\begin{align}
b_1^{\rm L}\deltalin(\bfq)\,\,& \longrightarrow\,\,b_1^{\rm L}\,\bigl\{\deltas(\bfq)+\deltab\bigr\},
\label{eq:1st_2nd_terms}
\\
-\partial_{q_j}\{ R_{jk}^{(1)}\psi_k^{(1)}(\bfq) \} \,\,& \longrightarrow\,\,\Bigl(1+f\,\frac{\partial_{q_z}^2}{\partial_q^2}\Bigr)\deltas(\bfq)+\Bigl(1+\frac{f}{3}\Bigr)\deltab+f\,\tau_{zz},
\label{eq:3rd_term}
\\
\frac{1}{2}R_{i\ell}^{(1)}R_{jm}^{(1)} \partial_{q_i}\partial_{q_j}\Bigl\{\psi_\ell^{(1)}(\bfq)\psi_m^{(1)}(\bfq)\Bigr\}
\,\,& \longrightarrow\,\, 
\Biggl[\frac{\deltab}{3}\,\Bigl\{
4+f+\bfq\cdot\nabla_q+f\,q_z\partial_{q_z}
+\bigl(2f^2+5f\bigr)\frac{\partial_{q_z}^2}{\partial_q^2}\Bigr\}
\nonumber
\\
&\quad +q_k\Bigl(\tau_{\ell k}\partial_{q_\ell}+f\,\tau_{z k}\partial_{q_z}\Bigr)\Bigl(1+f\,\frac{\partial^2_{q_z}}{\partial_q^2}\Bigr)
+\tau_{ij}\frac{\partial_{q_i}\partial_{q_j}}{\partial^2_q}
\nonumber
\\
&\quad 
+\tau_{zz}\Bigl(f+2f^2\frac{\partial_{q_z}^2}{\partial_q^2}\Bigr)+2f\,\tau_{iz}\frac{\partial_{q_z}\partial_{q_i}}{\partial_q^2}
\Biggr]\deltas(\bfq),
\label{eq:4th_term}
\\
-b_1^{\rm L}\,\partial_{q_j}\{ R_{jk}^{(1)}\psi_k^{(1)}(\bfq) \deltalin(\bfq)\} 
\,\,& \longrightarrow\,\, 
b_1^{\rm L}\,\Biggl[\deltab\Bigl\{2+\frac{1}{3}(\bfq\cdot\nabla_q)+\frac{f}{3}\Bigl(1+3\frac{\partial^2_{q_z}}{\partial_q^2}+\,q_z \partial_{q_z}\Bigr)\Bigr\}
\nonumber
\\
&\quad  + 
q_m\bigl(\tau_{jm}\partial_{q_j}+f\,\tau_{zm}\partial_{q_z}\bigr) +f\,\tau_{zz}\,\Biggr]\deltas(\bfq),
\label{eq:5th_term}
\\
-\partial_{q_j}\{ R_{jk}^{(2)}\psi_k^{(2)}(\bfq) \} 
\,\,& \longrightarrow\,\, 
 \frac{3}{7}\Bigl[\Bigl(\frac{2}{3}\deltab-\tau_{ij}
\frac{\partial_{q_i}\partial_{q_j}}{\partial_q^2}\Bigr)
\Bigl(1+2f\frac{\partial_{q_z}^2}{\partial_q^2}\Bigr)
\Bigr]\deltas(\bfq),
\label{eq:6th_terms}
\\
\frac{1}{2}b_2^{\rm L}\,\{\deltalin(\bfq)\}^2
\,\,& \longrightarrow\,\, 
b_2^{\rm L}\,\deltab\,\deltas(\bfq),
\label{eq:7th_term}
\\
\frac{1}{2}b_{s^2}^{\rm L}\,C_{ij}(\bfq)C_{ij}(\bfq)
\,\,& \longrightarrow\,\, 
b_{s^2}^{\rm L}\, \tau_{ij}\, C_{ij}^{\rm short}(\bfq).
\label{eq:8th_term}
\end{align}
Here, we only retain the linear-order contributions, and ignored all the higher-order terms of $\mathcal{O}(\deltas^2)$, $\mathcal{O}(\deltab^2)$, $\mathcal{O}(\tau_{ij}^2)$ and $\mathcal{O}(\deltab\tau_{ij})$. Putting these terms back to Eq.~(\ref{eq:deltaz_expanded_2nd-order}), the integrals over $\bfk$ and $\bfq$ are analytically performed, and we obtain
\begin{align}
\deltag^{\rm(S)}(\bfs) &=\Bigl(1+b_1^{\rm L}+f\,\frac{\partial_{s_z}^2}{\partial_s^2}\Bigr)\deltas(\bfs)
+\,\Bigl(1+b_1^{\rm L}+\frac{f}{3}\Bigr)\deltab+f\,\tau_{zz}
\nonumber
\\
&+\Biggl[\frac{\deltab}{3}\,\Bigl\{
4+f+\bfs\cdot\nabla_s+f\,s_z\partial_{s_z}
+\bigl(2f^2+5f\bigr)\frac{\partial_{s_z}^2}{\partial_s^2}\Bigr\}
\nonumber
\\
&+s_i\Bigl(\tau_{ij}\partial_{s_j}+f\,\tau_{iz}\partial_{s_z}\Bigr)\Bigl(1+f\,\frac{\partial^2_{s_z}}{\partial_s^2}\Bigr)
+\tau_{ij}\frac{\partial_{s_i}\partial_{s_j}}{\partial^2_s}
+\tau_{zz}\Bigl(f+2f^2\frac{\partial_{s_z}^2}{\partial_s^2}\Bigr)
\nonumber
\\
&+2f\,\tau_{iz}\frac{\partial_{s_z}\partial_{s_i}}{\partial_s^2}
+\deltab\,b_1^{\rm L}\,\Bigl\{2+\frac{1}{3}(\bfs\cdot\nabla_s)+\frac{f}{3}\Bigl(1+3\frac{\partial^2_{s_z}}{\partial_s^2}+\,s_z \partial_{s_z}\Bigr)\Bigr\}
\nonumber
\\
&+ 
b_1^{\rm L}\Bigl\{s_i\bigl(\tau_{ij}\partial_{s_j}+f\,\tau_{iz}\partial_{s_z}\bigr) +f\,\tau_{zz}\,\Bigr\}
+ \frac{3}{7}\Bigl\{\Bigl(\frac{2}{3}\deltab-\tau_{ij}
\frac{\partial_{s_i}\partial_{s_j}}{\partial_s^2}\Bigr)
\Bigl(1+2f\frac{\partial_{s_z}^2}{\partial_s^2}\Bigr)
\Bigr\}
\nonumber
\\
&+b_2^{\rm L}\,\deltab+b_{s^2}^{\rm L}\,\tau_{ij}\,\frac{\partial_{s_i}\partial_{s_j}}{\partial^2_s}\Biggr]\,\deltas(\bfs).
\label{eq:redshift_space_density_with_DC_modes}
\end{align}
Rewriting the Lagrangian bias parameters with Eulerian counterparts through the relation at Eq.~(\ref{eq:bias_delta_Lagrangian_Eulerian}), the above expression is recast as 
\begin{align}
& \delta_{\rm g}^{\rm(S)}(\bfs)=
\Bigl(b_1+f\,\frac{\partial_{s_z}^2}{\partial^2}\Bigr)\deltas(\bfs)
\nonumber
\\
&\quad+\Biggl[\deltab\,\Bigl\{b_1\Bigl(\frac{34}{21}+\frac{1}{3}(\bfs\cdot\nabla_s)\Bigr)+b_2\Bigr\}
+\tau_{ij} \Bigl\{b_1\Bigl(\frac{4}{7}\frac{\partial_{s_i}\partial_{s_j}}{\partial^2_s}+s_i\partial_{s_j}\Bigr)+b_{s^2}\,\,\frac{\partial_{s_i}\partial_{s_j}}{\partial^2_s}\Bigr\}\Biggr]\deltas(\bfs)
\nonumber
\\
&\quad+f\,\deltab\Biggl[b_1\Bigl(\frac{1}{3}+\frac{1}{3}s_z\partial_{s_z}+\frac{\partial_{s_z}^2}{\partial^2_s}\Bigr)+\frac{2}{3}\Bigl(\frac{13}{7}+f\Bigr)\frac{\partial_{s_z}^2}{\partial_s^2}+\frac{1}{3}\Bigl\{(\bfs\cdot\nabla_s)+f\,s_z\partial_{s_z}\Bigr\}\frac{\partial_{s_z}^2}{\partial_s^2}\Biggr]\deltas(\bfs)
\nonumber
\\
&\quad +f\,\tau_{ij}\Biggl[
b_1\Bigl(\hatz_i\hatz_j+\hatz_i\,s_j\partial_{s_z}\Bigr)+
s_j\Bigl(\partial_{s_i}+f\,\hatz_i\partial_{s_z}\Bigr)\frac{\partial_{s_z}^2}{\partial_s^2}
+
2\Bigl(\hatz_j\frac{\partial_{s_z}\partial_i}{\partial_s^2}+f\,\hatz_i\hatz_j\frac{\partial^2_{s_z}}{\partial^2_s}\Bigr)
\nonumber
\\
&\quad
-\frac{6}{7}\frac{\partial_{s_i}\partial_{s_j}}{\partial^2_s}\,\frac{\partial_{s_z}^2}{\partial_s^2}\Biggr]\deltas(\bfs),
\label{eq:deltaz_tau_config}
\end{align}
where we have dropped all the terms independent of the short-mode fluctuations $\deltas$, whose contributions act as the DC mode. Eq.~(\ref{eq:deltaz_tau_config}) describes the field-level response to the super-sample modes in configuration space. To further obtain the Fourier-space expression, the terms involving the operators are transformed, using the following relations:
\begin{align}
 \frac{\partial_{s_i}\partial_{s_j}}{\partial^2_s}\,\deltas(\bfs)&=-\int\frac{d^3\bfk}{(2\pi)^3}\,e^{i\,\bfk\cdot\bfs}\,\hatk_i\hatk_j\,\deltas(\bfk),
\label{eq:real-Fourier_si_sj}
\\
s_j\partial_{s_i}\deltas(\bfs)&=-\int\frac{d^3\bfk}{(2\pi)^3}\,e^{i\,\bfk\cdot\bfs}\Bigl(\deltaK_{ij}+k_i\partial_{k_j}\Bigr)\deltas(\bfk),
\label{eq:real-Fourier_si_del_sj}
\\
 s_i\partial_{s_j}\frac{\partial_{s_z}^2}{\partial^2_s}\,\deltas(\bfs)&=-\int\frac{d^3\bfk}{(2\pi)^3}\,e^{i\,\bfk\cdot\bfs}\mu_k^2\,\Bigl\{\,2\,\Bigl(\frac{k_j}{k_z}\deltaK_{iz}-\frac{k_ik_j}{k^2}\Bigr)+\deltaK_{ij}+k_j\partial_{k_i}\,\Bigr\}\deltas(\bfk)
\label{eq:real-Fourier_s_i_s_j_del2}
\end{align}
with the quantity $\mu_k$ being $\hatk_z$. Note that the latter two relations are obtained by repeating the integration by part. With these relations, the Fourier transform of Eq.~(\ref{eq:deltaz_tau_config}) is shown to give Eq.~(\ref{eq:deltag_redshift_SS-modes}).

\section{Power spectrum response to the super-sample modes in redshift space}
\label{Appendix:pkred_SS-mode}

In this Appendix, we show that the expression of the redshift-space density field, given at Eq.~(\ref{eq:deltag_redshift_SS-modes}), consistently reproduces the power spectrum response to the super-sample modes known in the literature.

Let us substitute the expression at Eq.~(\ref{eq:deltag_redshift_SS-modes}) into the definition of galaxy power spectrum: 
\begin{align}
 \langle\deltag^{\rm(S)}(\bfk)\deltag^{\rm(S)}(\bfk')\rangle=(2\pi)^3\delta_{\rm D}(\bfk+\bfk')\,P_{\rm gg}^{\rm(S)}(\bfk).
\end{align}
Then, the expression at left-hand side leads to
\begin{align}
& \langle\deltag^{\rm(S)}(\bfk)\deltag^{\rm(S)}(\bfk')\rangle
\simeq (2\pi)^3\,\delta_{\rm D}(\bfk+\bfk')\,(b_1+f\,\mu_k^2)^2\,P_{\delta\delta}(k)
\nonumber
\\
&\qquad+\deltab\,\Bigl[(b_1+f\,\mu_k^2)\Bigl\{\widehat{a}_\delta(\bfk')+f\,
\widehat{a}_\delta^{\rm(S)}(\bfk')\Bigr\}+
(b_1+f\,\mu_{k'}^2)\Bigl\{\widehat{a}_\delta(\bfk)+f\,
\widehat{a}_\delta^{\rm(S)}(\bfk)\Bigr\}\Bigr]
\nonumber
\\
&\qquad\qquad\qquad\qquad\qquad\qquad\qquad\qquad\qquad\qquad
\times\,\langle\deltas(\bfk)\deltas(\bfk')\rangle
\nonumber
\\
&\qquad+\tau_{ij}\,\Bigl[(b_1+f\,\mu_k^2)\Bigl\{\widehat{b}_{\tau,ij}(\bfk')+f\,
\widehat{b}_{\tau,ij}^{\rm(S)}(\bfk')\Bigr\}+
(b_1+f\,\mu_{k'}^2)\Bigl\{\widehat{b}_{\tau,ij}(\bfk)+f\,
\widehat{b}_{\tau,ij}^{\rm(S)}(\bfk)\Bigr\}\Bigr]
\nonumber
\\
&\qquad\qquad\qquad\qquad\qquad\qquad\qquad\qquad\qquad\qquad
\times\,\langle\deltas(\bfk)\deltas(\bfk')\rangle.
\label{eq:ensemble_pkred}
\end{align}
In Eq.~(\ref{eq:ensemble_pkred}), the first line at right-hand side includes the linear-order power spectrum, while the second and third lines have the contributions arising from the super-sample modes, $\deltab$ and $\tau_{ij}$, respectively. Here, we have ignored the higher-order contributions of $\mathcal{O}(\deltab^2)$, $\mathcal{O}(\tau_{ij}^2)$, and $\mathcal{O}(\deltab\tau_{ij})$.

To further proceed to the calculations in the second and third lines, one has to deal with the operators acting on the ensemble average, $\langle\cdots\rangle$, in which the Dirac delta function is implicitly contained. To do this, we use the relations summarized in Appendix \ref{Appendix:formulas}. Then, the second line leads to 
\begin{align}
& \Bigl[(b_1+f\,\mu_k^2)\Bigl\{\widehat{a}_\delta(\bfk')+f\,
\widehat{a}_\delta^{\rm(S)}(\bfk')\Bigr\}+
(b_1+f\,\mu_{k'}^2)\Bigl\{\widehat{a}_\delta(\bfk)+f\,
\widehat{a}_\delta^{\rm(S)}(\bfk)\Bigr\}\Bigr]\langle\deltas(\bfk)\deltas(\bfk')\rangle
\nonumber
\\
&=(2\pi)^3\delta_{\rm D}(\bfk+\bfk')\,(b_1+f\,\mu_k^2)\,P_{\delta\delta}(k)
\nonumber
\\
&\quad\times\Biggl[
b_1\Bigl\{\frac{47}{21}+f\Bigl(\frac{1}{3}+2\mu^2\Bigr)-\frac{1}{3}(1+f\,\mu^2)
\frac{\partial\ln P_{\delta\delta}(k)}{\partial \ln k}
\Bigr\}
\nonumber
\\
&\qquad\qquad\qquad
+2\,b_2+f\,\mu^2\Bigl\{\frac{31}{21}-\frac{f}{3}(1-4\mu^2)-\frac{1}{3}(1+f\,\mu^2)\frac{\partial\ln P_{\delta\delta}(k)}{\partial \ln k}\Bigr\}
\Biggr].
\label{eq:pkred_deltab_results}
\end{align}
Also, the third line is calculated as 
\begin{align}
& \tau_{ij}\Bigl[(b_1+f\,\mu_k^2)\Bigl\{\widehat{b}_{\tau,ij}(\bfk')+f\,
\widehat{b}_{\tau,ij}^{\rm(S)}(\bfk')\Bigr\}+
(b_1+f\,\mu_{k'}^2)\Bigl\{\widehat{b}_{\tau,ij}(\bfk)+f\,
\widehat{b}_{\tau,ij}^{\rm(S)}(\bfk)\Bigr\}\Bigr]
\,\langle\deltas(\bfk)\deltas(\bfk')\rangle
\nonumber
\\
&=(2\pi)^3\delta_{\rm D}(\bfk+\bfk')\,(b_1+f\,\mu_k^2)\,P_{\delta\delta}(k)
\nonumber
\\
&\quad\times
\Biggl[
\Bigl\{\frac{8}{7}b_1+2b_{s^2}+\frac{16}{7}\,f\,\mu_k^2-(b_1+f\,\mu_k^2)\frac{\partial P_{\delta\delta}(k)}{\partial \ln k}\Bigr\}\hatk_i\hatk_j\,\tau_{ij}\,
\nonumber
\\
&\qquad\qquad\quad-f\,\mu_k\,\Bigl\{(b_1+f\,\mu_k^2)\frac{\partial\ln P_{\delta\delta}(k)}{\partial\ln k}-4f\,\mu_k^2\Bigr\}\tau_{iz}\hatk_i
+f\,(b_1-f\,\mu_k^2)\tau_{zz}
\Biggr].
\end{align}

Summing up the contributions from super-sample modes given above, the leading-order expression of the power spectrum in redshift space becomes
\begin{align}
 P_{\rm gg}^{\rm(S)}(\bfk)=(b_1+f\,\mu_k^2)^2\,P_{\delta\delta}(k)+
\Bigl(\frac{\partial P^{\rm(S)}(\bfk)}{\partial \deltab}\Bigr)\,\deltab + 
\Bigl(\frac{\partial P^{\rm(S)}(\bfk)}{\partial \tau_{ij}}\Bigr)\,\tau_{ij}
\end{align}
with the response to the super-sample modes, $(\partial P^{\rm(S)}(\bfk)/\partial \deltab)$ and $(\partial P^{\rm(S)}(\bfk)/\partial \tau_{ij})$, respectively given by
\begin{align}
\Bigl(\frac{\partial P_{\rm gg}^{\rm(S)}(\bfk)}{\partial \deltab}\Bigr)&=
 \Bigl\{\frac{47}{21}b_1+2b_2-\frac{b_1}{3}\frac{d\ln P_{\delta\delta}(k)}{d\ln k}\Bigr\}\,b_1\,P_{\delta\delta}(k)
\nonumber
\\
&+\Bigl\{\frac{b_1^2}{3}+\mu_k^2\Bigl(\frac{26}{7}b_1+2b_1^2+2b_2\Bigr)-\frac{\mu_k^2}{3}b_1(2+b_1)\frac{d\ln P_{\delta\delta}(k)}{d\ln k}\Bigr\}\,f\,P_{\delta\delta}(k)
\nonumber
\\
&+\Bigl\{\frac{1}{21}\bigl(31+70\,b_1\bigr)-\frac{1}{3}(1+2b_1)\frac{d\ln P_{\delta\delta}(k)}{d\ln k}\Bigr\}\,f^2\,\mu_k^4\,P_{\delta\delta}(k)
\nonumber
\\
&+\Bigl\{\frac{1}{3}(4\mu_k^2-1)-\frac{\mu_k^2}{3}\frac{d\ln P_{\delta\delta}(k)}{d\ln k}\Bigr\}\,f^3\,\mu_k^4\,P_{\delta\delta}(k),
\label{eq:pkred_deltab_Akitsu}
\end{align}
and
\begin{align}
\Bigl(\frac{\partial P_{\rm gg}^{\rm(S)}(\bfk)}{\partial \tau_{ij}}\Bigr) &= 
\Bigl\{\frac{8}{7}b_1+2b_{s^2}-b_1\frac{\partial \ln P_{\delta\delta}(k)}{\partial \ln k}\Bigr\}\hatk_i\hatk_j\,\,b_1\,P_{\delta\delta}(k)
\nonumber
\\
&+ \Bigl\{b_1^2 \hatz_i\hatz_j+
\Bigl(\frac{24}{7}b_1\,+2 b_{s^2}\,-2b_1\frac{\partial\ln P_{\delta\delta}(k)}{\partial \ln k} \Bigr)\mu_k^2\,\hatk_i\hatk_j\,-b_1^2\frac{\partial\ln P_{\delta\delta}(k)}{\partial\ln k}\mu_k\hatk_{(i}\hatz_{j)}
\Bigr\}\,f\,P_{\delta\delta}(k)
\nonumber
\\
&
+\Bigl\{\mu_k\Bigl(\frac{16}{7}-\frac{\partial\ln P_{\delta\delta}(k)}{\partial \ln k}  \Bigr) \hatk_i\hatk_j\,+2\,b_1\Bigl(2-\frac{\partial\ln P_{\delta\delta}(k)}{\partial \ln k}\Bigr)\hatk_{(i}\hatz_{j)} \Bigr\}\,f^2\,\mu_k^3\,P_{\delta\delta}(k)
\nonumber
\\
&+\Bigl\{-\hatz_i\hatz_j+\mu_k\Bigl(4-\frac{\partial\ln P_{\delta\delta}(k)}{\partial \ln k} \Bigr)\hatk_{(i}\,\hatz_{j)} \Bigr\}\,f^3\,\mu_k^4\,P_{\delta\delta}(k),
\end{align}
where the subscripts enclosed by parenthesis indicate the symmetrization of the indices, i.e., $\hatk_{(i}\hatz_{j)}=(\hatk_i\hatz_j+\hatz_i\hatk_j)/2$. In the above, the resultant expressions are presented in powers of $f$ for ease of comparison. These expressions coincide with those given at Eqs.~(13) and (14) of Ref.~\cite{Akitsu_Sugiyama_Shiraishi2019}.

\section{Galaxy shape bias in Eulerian and Lagrangian space}
\label{Appendix:shape_bias}

In this Appendix, we consider the perturbative description of the galaxy shape, and discuss its relation between Eulerian and Lagrangian space, given at Eqs.~(\ref{eq:gamma_expansion}) and (\ref{eq:gammaL_2nd_order}).

Let us see how the ellipticity field defined in Lagrangian space is mapped into that in Eulerian space. To do this, we adopt the simple linear alignment model \cite{Catelan_etal2001,Hirata_Seljak2004,Okumura_Taruya2019,Okumura_Taruya_Nishimichi2020,Kurita_etal2021} in Lagrangian space:  
\begin{align}
 \gammaL_{ij}(\bfq)=-\cK\,\Bigl\{1+\deltag^{\rm L}(\bfq)\Bigr\}\,C_{ij}(\bfq)
\label{eq:gammaL_LA_model}
\end{align}
with $\cK$ being the linear-order shape bias parameter. The field $C_{ij}$ is given at Eq.~(\ref{eq:def_C_ij}).

In order to relate the Lagrangian ellipticity field at Eq.~(\ref{eq:gammaL_LA_model}) with the Eulerian counterpart, $\gamma_{ij}$, we use the following relations:
\begin{align}
& \Bigl\{1+\delta_{\rm g}(\bfx)\Bigr\}\,\,d^3\bfx=
 \Bigl\{1+\deltag^{\rm L}(\bfq)\Bigr\}\,d^3\bfq, 
\label{eq:Euler_Lagrangian_deltag}
\\
& \gamma_{ij}(\bfx)\,d^3\bfx=\gammaL_{ij}(\bfq)\,d^3\bfq.
\label{eq:Euler_Lagrangian_gamma}
\end{align}
Here, the first line comes from Eq.~(\ref{eq:galaxy_Eulerian_Lagrangian}). The second line is analogous to the first line, and gives an expression consistent with Eq.~(\ref{eq:gamma_redshift_integral_form}) if one goes to real space\footnote{The real-space counterpart of Eq.~(\ref{eq:gamma_redshift_integral_form}) is obtained by setting the growth factor $f$ to $0$ in the displacement field $\bfpsi^{\rm S}$ (see Eqs.~(\ref{eq:displacement_z-space}) and (\ref{eq:n-th_order_displacement_z-space})).}.

Let us substitute Eqs.~(\ref{eq:gamma_expansion}) and (\ref{eq:gammaL_LA_model}) into the left- and right-hand sides of Eq.~(\ref{eq:Euler_Lagrangian_gamma}), respectively. With a help of Eq.~(\ref{eq:Euler_Lagrangian_deltag}),  the factor of the density weight, $\{1+\deltag\}$ or $\{1+\deltag^{\rm L}\}$, is eliminated, and one obtains
\begin{align}
& \bK K_{ij}(\bfx)+\bdeltaK\,\deltaE(\bfx)K_{ij}(\bfx) + 
 \bKK \Bigl\{K_{i\ell}(\bfx)K_{\ell j}(\bfx)-\frac{1}{3}\deltaK_{ij}\,
\bigl[{\rm Tr}\,K_{\ell m}(\bfx)\bigr]^2\Bigr\}
\nonumber
\\
& +\bt\,t_{ij}(\bfx) + \mathcal{O}(\delta^3)
=-\cK\,C_{ij}(\bfq).
\label{eq:gamma_Lagrangian_Eulerian}
\end{align}
Here, the left-hand side is given as a function of Lagrangian-space position, while the right-hand side is a function of Eulerian-space position. For more explicit comparison between both sides, we express the Lagrangian-space position $\bfq$ in terms of the Eulerian-space position, $\bfx$. Based on the Lagrangian perturbation theory, we have (see Eqs.~(\ref{eq:Eulerian_Lagrangian})-(\ref{eq:psi_1st})):
\begin{align}
 \bfx = \bfq+\bfpsi(\bfq),\quad \psi_i(\bfq)\simeq -\Bigl(\frac{\partial_{q_i}}{\partial^2_q}\Bigr)\,\deltalin(\bfq).
\label{eq:displacement_x_q_ZA}
\end{align}
The inverse mapping relation valid at leading order then leads to 
\begin{align}
 \bfq \simeq \bfx-\bfpsi(\bfx).
\end{align}
Using this, the right-hand side of Eq.~(\ref{eq:gamma_Lagrangian_Eulerian}) is rewritten with 
\begin{align}
\mbox{RHS of (\ref{eq:gamma_Lagrangian_Eulerian})}\simeq\, -\cK\,
\Bigl\{1-\bfpsi(\bfx)\cdot \nabla_x \Bigr\}C_{ij}(\bfx).
\label{eq:RHS_of_gamma_ij}
\end{align}
Note that the field $C_{ij}$ is now given as function of $\bfx$.

Going back to Eq.~(\ref{eq:gamma_Lagrangian_Eulerian}),  we see that the left-hand side is expressed in terms of the evolved fields, $\delta$, $K_{ij}$ and $t_{ij}$, the latter two of which are given by the traceless Hessian matrix of $\delta$ and $\theta-\delta$. These are to be evaluated perturbatively by using the (Eulerian) standard perturbation theory \cite{Bernardeau:2001qr}. That is, the density and velocity-divergence fields, $\delta$ and $\theta$, are expanded as $\delta=\delta_1+\delta_2+\cdots$ and $\theta=\theta_1+\theta_2+\cdots$. Note that at first order, the field $\delta_1$ is identified with the linear density field $\deltalin$, and we have $\delta_1=\theta_1$ for the initial conditions dominated by the growing-mode solution. Then, the right-hand side of Eq.~(\ref{eq:gamma_Lagrangian_Eulerian}), valid at the second order, is recast as
\begin{align}
&\mbox{LHS of (\ref{eq:gamma_Lagrangian_Eulerian})}\simeq 
\bK \,\Bigl(\frac{\partial_{x_i}\partial_{x_j}}{\partial^2_x}-\frac{1}{3}\deltaK_{ij}\Bigr)\,\Bigl\{\deltalin(\bfx)+\delta_2(\bfx)\Bigr\}
+\bdeltaK\,\deltalin(\bfx)C_{ij}(\bfx)
\nonumber
\\
&\qquad
+ \bKK \Bigl\{C_{i\ell}(\bfx)C_{\ell j}(\bfx)-\frac{1}{3}\deltaK_{ij}\,
\bigl[\mbox{Tr}\,C_{\ell m}(\bfx)\bigr]^2\Bigr\}
+\bt\,\Bigl(\frac{\partial_{x_i}\partial_{x_j}}{\partial^2_x}-\frac{1}{3}\deltaK_{ij}\Bigr)\,\Bigl\{\theta_2(\bfx)-\delta_2(\bfx)\Bigr\}.
\label{eq:LHS_of_gamma_ij}
\end{align}

The order-by-order comparison between Eqs.~(\ref{eq:RHS_of_gamma_ij}) and (\ref{eq:LHS_of_gamma_ij}) leads to 
\begin{align}
\mathcal{O}(\deltalin)\,\,&:\quad -\cK\,C_{ij}(\bfx) = \bK\,C_{ij}(\bfx),
\label{eq:gamma_relation_1st_order}
\\
\mathcal{O}(\deltalin^2)\,\,&:\quad \cK\,(\bfpsi(\bfx)\cdot\nabla_x)\,C_{ij}(\bfx)
=  \bK 
\Bigl(\frac{\partial_{x_i}\partial_{x_j}}{\partial^2_x}-\frac{1}{3}\deltaK_{ij}\Bigr)\delta_2(\bfx)+\bdeltaK\,\deltalin(\bfx)C_{ij}(\bfx)
\nonumber
\\
&\quad + \bKK \Bigl\{C_{i\ell}(\bfx)C_{\ell j}(\bfx)-\frac{1}{3}\deltaK_{ij}\,
\bigl[\mbox{Tr}\, C_{\ell m}(\bfx)\bigr]^2\Bigr\}
 +\bt\,\Bigl(\frac{\partial_{x_i}\partial_{x_j}}{\partial^2_x}-\frac{1}{3}\deltaK_{ij}\Bigr)\,\Bigl\{\theta_2(\bfx)-\delta_2(\bfx)\Bigr\}.
\label{eq:gamma_relation_2nd_order}
\end{align}
From Eq.~(\ref{eq:gamma_relation_1st_order}), we immediately get
\begin{align}
 \bK=-\cK.
\label{eq:bias_gamma_linear}
\end{align}
On the other hand, it is shown by Ref.~\cite{Schmitz_etal2018} that making use of the Fourier-space expressions of the perturbation theory kernels, 
the left-hand side of Eq.~(\ref{eq:gamma_relation_2nd_order}), with the displacement field $\bfpsi$ given at Eq.~(\ref{eq:displacement_x_q_ZA}), is rewritten in the following expressions (see Appendix A of their paper): 
\begin{align}
& (\bfpsi(\bfx)\cdot\nabla_x)\,C_{ij}(\bfx)=
 - \Bigl(\frac{\partial_{x_i}\partial_{x_j}}{\partial^2_x}-\frac{1}{3}\deltaK_{ij}\Bigr)\delta_2(\bfx)+\frac{2}{3}\,\deltalin(\bfx)C_{ij}(\bfx) 
\nonumber
\\
&\quad 
+\Bigl\{C_{i\ell}(\bfx)C_{\ell j}(\bfx)-\frac{1}{3}\deltaK_{ij}\,
\bigl[{\rm Tr}\,C_{\ell m}(\bfx)\bigr]^2\Bigr\}
 -\frac{5}{2}\,\Bigl(\frac{\partial_{x_i}\partial_{x_j}}{\partial^2_x}-\frac{1}{3}\deltaK_{ij}\Bigr)\Bigl\{\theta_2(\bfx)-\delta_2(\bfx)\Bigr\}.
\label{eq:math_relation_2nd_order}
\end{align}
Note that in deriving Eq.~(\ref{eq:math_relation_2nd_order}), we have used the following expressions:
\begin{align}
\delta_2(\bfx)&= \frac{17}{21}\bigl\{\deltalin(\bfx)\bigr\}^2 + \Bigl\{\frac{\partial_{x_k}}{\partial_x^2}\deltalin(\bfx)\Bigr\}\bigl\{ \partial_{x_k}\deltalin(\bfx)\bigr\}+\frac{2}{7}\Bigl[\mbox{Tr}\,C_{ij}(\bfx)\Bigr]^2,
\label{eq:delta2_SPT}
\\
\theta_2(\bfx)&= \frac{13}{21}\bigl\{\deltalin(\bfx)\bigr\}^2 + \Bigl\{\frac{\partial_{x_k}}{\partial_x^2}\deltalin(\bfx)\Bigr\}\bigl\{ \partial_{x_k}\deltalin(\bfx)\bigr\}+\frac{4}{7}\Bigl[\mbox{Tr}\,C_{ij}(\bfx)\Bigr]^2.
\label{eq:theta2_SPT}
\end{align}
Hence, comparing this expression with Eq.~(\ref{eq:gamma_relation_2nd_order}), the shape bias parameters defined in Lagrangian and Eulerian space have to be related with each other through
\begin{align}
 \Bigl(\bK,\,\bdeltaK,\, \bKK,\, \bt\Bigr)=\Bigl(-1,\frac{2}{3},1,-\frac{5}{2}\Bigr)\,\cK, 
\label{eq:shape_bias_Lagrangian_Eulerian}
\end{align}
which consistently reproduces the linear-order relation at Eq.~(\ref{eq:bias_gamma_linear}). This directly manifests the fact that the mapping from the Lagrangian to Eulerian space is nonlinear, and even the linear ellipticity field proportional to $\deltalin$ in Lagrangian space leads to non-vanishing higher-order terms of the perturbative bias expansion in Eulerian space.

In general, the IA of galaxy is generated through the galaxy formation process involving the baryon physics, and is intrinsically nonlinear. Thus, on top of the linear alignment model at Eq.~(\ref{eq:gammaL_LA_model}), one expects that there are some corrections to the ellipticity field. Since the expression given at Eq.~(\ref{eq:gamma_expansion}) is known to provide a relevant basis for a perturbative description of the evolved IAs \cite{Vlah_Chisari_Schmidt2020,Schmitz_etal2018,Blazek_etal2019}, one can use the same basis to describe the IA in Lagrangian space. That is, up to the second order in $\deltalin$, the ellipticity field $\gamma_{ij}^{\rm L}$ is generally described by introducing the new parameters, $\cdeltaK$, $\cKK$, and $\ct$, as follows:
\begin{align}
& \gammaL_{ij} (\bfq) = \Bigl\{1+\deltag^{\rm L}(\bfq)\Bigr\}\Biggl[-\cK C_{ij}(\bfq) + \cdeltaK\,\deltalin(\bfq) C_{ij}(\bfq)
+ \cKK \Bigl\{C_{i\ell}(\bfq)C_{\ell j}(\bfq)-\frac{1}{3}\deltaK_{ij}
\bigl[{\rm Tr}\,C_{\ell m}(\bfq)\bigr]^2\Bigr\} 
\nonumber
\\
&\qquad \qquad 
+\ct\,\Bigl(\frac{\partial_{q_i}\partial_{q_j}}{\partial^2_q}-\frac{1}{3}\deltaK_{ij}\Bigr)\Bigl\{-\frac{4}{21}\deltalin(\bfq)^2+\frac{2}{7}\bigl[{\rm Tr}\,C_{\ell m}(\bfq)\bigr]^2\Bigr\}+\mathcal{O}(\deltalin^3)\Biggr]. 
\nonumber
\end{align}
This is Eq.~(\ref{eq:gammaL_2nd_order}). Here, we have used Eqs.~(\ref{eq:delta2_SPT}) and (\ref{eq:theta2_SPT}) to rewrite the expression of the field $t_{ij}$ (see Eq.~(\ref{eq:def_t_ij}) for definition). Finally, note that adopting the perturbative expansion given above, the relation of the shape bias parameters between Lagrangian and Eulerian space is generalized, and it is given as follows (see Eq.~(\ref{eq:bias_delta_gamma})): 
\begin{align}
 \Bigl(\bK,\,\bdeltaK,\, \bKK,\, \bt\Bigr)=\Bigl(-\cK,\,\cdeltaK+\frac{2}{3}\cK,\,\cKK+\cK,\,\ct-\frac{5}{2}\cK\Bigr).
\nonumber
\end{align}

\section{Sketch of the derivation of Eq.~(\ref{eq:gamma_with_SS-mode_z-space})}
\label{Appendix:derivation_of_gamma_with_SS-modes}

In this Appendix, we present the key equations to derive the leading-order expression for the IA field including the super-sample modes, given at Eq.~(\ref{eq:gamma_with_SS-mode_z-space}). 

As we mentioned in Sec.~\ref{subsec:IA_with_SS-modes}, there are three steps in deriving Eq.~(\ref{eq:gamma_with_SS-mode_z-space}). The first step is to expand Eq.~(\ref{eq:gamma_redshift_integral_form}) up to the second order in linear density field. Performing partly the integration by parts yields
\begin{align}
 \gamma_{ij}^{\rm(S)}(\bfs)&\simeq\int d^3\bfq \int \frac{d^3\bfk}{(2\pi)^3}\,e^{i\,\bfk\cdot(\bfs-\bfq)}
\Biggl[\Bigl\{-\cK\,   + (- \cK\,\bL_1\,+\cdeltaK)\deltalin(\bfq)\,\Bigr\}C_{ij}(\bfq) 
\nonumber
\\
&\quad +\cK\,\partial_{q_k}\bigl\{\psi_k^{\rm S}(\bfq)\,C_{ij}(\bfq)\bigr\} + \cKK \Bigl\{C_{i\ell}(\bfq)C_{\ell j}(\bfq)-\frac{1}{3}\deltaK_{ij}
{\rm Tr}[C(\bfq)^2]\Bigr\} 
\nonumber
\\
&\quad
+\ct\,\Bigl(\frac{\partial_{q_i}\partial_{q_j}}{\partial_q^2}-\frac{1}{3}\deltaK_{ij}\Bigr)\Bigl\{-\frac{4}{21}\deltalin(\bfq)^2+\frac{2}{7}{\rm Tr}[C(\bfq)^2]\Bigr\}  \Biggr], 
\label{eq:gammaE_integrand}
\end{align}
where we used the expansion form of the Lagrangian density field $\deltag^{\rm L}$ at Eq.~(\ref{eq:Lagrangian_bias_expansion}), and collect relevant terms. Then, the second step is to decompose each term in the integrand into long- and short-mode contributions. Since we are particularly concerned with the leading-order expression involving the super-sample modes, we ignore the higher-order short-mode contributions, and retain the linear-order short-mode contributions of $C_{ij}^{\rm short}$ and $\deltas$, allowing also to collect the long-mode contributions at linear order. We have
\begin{align}
 -\cK\, C_{ij}(\bfq)  &\quad \longrightarrow\quad  -\cK\,\bigl( C_{ij}^{\rm short} + \tau_{ij} \bigr),
\nonumber
\\
(- \cK\,\bL\,+\cdeltaK)\deltalin(\bfq)\,C_{ij}(\bfq) &\quad \longrightarrow\quad
 (- \cK\,\bL\,+\cdeltaK)
\Bigl(\deltab\, C_{ij}^{\rm short} + \deltas\,\tau_{ij}
\Bigr),
\nonumber
\\
 \cK\, \nabla_q\cdot\Bigl\{\bfpsi^{\rm S}(\bfq)\,C_{ij}(\bfq)\Bigr\}
&\quad \longrightarrow \quad \cK\, \Bigl[ -\Bigl\{\Bigl(1+\frac{f}{3}\Bigr)\deltab+f\,\tau_{zz}\Bigr\}\,C_{ij}^{\rm short}
\nonumber
\\
&\qquad\qquad-\tau_{ij}\,\Bigl(1
+f\,\frac{\partial_{q_z}^2}{\partial_q^2}\Bigr)\deltas 
\nonumber
\\
&\qquad\qquad -R_{\ell k}^{(1)}\Bigl(\frac{1}{3}\deltab\,q_k +\tau_{km}\,q_m\Bigr)\partial_{q_\ell}\,C_{ij}^{\rm short}
\Bigr],
\nonumber
\\
\cKK \Bigl\{C_{i\ell}(\bfq)C_{\ell j}(\bfq)-\frac{1}{3}\deltaK_{ij}
{\rm Tr}[C(\bfq)^2]\Bigr\} 
&\quad\longrightarrow\quad \cKK \Bigl\{\tau_{i\ell}\,C_{\ell j}^{\rm short}+C_{i\ell}^{\rm short}\,\tau_{\ell j}-\frac{2}{3}\deltaK_{ij} C_{\ell m}^{\rm short}\tau_{\ell m}\Bigr\},
\nonumber
\\
\ct\,\hatPi_{ij}(\bfq)\Bigl\{-\frac{4}{21}\deltalin(\bfq)^2+\frac{2}{7}{\rm Tr}[C(\bfq)^2]\Bigr\}  
&\quad\longrightarrow\quad \ct\,\Bigl(\frac{\partial_{q_i}\partial_{q_j}}{\partial_q^2}-\frac{1}{3}\deltaK_{ij}\Bigr)\Bigl\{-\frac{8}{21}\deltas\,\deltab + \frac{4}{7}C_{\ell m}^{\rm short}\tau_{\ell m}\Bigr\}.
\nonumber
\end{align}
Plugging back these expressions into Eq.~(\ref{eq:gammaE_integrand}), 
the integration over $\bfk$ and $\bfq$ leads to the expression given as a function of $\bfs$. Further, we use Eq.~(\ref{eq:bias_delta_gamma}) to rewrite the Lagrangian (shape) bias parameters with their Eulerian counterparts. We obtain
\begin{align}
\gamma_{ij}^{\rm(S)} (\bfs)\
&\simeq \bK\, K_{ij}(\bfs) 
+ \Bigl\{\bK\Bigl(b_1+\frac{2}{3}\Bigr) + \bdeltaK\Bigr\}\,\Bigl\{\deltab\,K_{ij}(\bfs)+\tau_{ij}\,\deltas(\bfs)\Bigr\}
\nonumber
\\
&
+f\,\bK\Bigl\{\Bigl(\frac{1}{3}\,\deltab+\tau_{zz}\Bigr)K_{ij}(\bfs)+\tau_{ij}\frac{\partial_{s_z}^2}{\partial_s^2}\deltas(\bfs)\Bigr\}
\nonumber
\\
&
+\bK\, R_{\ell k}^{(1)}\Bigl(\frac{1}{3}\deltab\,s_k +\tau_{km}\,s_m\Bigr)\partial_{s_\ell}\,K_{ij}(\bfs)
\nonumber
\\
& + (\bKK+\bK)\,  \Bigl( \tau_{i\ell}\,K_{\ell j}(\bfs)+K_{i\ell}(\bfs)\,\tau_{\ell j}-\frac{2}{3}\,\deltaK_{ij} K_{\ell m}(\bfs)\,\tau_{\ell m} \Bigr)
\nonumber
\\
& + \Bigl(\bt-\frac{5}{2}\bK\Bigr)\,\Bigl\{-\frac{8}{21}\,\deltab\,K_{ij}(\bfs) + \frac{4}{7}\,\tau_{\ell m}\,\Bigl(\frac{\partial_{s_i}\partial_{s_j}}{\partial_s^2}-\frac{1}{3}\,\deltaK_{ij}\Bigr)K_{\ell m}(\bfs)\Bigr\}
\nonumber
\\
& +\bK\,\tau_{ij},  
\label{eq:gamma_red_with_SS-modes}
\end{align}
Here, the last term represents the DC mode, and is constant over the survey region. Since this does not affect our subsequent calculations in the main text (i.e., power spectrum and its response to the super-sample modes),  we shall drop the last term, and compute the Fourier transform of Eq.~(\ref{eq:gamma_red_with_SS-modes}) (but see Eq.~(\ref{eq:deltag_real_space}) below). With a help of the formulas given at Eqs.~(\ref{eq:real-Fourier_si_sj})--(\ref{eq:real-Fourier_s_i_s_j_del2}), we obtain
\begin{align}
\gamma_{ij}^{\rm(S)} (\bfk)\ 
&=\Biggl[\bK\tildePi_{ij}(\bfk)+
\Bigl\{\bK\Bigl(b_1+\frac{2}{3}\Bigr)+\bdeltaK\Bigr\}\,\Bigl\{\deltab\,\tildePi_{ij}(\bfk)+\tau_{ij}\Bigr\}
\nonumber
\\
&
+f\,\bK\Bigl\{\Bigl(\frac{\deltab}{3}+\tau_{zz}\Bigr)\tildePi_{ij}(\bfk)+\mu_k^2\,\tau_{ij}\Bigr\}
\nonumber
\\
& -\bK\,\Bigl\{\deltab\, \tildePi_{ij}(\bfk) \Bigl(1+\frac{1}{3}\,\bfk\cdot\nabla_k\Bigr)
\nonumber
\\
& \qquad\qquad+
\hatk_k\bigl(\tau_{ki}\hatk_j+\tau_{kj}\hatk_i)-2\hatk_i\hatk_j\hatk_k\hatk_m\tau_{km}
+\tildePi_{ij}(\bfk)\,\tau_{km}k_k\,\partial_{k_m}
 \Bigr\}
\nonumber
\\
&-\bK\,f\,\Bigl\{\frac{\deltab}{3}\,\tildePi_{ij}(\bfk)\,(1+k_z\partial_{k_z}) +\frac{\deltab}{3}\,\mu(\hat{z}_i\hatk_j+\hat{z}_j\hatk_i-2\mu\hatk_i\hatk_j)
\nonumber
\\
&\qquad\qquad
+\tau_{zz}\tildePi_{ij}(\bfk)+\mu_k(\tau_{zi}\hatk_j+\tau_{zj}\hatk_i-2\hatk_i\hatk_j\tau_{zm}\hatk_m)+\tildePi_{ij}\tau_{zm}k_z\partial_{k_m}
\Bigr\}
\nonumber
\\
&+(\bKK+\bK)\,\Bigl\{\tau_{i\ell}\tildePi_{\ell j}(\bfk)+\tildePi_{i\ell}\tau_{\ell j}-\frac{2}{3}\deltaK_{ij}\,\tildePi_{\ell m}(\bfk)\tau_{\ell m}\Bigr\}
\nonumber
\\
&+\Bigl(\bt-\frac{5}{2}\bK\Bigr)\Bigl\{-\frac{8}{21}\deltab+\frac{4}{7}\tau_{\ell m}\tildePi_{\ell m}(\bfk)\Bigr\}\tildePi_{ij}(\bfk)\Biggr]\,\deltas(\bfk)
\label{eq:gammaE_Fourier_with_SS_z-space}
\end{align}
with the quantity $\mu_k$ being the directional cosine defined by $\mu_k=\hatk\cdot\hat{z}$. Finally, reorganizing the above expressions in terms of the super-sample modes $\deltab$ and $\tau_{ij}$, Eq.~(\ref{eq:gammaE_Fourier_with_SS_z-space}) is recast as the expression given at Eq.~(\ref{eq:gamma_with_SS-mode_z-space}).

\section{Useful relations}
\label{Appendix:formulas}

Here, we present several relations used to compute the power spectrum involving the super-sample modes. 

In the presence of the long-wavelength modes, the response of the short-mode observable fluctuations is described by the shift and modulation in scales and amplitudes. These effects invoke the spatial derivative in the field-level representation, and in computing the power spectrum, one encounters the ensemble-averaged quantities on which the operators directly acts. A typical example is 
\begin{align}
\Bigl\{ A(\bfk)k_\ell'\partial_{k_m'} + A(\bfk') k_\ell \partial_{k_m}\Bigr\}\langle\deltas(\bfk)\deltas(\bfk')\rangle,
\label{eq:derivative_ensemble}
\end{align}
where the quantity $A(\bfk)$ is assumed to be an even function of the wavevector $\bfk$, i.e., $A(-\bfk)=A(\bfk)$. Since the term, $\langle\cdots\rangle$, involve the Dirac delta function, a care is necessary for an explicit calculation of the above equation.

Recall that the Fourier-space expressions involving the super-sample modes, given at e.g., Eqs.~(\ref{eq:deltag_real_SS-modes}) and (\ref{eq:deltag_redshift_SS-modes}), have been derived from their configuration-space counterparts, one simple way to evaluate Eq.~(\ref{eq:derivative_ensemble}) is to go back to the configuration space: 
\begin{align}
\int\frac{d^3\bfk d^3\bfk'}{(2\pi)^6}\,e^{i\,(\bfk\cdot\bfx+\bfk'\cdot\bfy)}\,\Bigl\{A(\bfk)k_\ell'\partial_{k_m'}+A(\bfk')k_\ell\partial_{k_m}\Bigr\}\langle\deltas(\bfk)\deltas(\bfk')\rangle.
\label{eq:integ_A_deriv_expikr_ensemble}
\end{align}
In the above, performing the integration by parts leads to
\begin{align}
\mbox{Eq.~(\ref{eq:integ_A_deriv_expikr_ensemble})}
&=-\int\frac{d^3\bfk d^3\bfk'}{(2\pi)^6}\,\,\Bigl[A(\bfk)\partial_{k_m'}\Bigl\{
k_\ell'\,e^{i\,(\bfk\cdot\bfx+\bfk'\cdot\bfy)}\Bigr\}
+A(\bfk')\partial_{k_m}\Bigl\{k_\ell\,e^{i\,(\bfk\cdot\bfx+\bfk'\cdot\bfy)}\Bigr\}\Bigr]
\nonumber
\\
&
\qquad\qquad\qquad\qquad\qquad\qquad\qquad\qquad\qquad\qquad\qquad\quad\,
\times\,\langle\deltas(\bfk)\deltas(\bfk')\rangle
\nonumber
\\
&=-\int\frac{d^3\bfk d^3\bfk'}{(2\pi)^6}\,\,e^{i\,(\bfk\cdot\bfx+\bfk'\cdot\bfy)}\,\Bigl[\deltaK_{\ell m}\,\bigl\{A(\bfk)+A(\bfk')\bigr\}+ i\bigl\{k_\ell'y_m\,A(\bfk)+k_\ell x_m\,A(\bfk')\bigr\}\Bigr]
\nonumber
\\
&
\qquad\qquad\qquad\qquad\qquad\qquad\qquad\qquad\qquad\qquad\qquad\quad\,
\times\,\langle\deltas(\bfk)\deltas(\bfk')\rangle
\nonumber
\\
&=-\int\frac{d^3\bfk}{(2\pi)^3}\,e^{i\,\bfk\cdot\bfr}\,A(\bfk)\,\Bigl[2\,\deltaK_{\ell m}\,+ i\,k_\ell r_m\,\Bigr]\,P_{\delta\delta}(k).
\label{eq:integ_pk_A_expikr}
\end{align}
In the third line,  we have used the definition, $\langle\deltas(\bfk)\deltas(\bfk')\rangle=(2\pi)^3\delta_{\rm D}(\bfk+\bfk')\,P_{\delta\delta}(k)$, and defined the vector $\bfr$ by $\bfr\equiv\bfx-\bfy$.

In the integrand of the last line at Eq.~(\ref{eq:integ_pk_A_expikr}), the factor in the second term, $i\,k_\ell r_m\,e^{i\,\bfk\cdot\bfr}$, is rewritten with 
$k_\ell \partial_{k_m}(e^{i\,\bfk\cdot\bfr})$. Then, repeating again the integration by parts, one obtains
\begin{align}
\mbox{Eq.~(\ref{eq:integ_pk_A_expikr})}
& =-\int\frac{d^3\bfk}{(2\pi)^3}\,e^{i\,\bfk\cdot\bfr}\,\Bigl[2\,\deltaK_{\ell m} A(\bfk)P_{\delta\delta}(k)\,- \partial_{k_m}\Bigl\{k_\ell\,A(\bfk)\,P_{\delta\delta}(k)\Bigr\}\Bigr]
\nonumber
\\
& =-\int\frac{d^3\bfk}{(2\pi)^3}\,e^{i\,\bfk\cdot\bfr}\,\Bigl[\,\deltaK_{\ell m} \,- k_\ell\,\partial_{k_m}\bigl\{\ln A(\bfk)\bigr\} -\hatk_\ell\hatk_m\,\frac{\partial \ln P_{\delta\delta}(k)}{\partial \ln k}\Bigr]A(\bfk)P_{\delta\delta}(k).
\label{eq:integ_formula1}
\end{align}
Comparing the integrands between Eqs.~(\ref{eq:integ_A_deriv_expikr_ensemble}) and (\ref{eq:integ_formula1}), one finds the following relation: 
\begin{align}
&\Bigl\{A(\bfk)k_\ell'\partial_{k_m'}+A(\bfk')k_\ell\partial_{k_m}\Bigr\}\langle\deltas(\bfk)\deltas(\bfk')\rangle
\nonumber
\\
&
\,\,\longrightarrow\,\,(2\pi)^3\delta_{\rm D}(\bfk+\bfk')\Bigl[-\deltaK_{\ell m}+k_\ell \partial_{k_m}\ln A(\bfk)+\hatk_\ell\hatk_m\frac{\partial\ln P_{\delta\delta}(k)}{\partial\ln k}\Bigr]\,A(\bfk)\,P_{\delta\delta}(k).
\label{eq:master_relation1}
\end{align}
In similar way, another relation is also obtained: 
\begin{align}
\Bigl\{ 
&A(\bfk)\mu_{k'}^2\,k_\ell'\partial_{k_m'}+A(\bfk')\mu_k^2\,k_\ell \partial_{k_m}\Bigr\}\langle\deltas(\bfk)\deltas(\bfk')
\nonumber
\\
&
\,\,\longrightarrow\,\,(2\pi)^3\delta_{\rm D}(\bfk+\bfk')\Bigl[-2\mu_k\,\hatk_\ell\deltaK_{m z}+\mu_k^2\,\Bigl\{2\hatk_\ell\hatk_m-\deltaK_{\ell m}+k_\ell\partial_{k_m}\ln A(\bfk)
\nonumber
\\
&\qquad\qquad\qquad\qquad\qquad\qquad\qquad\qquad\qquad
+\hatk_\ell\hatk_m\frac{\partial\ln P_{\delta\delta}(k)}{\partial\ln k}\Bigr\}\,\Bigr]\,A(\bfk)\,P_{\delta\delta}(k),
\label{eq:master_relation2}
\end{align}
where the quantity $\mu_k$ is the directional cosine between line-of-sight direction and wavevector, i.e., $\mu_k=\hat{k}\cdot\hat{z}=\hatk_z$. Again, the function $A(\bfk)$ is assumed to be an even function of $\bfk$.

Based on Eqs.~(\ref{eq:master_relation1}) and (\ref{eq:master_relation2}) as the key relations, one can obtain the relations in several specific cases, relevant to compute the power spectrum expressions in Sec.~\ref{subsubsec:real_space} and \ref{subsec:pk_IA}, and Appendix \ref{Appendix:pkred_SS-mode}:  
\begin{align}
&\Bigl\{ A(\bfk)(\bfk'\cdot\nabla_{k'})+ A(\bfk')(\bfk\cdot\nabla_{k})\Bigr\}
 \langle\deltas(\bfk)\deltas(\bfk')\rangle
\nonumber
\\
&\,\,\longrightarrow\,\,(2\pi)^3\delta_{\rm D} (\bfk+\bfk')\Bigl[-3+(\bfk\cdot\nabla_k)\ln A(\bfk)+\frac{\partial\ln P_{\delta\delta}(k)}{\partial \ln k}\Bigr]A(\bfk)P_{\delta\delta}(k),
\label{eq:relation_veck_nabla}
\\
&\Bigl\{ A(\bfk)\,k_z'\partial_{k_i'}+ A(\bfk')\,k_z\partial_{k_i}\Bigr\}
 \langle\deltas(\bfk)\deltas(\bfk')\rangle
\nonumber
\\
&\,\,\longrightarrow\,\,(2\pi)^3\delta_{\rm D} (\bfk+\bfk')\Bigl[-\deltaK_{zi}+k_z \partial_{k_i}\ln A(\bfk)+\mu_k\hatk_i\frac{\partial\ln P_{\delta\delta}(k)}{\partial \ln k}\Bigr]A(\bfk)P_{\delta\delta}(k),
\label{eq:relation_kz_deriv_ki}
\\
&\Bigl\{ A(\bfk)\,k_z'\partial_{k_z'}+ A(\bfk')\,k_z\partial_{k_z}\Bigr\}
 \langle\deltas(\bfk)\deltas(\bfk')\rangle
\nonumber
\\
&\,\,\longrightarrow\,\,(2\pi)^3\delta_{\rm D} (\bfk+\bfk')\Bigl[-1+\frac{\partial}{\partial \ln k_z}\ln A(\bfk)+\mu_k^2\frac{\partial\ln P_{\delta\delta}(k)}{\partial \ln k}\Bigr]A(\bfk)P_{\delta\delta}(k),
\label{eq:relation_kz_deriv_kz}
\\
&\Bigl\{ A(\bfk)\mu_{k'}^2(\bfk'\cdot\nabla_{k'})+ A(\bfk')\mu_k^2(\bfk\cdot\nabla_{k})\Bigr\}
 \langle\deltas(\bfk)\deltas(\bfk')\rangle
\nonumber
\\
&\,\,\longrightarrow\,\,(2\pi)^3\delta_{\rm D} (\bfk+\bfk')\Bigl[-3+(\bfk\cdot\nabla_k)\ln A(\bfk)+\frac{\partial\ln P_{\delta\delta}(k)}{\partial \ln k}\Bigr]\,\mu_k^2\,A(\bfk)P_{\delta\delta}(k),
\label{eq:relation_mu2_veck_nable}
\\
&\Bigl\{ A(\bfk)\,\mu_{k'}^2\,k_z'\partial_{k_i'}+ A(\bfk')\,\mu_k^2\,k_z\partial_{k_i}\Bigr\}
 \langle\deltas(\bfk)\deltas(\bfk')\rangle
\nonumber
\\
&\,\,\longrightarrow\,\,(2\pi)^3\delta_{\rm D} (\bfk+\bfk')\Bigl[-3\deltaK_{iz}+k_z \partial_{k_i}\ln A(\bfk)+\mu_k\,\hatk_i\Bigl\{2+\frac{\partial\ln P_{\delta\delta}(k)}{\partial \ln k}\Bigr\}\Bigr]\,\mu_k^2\,A(\bfk)P_{\delta\delta}(k),
\label{eq:relation_mu2_kz_deriv_ki}
\\
&\Bigl\{ A(\bfk)\,\mu_{k'}^2\,k_z'\partial_{k_z'}+ A(\bfk')\,\mu_k^2\,k_z\partial_{k_z}\Bigr\}
 \langle\deltas(\bfk)\deltas(\bfk')\rangle
\nonumber
\\
&\,\,\longrightarrow\,\,(2\pi)^3\delta_{\rm D} (\bfk+\bfk')\Bigl[-1-2(1-\mu_k^2)+\frac{\partial}{\partial \ln k_z}\ln A(\bfk)+\mu_k^2\frac{\partial\ln P_{\delta\delta}(k)}{\partial \ln k}\Bigr]\,\mu_k^2\,A(\bfk)P_{\delta\delta}(k).
\label{eq:relation_mu2_kz_deriv_kz}
\end{align}

Note that the function $A(\bfk)$ in the above expressions is either the constant or a specific function of the form, $a+b\,\mu_k^2$,  with the coefficients $a$ and $b$ being scale-independent. In the latter case, the logarithmic derivative of the function $A$ becomes 
\begin{align}
& k_i\partial_{k_j}\ln(a+b\,\mu_k^2)=\frac{2b\,\mu_k}{a+b\,\mu_k^2}\bigl(\hatk_i\hatz_j-\mu_k\hatk_i\hatk_j\bigr).
\end{align}
This is reduced to the following specific relations:
\begin{align}
&(\bfk\cdot\nabla_k)\ln(a+b\,\mu_k^2)=0,
\nonumber
\\
&k_z \partial_{k_j}\ln(a+b\,\mu_k^2)=\frac{2b\,\mu_k^2}{a+b\,\mu_k^2}(\hatz_j-\mu_k\hatk_j),
\nonumber
\\
& \frac{\partial}{\partial \ln k_z}\ln(a+b\,\mu_k^2)=\frac{2b\,\mu_k^2(1-\mu_k^2)}{a+b\,\mu_k^2}.
\nonumber
\end{align}

\section{Integral formulas}
\label{Appendix:integrals}

In this Appendix, we summarize the formulas for integrals involving the unit vector $\hatk$ and tidal field $\tau_{ij}$, which are used to derive the angle-averaged power spectra in Sec.~\ref{subsec:pk_IA}. Writing the vector $\hatk$ explicitly as $\hatk=\{\sqrt{1-\mu^2}\cos\phi_k,\,\sqrt{1-\mu^2}\sin\phi_k,\,\mu\}$ [see Eq.~(\ref{eq:wavevector_phi_k})], we have
\begin{align}
& \int_0^{2\pi} \frac{d\phi_k}{2\pi}\, \tau_{\ell m}\hatk_\ell\hatk_m=\,\frac{3\mu^2-1}{2}\,\tau_{zz},
\label{eq:integral_1}
\\
& \int_0^{2\pi}\frac{d\phi_k}{2\pi}\,\hatk_i \tau_{iz}=\mu\,\tau_{zz},
\label{eq:integral_2}
\\
& \int_0^{2\pi} \frac{d\phi_k}{2\pi}\,\Bigl\{ (\hatk_x\hatk_\ell \tau_{\ell x}-\hatk_y\hatk_\ell \tau_{\ell y})\cos(2\phi_k)+
(\hatk_\ell\hatk_y \tau_{\ell x}+\hatk_x\hatk_\ell\tau_{\ell y})\sin(2\phi_k) 
\Bigr\} =\frac{\mu^2-1}{2}\,\tau_{zz},
\label{eq:integral_3}
\\
& \int_0^{2\pi} \frac{d\phi_k}{2\pi}\,\Bigl\{ -(\hatk_x\hatk_\ell\tau_{\ell x}-\hatk_y\hatk_\ell\tau_{\ell y})\sin(2\phi_k)+
(\hatk_\ell\hatk_y\tau_{\ell x}+\hatk_x\hatk_\ell\tau_{\ell y})\cos(2\phi_k) 
\Bigr\} =0,
\label{eq:integral_4}
\\
& \int_0^{2\pi}\frac{d\phi_k}{2\pi}\Bigl\{
(\hatk_x\tau_{xz}-\hatk_y\tau_{yz})\cos(2\phi_k)+
(\hatk_y\tau_{xz}-\hatk_x\tau_{yz})\sin(2\phi_k)
\Bigr\}=0,
\label{eq:integral_5}
\\
& \int_0^{2\pi}\frac{d\phi_k}{2\pi}\Bigl\{
(\tau_{xx}-\tau_{yy})\cos(2\phi_k)+
2\,\tau_{xy}\,\sin(2\phi_k)=0,
\label{eq:integral_6}
\end{align}
where we have used the fact that $\tau_{ij}$ is the symmetric and trace-free tensor, i.e., $\tau_{ij}=\tau_{ji}$ and $\tau_{kk}=0$.

\section{Multipole expansion}
\label{Appendix:multipoles}

In this Appendix, we present the analytical expressions for the power spectrum multipoles, $\overline{P}^{\rm (S)}_{\ell,{\rm EE}}$ and $\overline{P}^{\rm (S)}_{\ell,{\rm gE}}$, and  their Gaussian covariances, $^{\rm G}\cov_{\ell,\ell'}^{\rm EE,EE}$ and $^{\rm G}\cov_{\ell,\ell'}^{\rm gE,gE}$. Also, we derive the analytical expressions for the multipole expansion of the power spectrum response to the super-sample modes. These are used to compute the signal-to-noise ratios including the super-sample covariance in Sec.~\ref{subsec:SSC_IA}.

\subsection{Power spectrum multipoles and Gaussian covariance}
\label{Appendix:pk_cov_Gauss}

Let us first summarize the analytical expressions for the power spectrum multipoles, ignoring the contributions from the super-sample modes (i.e., $\deltab=0=\tau_{zz}$), which will be considered later.  Substituting the leading-order expressions at Eq.~(\ref{eq:pk_EE_w_SS-modes}) and (\ref{eq:pk_gE_w_SS-modes}) into Eq.~(\ref{eq:power_spectrum_multipoles}), non-zero multipoles are obtained at $\ell\leq 4$, and are given in a concise form: 
\begin{align}
& \overline{P}_{\ell,{\rm EE}}^{\rm(S)}(k)=c_\ell\,\bK^2\,P_{\delta\delta}(k),
\quad
 \overline{P}_{\ell,{\rm gE}}^{\rm(S)}(k)=d_\ell\,\bK\,P_{\delta\delta}(k)
\label{eq:EE_gE_spectrum_multpoles}
\end{align}
with the non-vanishing coefficients $c_\ell$ and $d_\ell$ respectively given by 
\begin{align}
 \bigl(c_0,c_2,c_4\bigr)=\Bigl(\frac{8}{15},\,-\frac{16}{21},\,\frac{8}{35}\Bigr),
\quad
 \bigl(d_0,d_2,d_4\bigr)=\Bigl(\frac{2}{3}\bigl(b_1+\frac{f}{5}\bigr),\,
-\frac{2}{3}\bigl(b_1-\frac{f}{7}\bigr),\,-\frac{8}{35}f\Bigr).
\label{eq:EE_gE_multpole_coefficients}
\end{align} 

Next consider the Gaussian covariance $^{\rm G}\cov_{\ell,\ell'}^{\rm X,X'}$ for their power spectrum multipoles. Below, we present the expressions for the auto covariance with ${\rm X}={\rm X'}$, but taking the shot noise and/or shape noise contributions into account. Given the estimators defined at Eq.~(\ref{eq:estimator_pk_EE}) and (\ref{eq:estimator_pk_gE}), the effect of survey window function can be safely ignored for the sub-survey modes of $k\gg 2\pi/V_{\rm W}^{1/3}$, and in this case, the Gaussian covariance is given by
\begin{align}
^{\rm G}\cov_{\ell,\ell'}^{\rm EE,EE}(k,k')\simeq\frac{2}{N_k}\,\deltaK_{k,k'}\frac{(2\ell+1)(2\ell'+1)}{2}
\int_{-1}^1 d\mu_k \mathcal{P}_\ell(\mu_k)\mathcal{P}_{\ell'}(\mu_k)
\Bigl\{\overline{P}_{\rm EE}^{\rm (S)}(\bfk)+\frac{\sigma_\gamma^2}{\ngal}\Bigr\}^2
\label{eq:Gauss_Cov_EE_EE}
\end{align}
for ${\rm X}={\rm EE}$, and  
\begin{align}
^{\rm G}\cov_{\ell,\ell'}^{\rm gE,gE}(k,k')&\simeq\frac{1}{N_k}\,\deltaK_{k,k'}\frac{(2\ell+1)(2\ell'+1)}{2}
\int_{-1}^1 d\mu_k \mathcal{P}_\ell(\mu_k)\mathcal{P}_{\ell'}(\mu_k)
\nonumber
\\
&
\times \Biggl[\Bigl\{P_{\rm gg}^{\rm (S)}(\bfk)+\frac{1}{\ngal}\Bigr\}\Bigl\{P_{\rm EE}^{\rm (S)}(\bfk)+\frac{\sigma_\gamma^2}{\ngal}\Bigr\}+\Bigl\{P^{\rm(S)}_{gE}(\bfk)\Bigr\}^2\Biggr]
\label{eq:Gauss_Cov_gE_gE}
\end{align}
for ${\rm X}={\rm gE}$. Here, the quantity $N_k$ is the effective number of Fourier modes in the bin, and is related to the bin size $\Delta k$ through $N_k=V_k V_{\rm W}/(2\pi)^3=4\pi\,k^2\,\Delta k/k_{\rm f}^3$, with $k_f$ being the fundamental mode given by $k_{\rm f}=2\pi/V_{\rm W}^{1/3}$. The quantities $\overline{n}_{\rm g}$ and $\sigma_\gamma$ are the number density of galaxies and scatter in the intrinsic shape per component, respectively. In the above, the terms inversely proportional to $\overline{n}_{\rm g}$ describe the shot-noise contributions.

Plugging the linear-order power spectra ignoring the long-mode contributions into the above, we obtain the following analytical expressions:    
\begin{align}
 ^{\rm G}\cov_{0,0}^{\rm EE,EE}(k,)&=\frac{2}{N_k}\,
\Biggl[\frac{128}{315}\,\{\bK^2\,P_{\delta\delta}(k)\}^2+\frac{16}{15}\bK^2\,P_{\delta\delta}(k)\frac{\sigma_\gamma^2}{\ngal}+ 
\Bigl(\frac{\sigma_\gamma^2}{\ngal}\Bigr)^2
\Biggr],
\label{eq:cov_EE_00}
\\
 ^{\rm G}\cov_{2,2}^{\rm EE,EE}(k)&=\frac{2}{N_k}\,
\Biggl[\frac{14720}{9009}\,\{\bK^2\,P_{\delta\delta}(k)\}^2+\frac{80}{21}\bK^2\,P_{\delta\delta}(k)\frac{\sigma_\gamma^2}{\ngal}+ 
5\,\Bigl(\frac{\sigma_\gamma^2}{\ngal}\Bigr)^2
\Biggr],
\label{eq:cov_EE_22}
\\
 ^{\rm G}\cov_{4,4}^{\rm EE,EE}(k)&=\frac{2}{N_k}\,
\Biggl[\frac{208512}{85085}\,\{\bK^2\,P_{\delta\delta}(k)\}^2+\frac{33552}{5005}\bK^2\,P_{\delta\delta}(k)\frac{\sigma_\gamma^2}{\ngal}+ 
9\,\Bigl(\frac{\sigma_\gamma^2}{\ngal}\Bigr)^2
\Biggr],
\label{eq:cov_EE_44}
\end{align}
for the EE auto-power spectrum, and  
\begin{align}
 ^{\rm G}\cov_{0,0}^{\rm gE,gE}(k)&=\frac{1}{N_k}\Biggl[
\frac{16}{315}\,\bK^2\,\bigl(21\,b_1^2+6\,b_1\,f+f^2\bigr)\bigl\{P_{\delta\delta}(k)\bigr\}^2
\nonumber
\\
&\quad+\frac{1}{15}\Bigl\{
\frac{8}{\ngal}\,\bK^2+(15\,b_1^2+10\,b_1\,f+3\,f^2)\frac{\sigma_\gamma^2}{\ngal}
\Bigr\}P_{\delta\delta}(k)
+\frac{\sigma_\gamma^2}{\ngal^2}
\Biggr],
\label{eq:cov_gE_00}
\\
 ^{\rm G}\cov_{2,2}^{\rm gE,gE}(k)&=\frac{1}{N_k}\Biggl[
\frac{80}{9009}\,\bK^2\,\bigl(429\,b_1^2+78\,b_1\,f+17\,f^2\bigr)\bigl\{P_{\delta\delta}(k)\bigr\}^2
\nonumber
\\
&\quad+\frac{5}{21}\Bigl\{
\frac{8}{\ngal}\bK^2+(21\,b_1^2+22\,b_1\,f+9\,f^2)\frac{\sigma_\gamma^2}{\ngal}
\Bigr\}P_{\delta\delta}(k)
+5\,\frac{\sigma_\gamma^2}{\ngal^2}
\Biggr],
\label{eq:cov_gE_22}
\\
 ^{\rm G}\cov_{4,4}^{\rm gE,gE}(k)&=\frac{1}{N_k}\Biggl[
\frac{144}{85085}\,\bK^2\,\bigl(3691\,b_1^2+1326\,b_1\,f+261\,f^2\bigr)\bigl\{P_{\delta\delta}(k)\bigr\}^2
\nonumber
\\
&\quad+\frac{9}{5005}\Bigl\{
\frac{1864}{\ngal}\bK^2+(5005\,b_1^2+5070\,b_1\,f+1929\,f^2)\frac{\sigma_\gamma^2}{\ngal}
\Bigr\}P_{\delta\delta}(k)
+9\,\frac{\sigma_\gamma^2}{\ngal^2}
\Biggr],
\label{eq:cov_gE_44}
\end{align}
for the gE cross-power spectrum. The covariance at higher multipoles of $\ell\geq6$ vanishes. Note that the covariances between different multipoles $(\ell\ne\ell')$ also appear non-vanishing for even multipoles of $\ell,\ell'\leq4$. But they are not used to estimate the signal-to-noise ratios for individual power spectrum multipoles, and hence we not present their expressions.

\subsection{Power spectrum responses}
\label{Appendix:dpk_ddeltab_dtauzz}

The multipole moments of the power spectrum responses, $\partial \overline{P}^{\rm (S)}_{\ell,{\rm X}}/\partial\deltab$ and $\partial \overline{P}^{\rm (S)}_{\ell,{\rm X}}/\partial\tau_{zz}$, are obtained by substituting the expressions given at Eqs.~(\ref{eq:dpkEE_ddeltab}), (\ref{eq:dpkEE_dtauzz}), (\ref{eq:dpkgE_ddeltab}) and (\ref{eq:dpkgE_dtauzz}) into Eq.~(\ref{eq:power_spectrum_multipoles}). Then the multipoles at $\ell=0$, $2$, and $4$ are found to be non-zero in both real and redshift space, whereas the hexadecacontapole ($\ell=6$) moment appears non-vanishing only in redshift space. The analytical expressions for the EE auto-power spectrum responses are summarized as follows:
\begin{align}
 \frac{\partial\overline{P}^{\rm (S)}_{\rm 0,EE}}{\partial\deltab}
&=\frac{8}{315}\,\bK\,\Biggl[
\bK\Bigl\{47+42b_1+11\,f-(7+f)\frac{\partial\ln P_{\delta\delta}(k)}{\partial\ln k}\Bigr\}-16\bt+42\bdeltaK\}
\Biggr]\,P_{\delta\delta}(k),
\label{eq:dpkddeltab_EE_0}
\\
 \frac{\partial\overline{P}^{\rm (S)}_{\rm 0,EE}}{\partial\tau_{zz}}
&=-\frac{8}{735}\,\bK\Biggl[
\bK\Bigl\{16-77\,f-7\,(2-f)\frac{\partial\ln P_{\delta\delta}(k)}{\partial\ln k}\Bigr\}+16\bt+98\bKK\}
\Biggr]\,P_{\delta\delta}(k),
\label{eq:dpkdtauzz_EE_0}
\\
 \frac{\partial\overline{P}^{\rm (S)}_{\rm 2,EE}}{\partial\deltab}
&=-\frac{16}{441}\,\bK\,\Biggl[
\bK\Bigl\{47+42b_1+7\,f-7\frac{\partial\ln P_{\delta\delta}(k)}{\partial\ln k}\Bigr\}-16\bt+42\bdeltaK\}
\Biggr]\,P_{\delta\delta}(k),
\label{eq:dpkddeltab_EE_2}
\\
 \frac{\partial\overline{P}^{\rm (S)}_{\rm 2,EE}}{\partial\tau_{zz}}
&=\frac{8}{147}\,\bK\,\Biggl[
\bK\Bigl\{8-14\,f-7\frac{\partial\ln P_{\delta\delta}(k)}{\partial\ln k}\Bigr\}+8\bt+28\bKK\}
\Biggr]\,P_{\delta\delta}(k),
\label{eq:dpkdtauzz_EE_2}
\\
 \frac{\partial\overline{P}^{\rm (S)}_{\rm 4,EE}}{\partial\deltab}
&=\frac{8}{8085}\,\bK\,\Biggl[
\bK\Bigl\{517+462b_1-119\,f-7(11-7f)\frac{\partial\ln P_{\delta\delta}(k)}{\partial\ln k}\Bigr\}-176\bt+462\bdeltaK\}
\Biggr]\,P_{\delta\delta}(k),
\label{eq:dpkddeltab_EE_4}
\\
 \frac{\partial\overline{P}^{\rm (S)}_{\rm 4,EE}}{\partial\tau_{zz}}
&=\frac{8}{2695}\,\bK\,\Biggl[
\bK\Bigl\{-128-119\,f+7(16+7\,f)\frac{\partial\ln P_{\delta\delta}(k)}{\partial\ln k}\Bigr\}-128\bt-154\bKK\}
\Biggr]\,P_{\delta\delta}(k),
\label{eq:dpkdtauzz_EE_4}
\\
 \frac{\partial\overline{P}^{\rm (S)}_{\rm 6,EE}}{\partial\deltab}
&=\frac{16}{693}\,\bK^2\,
\Bigl\{4-\frac{\partial\ln P_{\delta\delta}(k)}{\partial\ln k}\Bigr\}
\,P_{\delta\delta}(k),
\label{eq:dpkddeltab_EE_6}
\\
 \frac{\partial\overline{P}^{\rm (S)}_{\rm 6,EE}}{\partial\tau_{zz}}
&=\frac{8}{1617}\,\bK\,\Biggl[
\bK\Bigl\{24+56\,f-7(3+2\,f)\frac{\partial\ln P_{\delta\delta}(k)}{\partial\ln k}\Bigr\}+24\bt
\Biggr]\,P_{\delta\delta}(k).
\label{eq:dpkdtauzz_EE_6}
\end{align}
Likewise, the responses for the gE cross-power spectra are given below:
\begin{align}
  \frac{\partial\overline{P}^{\rm (S)}_{\rm 0,gE}}{\partial\deltab}
&= \frac{2}{63}\Bigl\{
\bK(21\,b_1^2+47\,b_1+21\,b_2)-b_1(8\,\bt-21\,\bdeltaK)
-7\,\bK \,b_1\, \frac{\partial \ln P_{\delta\delta}(k)}{\partial \ln k}
\Bigr\}\,P_{\delta\delta}(k)
\nonumber
\\
&+\frac{2}{315} \Bigl\{   \bK (39 + 91\, b_1+12f) -8\, \bt + 21\, \bdeltaK
\nonumber
\\
&\qquad\qquad\qquad\qquad\qquad
-\bK  (7+7b_1+3f) \frac{\partial \ln P_{\delta\delta}(k)}{\partial \ln k}\Bigr\}\,f\,P_{\delta\delta}(k),
\label{eq:dpkddeltab_gE_0}
\\
 \frac{\partial\overline{P}^{\rm (S)}_{\rm 0,gE}}{\partial\tau_{zz}}&=\frac{2}{105}\Bigl\{
-\bK\,(8\,b_1+7\,b_{s^2})-b_1(35\bKK+4\bt) +
7\,\bK\,b_1\,\frac{\partial \ln P_{\delta\delta}(k)}{\partial \ln k} \Bigr\}\,P_{\delta\delta}(k)
\nonumber
\\
&+\frac{2}{735} \Bigl\{\bK (12 + 343\, b_1 + 84\, f) -49\, \bKK + 4\,\bt 
\nonumber
\\
&\qquad\qquad\qquad\qquad\qquad
-7\bK\,(1 + 7 b_1 + 3 f)\frac{\partial \ln P_{\delta\delta}(k)}{\partial \ln k} \Bigr\}\,f\,P_{\delta\delta}(k),
\label{eq:dpkdtauzz_gE_0}
\\
  \frac{\partial\overline{P}^{\rm (S)}_{\rm 2,gE}}{\partial\deltab}
&=\frac{2}{63}\Bigl\{
-\bK(21\,b_1^2+21\,b_2+47\,b_1)+b_1(8\,\bt-21\,\bdeltaK)+
7\,\bK\,b_1\,\frac{\partial \ln P_{\delta\delta}(k)}{\partial \ln k} \Bigr\}\,P_{\delta\delta}(k),
\nonumber
\\
&+\frac{2}{441}\,\Bigl\{\bK (39 + 7 b_1 + 28 f)-8 \bt + 21 \bdeltaK 
\nonumber
\\
&\qquad\qquad\qquad\qquad\qquad
 -7\, \bK\, (1 + b_1 + f) \,\frac{\partial \ln P_{\delta\delta}(k)}{\partial \ln k} \Bigr\}\,f\,P_{\delta\delta}(k),
\label{eq:dpkddeltab_gE_2}
\\
  \frac{\partial\overline{P}^{\rm (S)}_{\rm 2,gE}}{\partial\tau_{zz}}
&=\frac{2}{147}\Bigl\{
5\bK(8b_1+7\,b_{s^2})+b_1(49\,\bKK+20\,\bt)-
35\,\bK\,b_1\,\frac{\partial \ln P_{\delta\delta}(k)}{\partial \ln k}
\Bigr\}\,P_{\delta\delta}(k)
\nonumber
\\
&
+\frac{2}{147}\,\Bigl\{ \bK (12 - 35 b_1 + 28 f) + 4 \bt-7\,\bKK
- 7\bK(1 + b_1 + f)\frac{\partial \ln P_{\delta\delta}(k)}{\partial \ln k}
\Bigr\}\,f\,P_{\delta\delta}(k),
\label{eq:dpkdtauzz_gE_2}
\\
  \frac{\partial\overline{P}^{\rm (S)}_{\rm 4,gE}}{\partial\deltab}
&=\frac{8}{8085}\,\Bigl\{
- \bK (429 + 616 b_1 + 112 f) +88 \bt - 231 \bdeltaK 
\nonumber
\\
&\qquad\qquad\qquad\qquad+ 7\,\bK\,  (11 + 11 b_1 + 4 f)
\frac{\partial \ln P_{\delta\delta}(k)}{\partial \ln k}
\Bigr\}\,f\,P_{\delta\delta}(k),
\label{eq:dpkddeltab_gE_4}
\\
\frac{\partial\overline{P}^{\rm (S)}_{\rm 4,gE}}{\partial\tau_{zz}}
&=\frac{12}{245}\Bigl\{-\,\bK\,(8\,b_1+7\,b_{s^2})-4\,b_1\,\bt
+7\,\bK\,b_1\,\frac{\partial \ln P_{\delta\delta}(k)}{\partial \ln k}
\Bigr\}\,P_{\delta\delta}(k)
\nonumber
\\
&+\frac{8}{2695} \Bigl\{ -2\,\bK ( 3 +77 b_1 + 56 f) + 77 \bKK - 2 \bt
\nonumber
\\
&\qquad\qquad\qquad\qquad\qquad
+\frac{7}{2}\,\bK (1 + 22 b_1 + 8 f)
\frac{\partial \ln P_{\delta\delta}(k)}{\partial \ln k}
\Bigr\}\,f\,P_{\delta\delta}(k),
\label{eq:dpkdtauzz_gE_4}
\\
  \frac{\partial\overline{P}^{\rm (S)}_{\rm 6,gE}}{\partial\deltab}
&=\frac{16}{693}\bK\,\Bigl\{ -4 + \frac{\partial \ln P_{\delta\delta}(k)}{\partial \ln k}
\Bigr\}\,f^2\,P_{\delta\delta}(k),
\label{eq:dpkddeltab_gE_6}
\\
  \frac{\partial\overline{P}^{\rm (S)}_{\rm 6,gE}}{\partial\tau_{zz}}
&=\frac{32}{1617}\,\Bigl\{ 
-\bK (9 + 14 f) - 3 \bt 
+ \frac{7}{4}\,\bK (3 + 2 f) 
\frac{\partial \ln P_{\delta\delta}(k)}{\partial \ln k}
\Bigr\}\,f\,P_{\delta\delta}(k).
\label{eq:dpkdtauzz_gE_6}
\end{align}

Note that the expressions for the response of the gE cross-power spectrum are valid for the galaxy density field defined with the global mean. As we mentioned in Sec.~\ref{subsec:pk_IA}, if we instead use the galaxy density field defined with the local mean measured in the survey region, the multipole moments of the responses, $\partial\overline{P}^{\rm (S)}_{\ell,\rm gE}/\partial\deltab$ and $\partial\overline{P}^{\rm (S)}_{\ell,\rm gE}/\partial\tau_{zz}$, are to be modified according to Eq.~(\ref{eq:dpkgE_ddeltab_local}) and (\ref{eq:dpkgE_dtauzz_local}), and they are changed to
\begin{align}
 \frac{\partial\overline{P}^{\rm (S)}_{0,\rm gE}}{\partial\deltab} 
&\longrightarrow \frac{\partial\overline{P}^{\rm (S)}_{0,\rm gE}}{\partial\deltab}-\frac{2}{45}\,\bK(3b_1+f)(5b_1+f)\,P_{\delta\delta}(k),
\label{eq:dpkgE0_ddeltab_local}
\\
 \frac{\partial\overline{P}^{\rm (S)}_{0,\rm gE}}{\partial\tau_{zz}} 
&\longrightarrow \frac{\partial\overline{P}^{\rm (S)}_{0,\rm gE}}{\partial\tau_{zz}}-\frac{2}{15}\,\bK(5b_1+f)\,f\,P_{\delta\delta}(k),
\label{eq:dpkgE0_dtauzz_local}
\\
 \frac{\partial\overline{P}^{\rm (S)}_{2,\rm gE}}{\partial\deltab} 
&\longrightarrow \frac{\partial\overline{P}^{\rm (S)}_{2,\rm gE}}{\partial\deltab}+\frac{2}{63}\,\bK(7b_1-f)(3b_1+f)\,P_{\delta\delta}(k),
\label{eq:dpkgE2_ddeltab_local}
\\
 \frac{\partial\overline{P}^{\rm (S)}_{2,\rm gE}}{\partial\tau_{zz}} 
&\longrightarrow \frac{\partial\overline{P}^{\rm (S)}_{2,\rm gE}}{\partial\tau_{zz}}+\frac{2}{21}\,\bK(7b_1-f)\,f\,P_{\delta\delta}(k),
\label{eq:dpkgE2_dtauzz_local}
\\
 \frac{\partial\overline{P}^{\rm (S)}_{4,\rm gE}}{\partial\deltab} 
&\longrightarrow \frac{\partial\overline{P}^{\rm (S)}_{4,\rm gE}}{\partial\deltab}+\frac{8}{105}\,\bK(3b_1+f)\,f\,P_{\delta\delta}(k),
\label{eq:dpkgE4_ddeltab_local}
\\
 \frac{\partial\overline{P}^{\rm (S)}_{4,\rm gE}}{\partial\tau_{zz}} 
&\longrightarrow \frac{\partial\overline{P}^{\rm (S)}_{4,\rm gE}}{\partial\tau_{zz}}+\frac{8}{35}\,f^2\,P_{\delta\delta}(k).
\label{eq:dpkgE4_dtauzz_local}
\end{align}
For multipoles higher than $\ell=4$, there is no modification, and Eqs.~(\ref{eq:dpkddeltab_gE_6}) and (\ref{eq:dpkdtauzz_gE_6}) remain unchanged.

\section{On the stochastic contributions}
\label{Appendix:stochasticity}

In this Appendix, we consider the stochastic contributions to the biased tracer fields, and following the procedure in Sec.~\ref{sec:formalism} and \ref{sec:SSmodes_IA}, we present the extra terms in both the field-level expressions and power spectrum responses in redshift space, including the super-sample modes.

\subsection{Field-level responses}
\label{Appendix:stochasticity_field_responses}

Consider first the galaxy density field, and starting from the Lagrangian space, we derive the leading-order corrections involving the super-sample modes. Let us  denote the stochastic contributions in Lagrangian space by $\delta_{\rm g,\epsilon}^{\rm L}$. According to Ref.~\cite{Desjacques_Jeong_Schmidt2018}, the stochasticity valid at the second order in density produces two additional terms as 
\begin{align}
 \delta_{\rm g,\epsilon}^{\rm L}(\bfq)= \epsilon^{\rm L}(\bfq) + \epsilon_\delta^{\rm L}(\bfq)\,\deltalin(\bfq).
\label{eq:deltag_epsilon_Lag}
\end{align} 
Here, the field $\epsilon^{\rm L}$ and $\epsilon^{\rm L}_\delta$ are the stochastic fields in Lagrangian space, arising from the small-scale perturbations associated to the galaxy formation processes. Substituting Eq.~(\ref{eq:deltag_epsilon_Lag}) into the density field $\deltag^{\rm L}$ in Eq.~(\ref{eq:deltag_redshift}), the Eulerian stochastic contributions in redshift space become
\begin{align}
 \delta_{\rm g,\epsilon}^{\rm(S)}(\bfs)&\simeq \epsilon^{\rm L}(\bfs)
-\partial_{s_k}\Bigl\{\psi_k^{\rm S}(\bfs)\epsilon^{\rm L}(\bfs)\Bigr\} 
 + \deltalin(\bfs)\,\epsilon_\delta^{\rm L}(\bfs),  
\end{align}
which are valid at second order. Note that the quantity $\psi_k^{\rm S}$ is the redshift-space displacement field defined at Eq.~(\ref{eq:n-th_order_displacement_z-space}). We then apply the long- and short-mode decomposition to the above equation, as we described in Sec.~\ref{subsec:SSmodes_GG} (see Appendix \ref{sec:deltag_redshift_SS-mode} for explicit calculations in redshift space). In what follows, we suppose that the stochastic fields are made of only the short-mode contributions. Ignoring the higher-order terms of the short-mode density field $\deltas$ and stochastic fields, keeping the super-sample modes at linear order leads to 
\begin{align}
 \delta_{\rm g,\epsilon}^{\rm(S)}(\bfs)
&\simeq \Bigl\{1+f\,\Bigl( \frac{1}{3}\deltab+\tau_{zz}\Bigr)\Bigr\}\epsilon(\bfs)
+\deltab\,\epsilon_\delta(\bfs),
\label{eq:deltag_eps_leading} 
\end{align} 
where the stochastic fields $\epsilon$ and $\epsilon_\delta$ are those defined in Eulerian space, and are related to the Lagrangian counterparts through
\begin{align}
 \epsilon\equiv \epsilon^{\rm L}, \quad \epsilon_\delta\equiv \epsilon^{\rm L}_\delta+\epsilon^{\rm L}.  
\end{align}
In deriving Eq.~(\ref{eq:deltag_eps_leading}), we have assumed that the fields $\epsilon^{\rm L}$ and $\epsilon^{\rm L}_\delta$ are nearly constant over the scales larger than the scale below which the modes irrelevant to galaxy surveys are integrated out, and thus dropped those involving the spatial derivatives.

Next consider the stochastic contributions to the galaxy shapes, again starting from the Lagrangian space. We denote the Lagrangian stochasticity in the traceless part of the shape field by $\gamma_{ij}^\epsilon$. At the second order, it consists of the following three terms \cite{Vlah_Chisari_Schmidt2020}:
\begin{align}
 \gamma_{ij,\epsilon}^{\rm L}(\bfq)= \epsilon_{ij}^{\rm L}(\bfq) + \epsilon_{\delta,ij}^{\rm L}(\bfq)\,\deltalin(\bfq)+ \epsilon_{\rm K}^{\rm L}(\bfq)\,C_{ij}(\bfq),
\label{eq:gamma_ij_epsilon_Lag}
\end{align} 
where the Lagrangian stochastic fields $\epsilon_{ij}^{\rm L}$ and $\epsilon_{\delta,ij}^{\rm L}$ are the symmetric traceless tensor. Similarly to the density field, we substitute Eq.~(\ref{eq:gamma_ij_epsilon_Lag}) into the Lagrangian IA field $\gammaL_{ij}$ in Eq.~(\ref{eq:gamma_redshift_integral_form}). Then, up to the second order, the Eulerian stochastic contributions in redshift space, $\gamma_{ij,\epsilon}^{\rm(S)}$, become
\begin{align}
\gamma_{ij,\epsilon}^{\rm(S)}(\bfs) &
\simeq \epsilon_{ij}^{\rm L}(\bfs)-\partial_{s_k}\Bigl\{\psi_k^{\rm S}(\bfs)\epsilon_{ij}^{\rm L}(\bfs)\Bigr\} + \deltalin(\bfs)\,\epsilon_{\delta,ij}^{\rm L}(\bfs)+ C_{ij}(\bfs)\,\epsilon_{\rm K}^{\rm L}(\bfs).
\end{align}
Applying the long- and short-mode decomposition to the above, the leading-order stochastic contributions involving the super-sample modes at linear order become
\begin{align}
 \gamma_{ij,\epsilon}^{\rm(S)}(\bfs)  
 \simeq \Bigl\{1+f\,\Bigl( \frac{1}{3}\,\deltab+\tau_{zz}\Bigr)\Bigr\}\epsilon_{ij}(\bfs)
 + \deltab\,\epsilon_{\delta,ij}(\bfs)+ \tau_{ij}\,\epsilon_{\rm K}(\bfs),
\label{eq:gamma_ij_eps_leading} 
\end{align}
where the Eulerian stochastic fields $\epsilon_{ij}$, $\epsilon_{\delta,ij}$ and $\epsilon_{\rm K}$ are related to the Lagrangian counterparts as follows:
\begin{align}
 \epsilon_{ij}\equiv \epsilon_{ij}^{\rm L}, \quad \epsilon_{\delta,ij}\equiv \epsilon_{\delta,ij}^{\rm L}+\epsilon_{ij}^{\rm L}, \quad
\epsilon_{\rm K} \equiv \epsilon_{\rm K}^{\rm L}.
\end{align}
Again, we have dropped the stochastic fields involving the spatial derivatives. From Eq.~(\ref{eq:E-/B-mode_decomposition}), the leading-order contributions to the E-/B-mode ellipticity fields are also computed in Fourier space, and we obtain 
\begin{align}
\Biggl(
\begin{array}{c}
\gamma_{\rm E,\epsilon}^{\rm (S)} \\
\gamma_{\rm B,\epsilon}^{\rm (S)} 
\end{array} 
\Biggr)(\bfk) &=\,\mathbf{R}(\phi_k)\,\Biggl(
\begin{array}{c}
\gamma_{xx,\epsilon}^{\rm(S)}-\gamma_{yy,\epsilon}^{\rm(S)} \\
2\gamma_{xy,\epsilon}^{\rm(S)}
\end{array} 
\Biggr)(\bfk)
\label{eq:E-/B-mode_stocasticity}
\end{align}
with the matrix $\mathbf{R}(\phi_k)$ given by Eq.~(\ref{eq:rotation_matrix}).

Eqs.~(\ref{eq:deltag_eps_leading}) and (\ref{eq:gamma_ij_eps_leading}) or (\ref{eq:E-/B-mode_stocasticity}) represent the stochastic contributions to the density and IA fields involving the super-sample modes at the field level. At the leading order, they are just added to the deterministic part presented in the main text. In general, there might be other stochastic contributions arising from the matter density field. Also, we may consider the higher-derivative corrections which appear as the deterministic bias at higher order (e.g., Refs.~\cite{Desjacques2008,Lazeyras_Schmidt2019}). Although these corrections produce additional terms involving the super-sample modes, all of them are given to be additive contributions as long as the leading-order contributions are concerned. 

\subsection{Power spectrum responses}
\label{subsec:stochasticity_power_spectrum_responses}

Provided the filed-level expressions in previous subsection, we now compute the power spectrum responses to the super-sample modes. Since the stochastic fields are all uncorrelated with LSS (i.e., $\deltas$ in our notation), they do not produce any extra terms at leading order coupled with the deterministic bias part. To derive explicitly the stochastic contributions, we need to specify the statistical properties of the stochastic fields. As we mentioned above, the stochasticity is supposed to be determined by the local processes, and do not possess any scale- and directional-dependence involving the wave vector, $\bfk$. We thus impose the following conditions (see Ref.~\cite{Vlah_Chisari_Schmidt2020}):
\begin{align}
&    \langle X(\bfk)Y(\bfk')\rangle=(2\pi)^3\,\delta_{\rm D}(\bfk+\bfk')\,P_{\rm X,Y},
\label{eq:noise_spectra1}
\\
&    \langle X_{ij}(\bfk)Y_{k\ell}(\bfk')\rangle=(2\pi)^3\,\delta_{\rm D}(\bfk+\bfk')\,\bigl(\deltaK_{ik}\deltaK_{j\ell}+\deltaK_{i\ell}\deltaK_{jk}-\frac{2}{3}\deltaK_{ij}\deltaK_{k\ell}\bigr)\,P_{\rm X,Y}^\gamma,
\label{eq:noise_spectra2}
\\
&    \langle X_{ij}(\bfk)Y(\bfk')\rangle=0.
\label{eq:noise_spectra3}
\end{align}
Here, the scalar quantities $X$, $Y$ stand for either $\epsilon$, $\epsilon_\delta$ or $\epsilon_{\rm K}$. On the other hand, tensor fields $X_{ij}$ and $Y_{ij}$ imply either $\epsilon_{ij}$ or $\epsilon_{\delta,ij}$. As it has been advocated in Ref.~\cite{Desjacques_Jeong_Schmidt2018}, the stochastic spectra $P_{X,Y}$ and $P_{X,Y}^\gamma$ are, on large scales, constant, but taking into account the finite size of galaxies and formation processes that we integrate out, they would have a moderate scale-dependence characterized by a series expansion in $k^2$ (see also Refs.~\cite{Schmittfull_Simonovic_Zaldarriaga2019,Schmidt_Elsner_Jasche_Lavaux2019,Cabass_Schmidt2020}).

Using the field-level expressions and the statistical properties in Eqs.~(\ref{eq:noise_spectra1})-(\ref{eq:noise_spectra3}), 
the power spectrum responses arising from the stochasticity can be separately computed, and we obtain the following contributions on top of the results derived in the main text: 
\begin{align}
 \frac{\partial P_{\rm gg}^{\rm (S)}(\bfk)}{\partial \deltab} &\longrightarrow 
 \frac{\partial P_{\rm gg}^{\rm (S)}(\bfk)}{\partial \deltab} +\frac{2}{3}\,f\,P_{\epsilon,\epsilon}+2\,P_{\epsilon,\epsilon_\delta},
\label{eq:dpkgg_ddeltab_stochasticity}
\\
 \frac{\partial P_{\rm gg}^{\rm (S)}(\bfk)}{\partial \tau_{zz}} &\longrightarrow 
 \frac{\partial P_{\rm gg}^{\rm (S)}(\bfk)}{\partial \tau_{zz}} +2f\,P_{\epsilon,\epsilon}
\label{eq:dpkgg_dtauzz_stochasticity}
\end{align}
for the galaxy density power spectrum, 
\begin{align}
 \frac{\partial \overline{P}_{\rm EE}^{\rm (S)}(\bfk)}{\partial \deltab} 
&\longrightarrow 
 \frac{\partial \overline{P}_{\rm EE}^{\rm (S)}(\bfk)}{\partial \deltab} +8\,P_{\epsilon,\epsilon_\delta}^\gamma,
\label{eq:dpkEE_ddeltab_stochasticity}
\\
 \frac{\partial \overline{P}_{\rm EE}^{\rm (S)}(\bfk)}{\partial \tau_{zz}} 
&\longrightarrow 
 \frac{\partial \overline{P}_{\rm EE}^{\rm (S)}(\bfk)}{\partial \tau_{zz}} +8f\,P_{\epsilon,\epsilon}^\gamma
\label{eq:dpkEE_dtauzz_stochasticity}
\end{align}
for the angle-averaged E-mode auto power spectrum. For the gE cross power spectrum, taking the angle-average at Eq.~(\ref{eq:def_averaged_pkred}), there is no leading-order contribution involving the stochasticity. This is also the case for gB and EB cross spectra. However, for the B-mode auto power spectrum response, the stochasticity induces non-vanishing contribution and we have  
\begin{align}
 \frac{\partial \overline{P}_{\rm BB}^{\rm (S)}(\bfk)}{\partial \deltab} 
=
8\,P_{\epsilon,\epsilon_\delta}^\gamma,
\quad
 \frac{\partial \overline{P}_{\rm BB}^{\rm (S)}(\bfk)}{\partial \tau_{zz}} 
= 
8f\,P_{\epsilon,\epsilon}^\gamma.
\label{eq:dpkBB_stochasticity}
\end{align}

In the above, the power spectrum responses of the stochastic fields all appear as additive contributions. That is, all the terms arising from the stochasticity are expressed separately from the contribution computed in the main text, which is encapsulated with the first term at the right hand side. Note cautiously that the responses to the super-sample modes arise from the mode coupling between long and short modes. As a consequence, even at the leading order, the resultant responses given above show non-trivial dependence on the power spectra of stochastic fields, which is expressed in terms of not only the auto power spectra, but also the cross power spectra of stochastic fields like $P_{\epsilon,\epsilon_\delta}$ and $P_{\epsilon,\epsilon_\delta}^\gamma$.

\bibliography{ref}
\bibliographystyle{apsrev4-1}

\end{document}